\documentclass[preprint,superscriptaddress,preprintnumbers, showpacs, floatfix]{revtex4-1}

\usepackage{ifthen} 
\usepackage{graphicx}
\usepackage{grffile}
\usepackage{amsmath}
\usepackage{braket}
\usepackage{cancel}
\usepackage{xr-hyper} 
\usepackage{hyperref}
\hypersetup{backref,pdfpagemode=FullScreen,colorlinks=true,allcolors=blue}
\usepackage{color}
\usepackage{gensymb}
\usepackage{subfig}
\usepackage{ragged2e}
\usepackage{soul}
\usepackage{amssymb}

\def\bl#1{\textcolor{blue}{#1}}

\def\edit#1{\textcolor{black}{#1}}

\begin{document}

\title{Strain-induced Weyl and Dirac \edit{states} and direct-indirect gap transitions in group-V materials}

\author{Glenn Moynihan}
\email{omuinneg@tcd.ie}
\affiliation{School of Physics, CRANN and AMBER, 
Trinity College Dublin, Dublin 2, Ireland}

\author{Stefano Sanvito}
\affiliation{School of Physics, CRANN and AMBER, 
Trinity College Dublin, Dublin 2, Ireland}

\author{David D. O'Regan}
\affiliation{School of Physics, CRANN and AMBER, 
Trinity College Dublin, Dublin 2, Ireland}

\date{\today{}}

\begin{abstract}
We perform comprehensive 
density-functional theory calculations on 
strained two-dimensional 
phosphorus (P), arsenic (As) and antimony (Sb) 
in the monolayer, bilayer, and bulk  $\alpha$-phase, 
from which we compute the key 
mechanical and electronic properties 
of these materials.  
Specifically, we compute their electronic band structures, 
band gaps, and charge-carrier effective masses, 
and identify the qualitative electronic 
and structural transitions that may occur.
Moreover, 
we compute the elastic properties 
such as the Young's modulus $Y$; 
shear modulus $G$; 
bulk modulus $\mathcal{B}$; 
and Poisson ratio $\nu$ 
and present their isotropic averages of
as well as their dependence on the in-plane orientation, 
for which the relevant expressions are derived.
We predict strain-induced  
\edit{Dirac states} in the  
monolayers of As and Sb 
and  
the bilayers of P, As, and Sb, 
as well as the \edit{possible existence of Weyl states} 
in the bulk phases of P and As.
\edit{These phases are predicted to support charge velocities 
up to $10^6$~$\textrm{ms}^{-1}$ 
and, in some highly anisotropic cases, 
permit one-dimensional ballistic conductivity 
in the puckered direction.}
\edit{We also predict} 
numerous band gap transitions 
for moderate in-plane stresses.
Our results contribute to the 
mounting evidence for the 
utility of these materials, 
made possible by their broad range in tuneable properties, 
and facilitate the directed exploration of their 
potential application in next-generation electronics.
\end{abstract}


\maketitle

\section{Background}
Two-dimensional black phosphorus (BP), 
or {\it phosphorene}, 
is one of several predicted stable allotropes of 
few-layer phosphorus~\cite{doi:10.1021/acs.nanolett.5b01041,PhysRevLett.113.046804,PhysRevLett.112.176802,doi:10.1021/nn5059248}, 
and it has attracted considerable attention since its recent 
successful synthesis~\cite{Li2014,Xia2014,doi:10.1063/1.4868132,doi:10.1021/nn501226z,Ling14042015} 
that is now possible with liquid phase exfoliation~\cite{C4CC05752J,PMID:26469634}.
The excitement behind BP 
is driven by its growing list of 
technologically relevant anisotropic 
mechanical and electronic properties. 
The theoretically predicted  
properties include 
a tuneable band-gap~\cite{PhysRevLett.112.176801,PhysRevB.90.085402,PhysRevB.91.235118,doi:10.1021/jp508618t,PhysRevB.91.115412,0957-4484-25-45-455703,doi:10.1021/nn501226z}, 
a negative Poisson's ratio~\cite{Jiang2014}, 
anisotropic conduction~\cite{doi:10.1021/nl500935z,C4CS00257A}, 
and linear dichroism~\cite{Qiao2014,PhysRevB.89.235319}.
The properties that have been 
experimentally verified so far include 
a high hole-mobility 
between $300-1000$~cm\textsuperscript{2}/Vs~\cite{doi:10.1021/nn501226z,PhysRev.92.580,doi:10.1063/1.4868132,Qiao2014}, 
considerable mechanical flexibility~\cite{doi:10.1063/1.4885215}, 
and a layer-dependent band gap~\cite{9909120720141024,PhysRevB.89.235319,Zant2014} ranging from 0.3~eV in bulk to 2.0~eV in the monolayer.
The anisotropic crystal structure 
of BP is responsible for its unusual 
electro-mechanical properties, 
which are predicted to be strongly directional-dependent 
and highly responsive to 
mechanically strain~\cite{doi:10.1021/nl500935z,PMID:26469634}. 

\begin{figure}[!htb]
\begin{tabular}{cc}
\begin{tabular}{c}
\begin{subfloat}[]{
\centering
\includegraphics[width=0.505\textwidth]{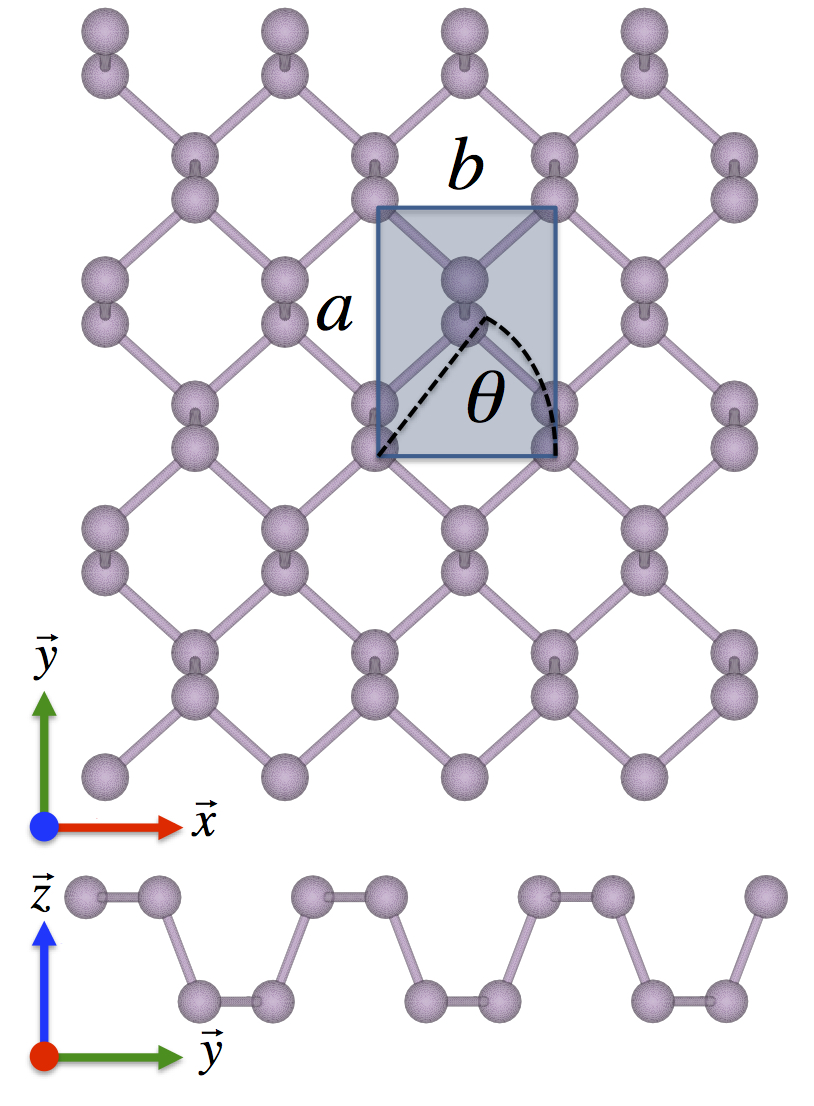}
  \label{fig:crystalstructure}}
\end{subfloat} 
\end{tabular}
&
\begin{tabular}{c}
\smallskip
\begin{subfloat}[]{
\centering
\includegraphics[width=0.40\textwidth]{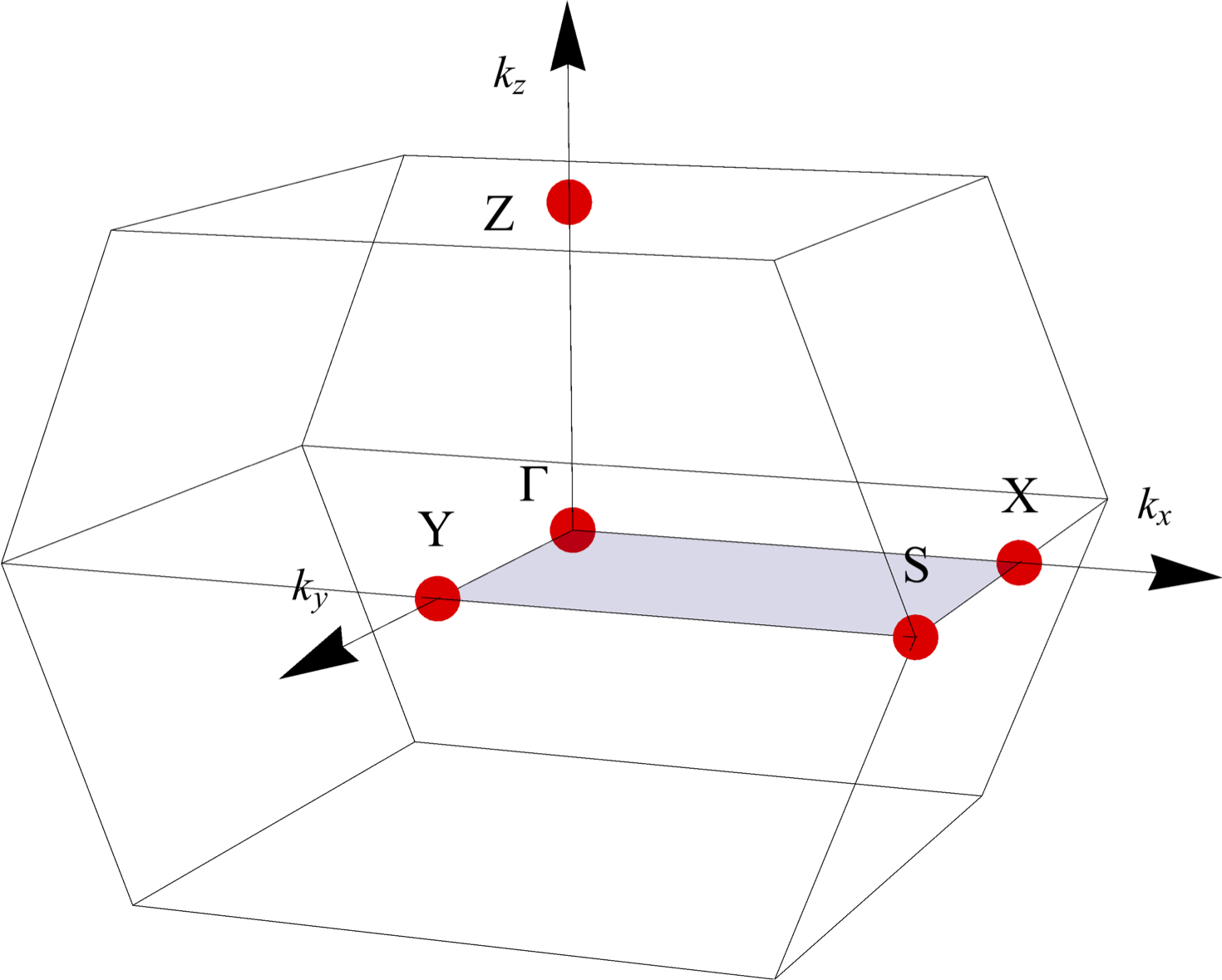}
 \label{fig:bz}}
\end{subfloat}
\\
\begin{subfloat}[]{
\centering
\includegraphics[width=0.45\textwidth]{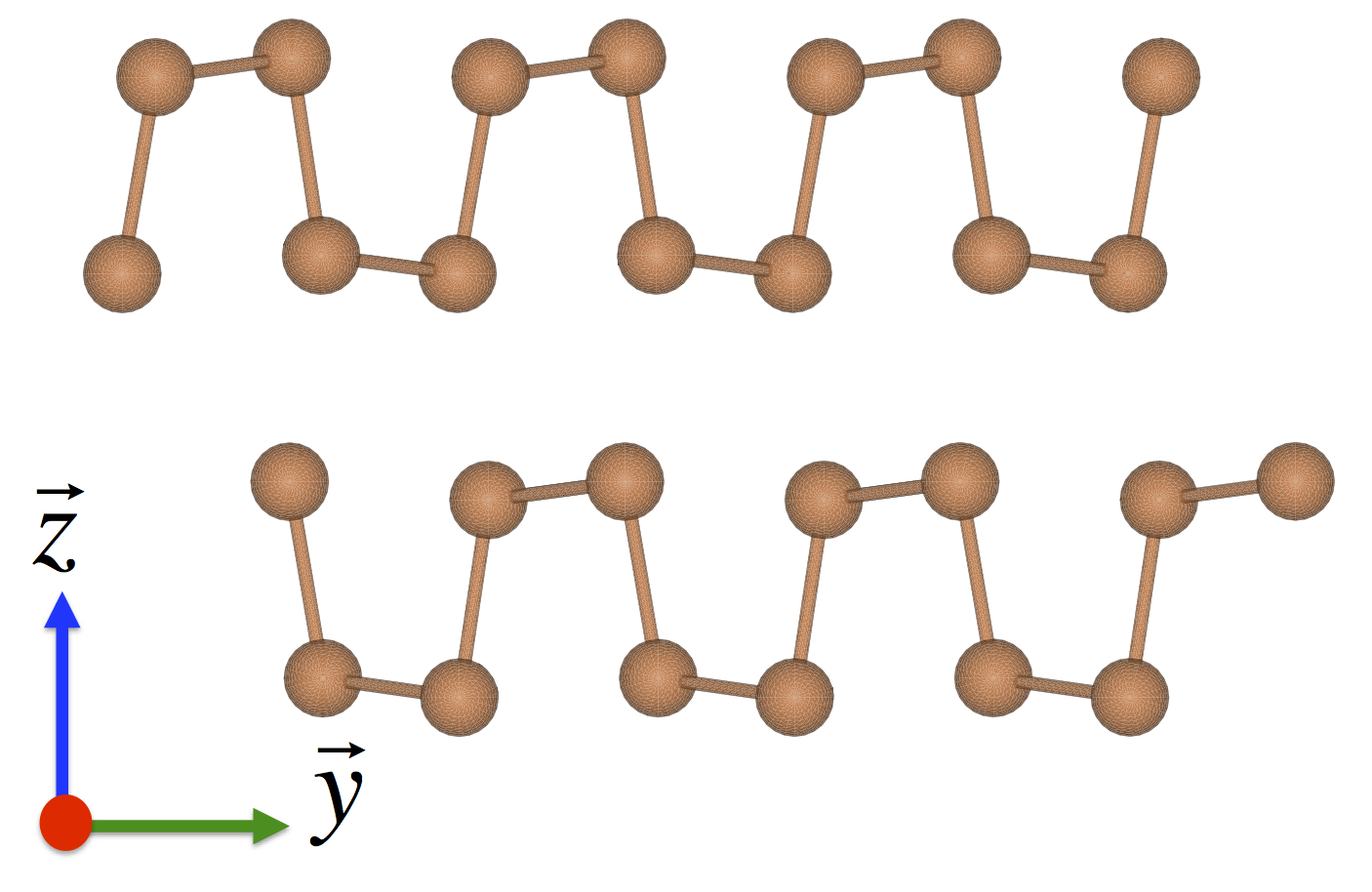}
\label{fig:buckled}}
\end{subfloat}
\end{tabular}\\
\end{tabular}
\caption{(Color online) 
a) Top and side view of the 
2D orthorhombic puckered structure 
(generated in VESTA~\cite{Momma:db5098}) 
with the primary vectors along the 
zigzag ($\vec x$) and puckered ($\vec y$) directions shown. 
The unit cell, 
given by the shaded region, 
is described by the lattice parameters $a$ and $b$ 
with the in-plane angle $\theta$ also defined.
b) 3D Brillouin zone with high-symmetry points 
$\Gamma$, $X$, $S$, $Y$, \edit{and $Z$}.
c) Side-view of the buckled Sb state  
at $\varepsilon_{yy}=-4\%$ compressive strain.
}
\label{fig:phos_structure}
\end{figure}

With this renewed interest in BP, 
focus has quickly turned to 
few-layer phases of the other pnictogens, namely 
arsenic~\cite{PhysRevB.91.085423} (As), 
antimony~\cite{doi:10.1021/acsami.5b02441} (Sb), 
bismuth~\cite{PhysRevB.91.075429,doi:10.1021/nl502997v} (Bi), 
and their alloys~\cite{doi:10.1021/acs.nanolett.5b02227,doi:10.1021/acs.jpcc.5b07323,Xie2016433,ADMA:ADMA201501758,doi:10.1021/acs.jpcc.5b02096}, 
which are attracting steadily increasing attention.
Moreover, the search for a bulk counterpart to graphene, 
capable of supporting ballistic electron transport, 
has recently driven the search for new 
Weyl and Dirac semi-metals~\cite{PhysRevLett.102.166803,Shekhar2015,PhysRevB.83.205101,NIELSEN1983389,turner2013beyond,PhysRevX.4.031035,PhysRevLett.108.140405,PhysRevB.85.195320,PhysRevB.88.125427,C3CP53257G,doi:10.1093/nsr/nwu080}.
\edit{In spite of the difficulty 
in attaining Dirac points 
in two-dimensional materials~\cite{C3CP53257G,doi:10.1093/nsr/nwu080}, 
the first experimental Dirac semi-metal in few-layer BP 
was observed by Kim \emph{et al.}~\cite{Kim723}, 
while As~\cite{doi:10.1063/1.4943548}, 
Sb~\cite{Yao2013,Lu2016,Zhao2015}, Bi~\cite{Lu2016}, 
and P~\cite{Lu2016,PhysRevB.93.245303,Zhao2015,2053-1583-4-2-025071} 
are also 
predicted to be potential candidates.
}
Indeed, many of the predicted strain-induced properties 
of these materials, such as 
direct-indirect band gap transitions~\cite{ANIE:ANIE201411246,PhysRevB.91.085423,doi:10.1021/acsami.5b02441,PhysRevB.90.085402,0957-4484-26-7-075701},
a negative Poisson's ratio~\cite{Jiang2014,1882-0786-8-4-041801},
as well as 
electronic~\cite{PhysRevLett.113.046804,PhysRevB.91.161404,ANIE:ANIE201411246}, 
structural~\cite{PhysRevB.92.064114}, 
and 
topological~\cite{Lu2016,PhysRevLett.115.186403,PhysRevB.93.195434,PhysRevB.91.195319} transitions, 
are already spurring their incorporation in emergent technologies 
such as 
field-effect transistors~\cite{doi:10.1021/nn501226z,C4CS00257A},
gas-sensors~\cite{doi:10.1021/acsnano.5b01961,doi:10.1021/jz501188k}, 
optical switches~\cite{PhysRevB.90.075434,1347-4065-28-11A-2104}, 
solar-cells~\cite{Xie2016433},
next-generation batteries~\cite{ADMA:ADMA200602592,doi:10.1021/jp302265n,Sun2015},
 reinforcing fillers~\cite{PMID:26469634,doi:10.1021/nl501617j}, 
 and topological insulators~\cite{PhysRevLett.115.186403,Lu2016,PhysRevB.93.195434,PhysRevB.91.195319,Yao2013}.

In this work, our aim is to provide 
a comprehensive analysis of the 
monolayer, bilayer and bulk phases of 
orthorhombic P, As and Sb, 
in order to 
identify and compare the qualitative 
strain-related properties of each structure 
from a consistent set of calculations,  
\edit{thus treating each 
material on the same footing}.
Our findings 
provide new insights into their 
electro-mechanical properties,  
especially regarding arsenic and antimony, 
which have been relatively unexplored to date.
Specifically, we identify  
qualitative transitions in band gaps, 
effective masses, 
structure, and topology  
that occur at various strains, 
and compute the elastic properties 
that determine the required stresses 
to attain these electronic states.

We predict the existence of 
strain-induced Dirac \edit{states}   
in monolayer As and Sb, 
bilayer P, As and Sb, 
as well as 
possible Weyl \edit{states} in bulk P and As, 
at moderate stress values.
\edit{
Our findings show that all of the predicted 
Dirac and Weyl points are indeed linear, 
at least in the $\Gamma-Y$ direction, 
i.e. the puckered direction.
Thus, following the convention of terminology 
found in the Refs.~\cite{doi:10.1021/acs.nanolett.5b04106,Kim723,Shekhar2015}, and other sources, 
we classify Dirac or Weyl states 
as those associated with regions of sustained linear dispersion 
in the band structure, 
at or near the Fermi level, in at least one direction.}
\edit{We predict these states 
to support ballistic conduction 
and are unaffected by the spin-orbit coupling (SOC).
In particular, few-layer P and As 
exhibit a strong indication 
of anisotropic conduction, 
dominated by ballistic conductivity 
along $\Gamma-Y$.}

The outline of this article is as follows: 
In Section~\ref{sec:methodology} we 
review the details of our calculations and discuss 
the Voigt-Reuss-Hill averaging scheme 
used to compute the 
experimentally-relevant elastic properties.
In Section~\ref{sec:results} we present 
the results of our calculations 
including the lattice constants, 
strain dependence 
of electronic properties, 
in particular the potential Dirac and Weyl \edit{states} identified, 
and computed elastic properties. 
We conclude in 
Section~\ref{sec:conclusion} 
with a discussion of  
the implications of our key results.

\section{Methodology}
\label{sec:methodology}
\subsection{Calculation details}
The calculations were performed with the 
{\sc QuantumEspresso} package~\cite{0953-8984-21-39-395502}
using the Perdew-Burke-Ernzerhof (PBE)  form of 
the generalized-gradient approximation (GGA) 
exchange-correlation functional~\cite{PhysRevLett.77.3865}. 
An ultrasoft pseudopotential~\cite{PhysRevB.41.7892}
from the SSSP Library~\cite{kucukbenli2014projector} 
(with 5 valence electrons) was used 
to represent the core electrons.
\edit{
Non-SOC calculations were initially performed 
and those that exhibited potential Dirac or Weyl states 
were reassessed including non-perturbative SOC.}
In all calculations, van der Waals (vdW) interactions 
were incorporated using the B97-D empirical 
dispersion correction functional~\cite{JCC:JCC20495}. 
In order to achieve 
an energy convergence of at least 1~meV/atom 
and force convergence of at least $1.3\times 10^{-4}$~eV/a$_0$, 
we found it sufficient to use a common 
plane-wave energy cutoff of 1100~eV
with `cold' smearing~\cite{PhysRevLett.82.3296} 
of $10^{-4}$~K for all elements.
To achieve the same convergence,  
the Brillouin zone sampling for bulk systems
was $15\times15\times15$, 
and $15\times15\times1$ 
for monolayers and bilayers.
Uniaxial and shear strains 
between $\pm$5\% were applied 
in increments of 1\% to the unit cell 
with internal relaxation subject 
to the same force convergence 
criterion as above.
Electronic band structures 
were calculated along the high-symmetry points 
of the Brillouin zone 
\edit{\{$\Gamma$, $X$, $S$, $Y$,$Z$\}} (Fig.~\ref{fig:bz}) 
for each value of in-plane strain 
 (Figs.~\bl{S1}~-~\bl{S9}
in the 
\href{http://iopscience.iop.org/2053-1583/4/4/045018/media/Supplementary_Information.pdf}{Supplemental Material}).
For shear strains the Brillouin zone 
deforms into an asymmetric honeycomb, 
yet we continued to sample along the original  
path since the deformation 
up to 5\% strain is negligible 
and the effective masses are all
calculated at the $\Gamma$-point.
We determine 
the Kohn-Sham band gap 
from the band structures 
and charge-carrier effective masses 
according to the nearly-free electron model 
$m_{ij}^\star=\hbar^2\left(\partial^2E/\partial k_i\partial k_j \right)^{-1}$ 
using a cubic spline fit to 
9 data points about the $\Gamma$-point.
\edit{
The charge velocities were similarly 
determined according to the dispersion relation 
$v=\hbar^{-1}dE(k)/dk$ 
from the linear fit to the Dirac or Weyl lines.}
The elements of the stiffness matrix $C$ 
were derived from the gradients of the 
resultant stress-strain profiles 
$c_{ij} = \partial \sigma_i/\partial\varepsilon_j$, 
from which all elastic properties were derived.
In practice, however, 
the computed stiffness tensors are not exactly symmetric 
due to numerical noise 
but we make them so by taking 
the average of $C$ and its transpose $C^T$ 
as the \emph{effective} stiffness tensor.
We begin our discussion with 
a brief overview of the the 
Voigt-Reuss-Hill scheme, 
which is a popular model 
used for computing effective isotropic 
elastic properties.  

\subsection{The Voigt-Reuss-Hill scheme}
In order to effectively preserve, study and strain-engineer
few-layer nano-structures, 
such as BP~\cite{2053-1583-4-2-021032}, 
graphene~\cite{R.2010}, 
or molybdenum disulfide~\cite{doi:10.1021/nl2043612} (MoS\textsubscript{2}), 
the nano-flakes are typically 
deposited onto a suitable substrate.
The cumulative contribution of 
dispersed nano-flakes distributed 
on or within a bulk medium results 
in the macroscopic elastic properties 
that are measured by experiments.
The theoretical calculation of 
these elastic properties 
requires an appropriate mixture model 
(such as the rule-of-mixtures~\cite{PMID:26469634} (ROM) 
or the Halpin-Tsai~\cite{PEN:PEN760160512} models)
that require the (typically averaged)
elastic properties of the interstitial nano-flakes.

In the theory of effective media, 
isotropic bulk properties are 
computed by averaging the stiffness tensor $C$ 
over all possible rotated reference frames~\cite{MULLEN19972247,hearmon1969elastic,0965-0393-7-6-301}, 
as outlined in the Supplemental Material.
The result is called the Voigt average~\cite{ZAMM:ZAMM19290090104,cook1999advanced} 
and it gives isotropic averages for the bulk 
Young's modulus $Y_V$, 
and shear modulus $G_V$, 
given in Eq.~\href{http://iopscience.iop.org/2053-1583/4/4/045018/media/Supplementary_Information.pdf}{\bl{S7}}.
The same scheme applied 
to the compliance tensor $S$ results in 
the corresponding Reuss averages~\cite{ANDP:ANDP18892741206}, 
$Y_R$ and $G_R$, 
given in Eq.~\href{http://iopscience.iop.org/2053-1583/4/4/045018/media/Supplementary_Information.pdf}{\bl{S9}}.
The Voigt scheme assumes that 
the material undergoes constant strain 
and it returns over-estimated elastic constants.
Conversely, 
the Reuss scheme assumes constant stress 
and it tends to under-estimate the elastic constants.
The Hill averages~\cite{0370-1298-65-5-307} 
\begin{equation}
Y_{H}=\frac{Y_V+Y_R}{2},
\quad\mbox{and}\quad
G_{H}=\frac{G_V+G_R}{2},
\label{eq:hillaverage}
\end{equation}
from which the isotropic Poisson's ratio $\nu_H$ and 
bulk modulus  $\mathcal{B}_H$ are expressed as 
\begin{equation}
\nu_H=\frac{Y_H}{2G_H}-1, 
\quad\mbox{and}\quad
\mathcal{B}_H=\frac{Y_HG_H}{3(3G_H-Y_H)}.
\label{eq:poisson_ratio}
\end{equation}
These are widely considered as reliable estimates of 
the actual physical values~\cite{0965-0393-7-6-301}.
The Voigt-Reuss-Hill approach described above 
is used in the present work to 
determine the isotropic averages of the 
Young's, shear and bulk moduli, 
and the Poisson ratio  
of the bulk P, As, and Sb structures using 
the elements of the elastic tensors.
Let us now discuss how the above 
approach may be adapted 
to derive the relevant equations 
for the specific case 
of two-dimensional materials.

\subsection{In-plane Voigt-Reuss-Hill average}
If the interstitial nano-flakes in a bulk medium 
form high-quality planar sediments~\cite{PMID:26469634,PMID:23203296,Mahmoud20111534,ADFM:ADFM200801776,APP:APP38645,Young20121459,Yang2013,C4NR01208A}, 
the random orientation occurs instead 
in the plane of the flakes 
and we must calculate 
isotropic-averages {\it in-plane}. 
Due to the weak vdW bonds 
between layers, 
strains related to out-of-plane directions 
can be ignored 
resulting in the reduced-stiffness tensor 
(Eq.~\href{http://iopscience.iop.org/2053-1583/4/4/045018/media/Supplementary_Information.pdf}{\bl{S3}}).
In the Supplemental Material, 
we re-derive the angular dependence  
of the rotated tensor-elements 
$C_{ij}\left(\theta\right)$ and $S_{ij}\left(\theta\right)$
about the $\vec{z}$-axis 
(Eq.~\href{http://iopscience.iop.org/2053-1583/4/4/045018/media/Supplementary_Information.pdf}{\bl{S6}}) 
as a function of the elements in the original 
reference frame, 
similar to the general Voigt-Reuss scheme.
The angular-dependence of the in-plane  
elastic constants are then expressed as    
\begin{equation}
\begin{aligned}
&Y_{V}\left(\theta\right)=\frac{C_{11}^2-C_{12}^2}{C_{11}},
\qquad
&G_{V}\left(\theta\right)=C_{66},
\qquad\qquad
&\nu_{V}\left(\theta\right)=\frac{C_{12}}{C_{11}},
\\
&Y_{R}\left(\theta\right)=\frac{1}{S_{11}},
\qquad
&G_{R}\left(\theta\right)=\frac{1}{S_{66}},
\qquad\qquad
&\nu_{R}\left(\theta\right)=-\frac{S_{12}}{S_{11}},
\end{aligned}
\label{eq:elasticconsts2d}
\end{equation}
with the Hill-average taken as in Eq.~\ref{eq:hillaverage}.
By integrating the elastic tensors 
$C_{ij}\left(\theta\right)$ and $S_{ij}\left(\theta\right)$ 
over $2\pi$, 
the in-plane averages are then 
computed analogously.
%

\section{results}
\label{sec:results}
\subsection{Lattice Constants}
The lattice constants $a$, $b$, $c$ 
of the fully-relaxed structures 
are presented in Table~\ref{table:lattice_consts}, 
where, in the monolayer and bilayer cases,  
we quote the layer thickness $c^\prime$ 
instead of the unit cell height $c$.
Our computed lattice parameters  compare well 
with other recent theoretically predicted values~\cite{PhysRevB.92.064114,PhysRevB.86.035105,PhysRevB.94.205410,PhysRevB.91.085423,1882-0786-8-5-055201,doi:10.1021/acsami.5b02441,ANDP:ANDP201600152,Lu2016}
and the available experimental data~\cite{doi:10.1063/1.438523,doi:10.1080/14786437508229285}.
\begin{table}[h!]
\begin{ruledtabular}
\begin{tabular}{lllll}
&$a$ (\AA) &$b$ (\AA)&$c$ (\AA)\\
\hline
P\textsubscript{mono}	&4.57	&3.31	&2.11\\
P\textsubscript{bi}		&4.51	&3.31 	&7.34\\
P\textsubscript{bulk} 		&4.43 (4.37~\footnote{\label{f1:lattice_consts}Ref.~\cite{doi:10.1063/1.438523}})	&3.32 (3.31~\textsuperscript{\ref{f1:lattice_consts}})	&10.47 (10.47~\textsuperscript{\ref{f1:lattice_consts}})\\
\hline
As\textsubscript{mono}	&4.70	&3.67	&2.39\\
As\textsubscript{bi}		&4.64	&3.69	&7.86\\
As\textsubscript{bulk} 	&4.56 (4.47~\footnote{\label{f2:lattice_consts}Ref.~\cite{doi:10.1080/14786437508229285}})	&3.71 (3.65~\textsuperscript{\ref{f2:lattice_consts}})	&10.94 (11.0~\textsuperscript{\ref{f2:lattice_consts}})\\
\hline
Sb\textsubscript{mono}	&5.02	&4.23	&2.79\\
Sb\textsubscript{bi}		&4.88	&4.26	&8.83&\\
Sb\textsubscript{bulk} 	&4.73	&4.29 (4.3~\footnote{\label{f3:lattice_consts}Ref.~\cite{Barrett:a03833}}) 	&12.09 (11.2~\textsuperscript{\ref{f3:lattice_consts}})
\end{tabular}
\end{ruledtabular}
\caption{
Lattice parameters (\AA)  for monolayer, bilayer and bulk structures of P, As, Sb 
compared to experimental data~\cite{doi:10.1063/1.438523,doi:10.1080/14786437508229285,Barrett:a03833} quoted in parentheses. 
For the monolayers and bilayers the layers thickness $c^\prime$ is given.
}
\label{table:lattice_consts}
\end{table}
For a given element, 
we find that the lattice parameter 
along the puckered direction, `$a$', shortens 
as the number of layers increases, 
which agrees with observations in other studies.
This is attributed to the increased 
vdW forces between layers, 
which leads to increased binding primarily 
in the softer puckered direction.
%

\subsection{Electronic properties}
All of the band structures 
pertaining to the following analysis 
are presented in 
Figs.~\href{http://iopscience.iop.org/2053-1583/4/4/045018/media/Supplementary_Information.pdf}{\bl{S1}}~-~\href{http://iopscience.iop.org/2053-1583/4/4/045018/media/Supplementary_Information.pdf}{\bl{S9}} 
of the Supplemental Material.
\edit{
Where we identify possible Dirac or Weyl states, 
high resolution, three-dimensional band structures 
with SOC at representative strains are recalculated.
To confirm the existence of linear-dispersion, 
we also plot lines along the surface of the Dirac and Weyl points  
at $0^\circ$, $30^\circ$ , $60^\circ$  and $90^\circ$ 
with respect to the $\Gamma-X$ line.
A representative sample of these results are presented in 
Figs.~\ref{fig:pbixxdiracbs}~-~\ref{fig:sbmonoxxdiracbs}, 
while the rest can be found in 
Figs.~\href{http://iopscience.iop.org/2053-1583/4/4/045018/media/Supplementary_Information.pdf}{\bl{S10}}~-~\href{http://iopscience.iop.org/2053-1583/4/4/045018/media/Supplementary_Information.pdf}{\bl{S13}} 
of the Supplemental Material.
}

In general we find the band gap to be very  
responsive to uniaxial in-plane 
strain but significantly less so 
with respect to shear strain.
We identify several direct-indirect band gap transitions 
as well as the opening and closing of band gaps, 
summarized in Table~\ref{table:transitions}.
We find the charge-carrier effective masses 
vary approximately linearly 
with respect to the uniaxial strain in general 
but with notable exceptions that will be discussed.
This section is divided 
into three parts discussing 
each of the species - 
P, As and Sb - 
for which we review  
the qualitative calculation results 
including band gap transitions, 
effective mass behavior 
and linearly-dispersive bands.

\subsubsection{Phosphorus}
\begin{figure*}
\begin{subfloat}[Monolayer P band gap (eV)]{
  \centering
\includegraphics[width=0.3\textwidth]{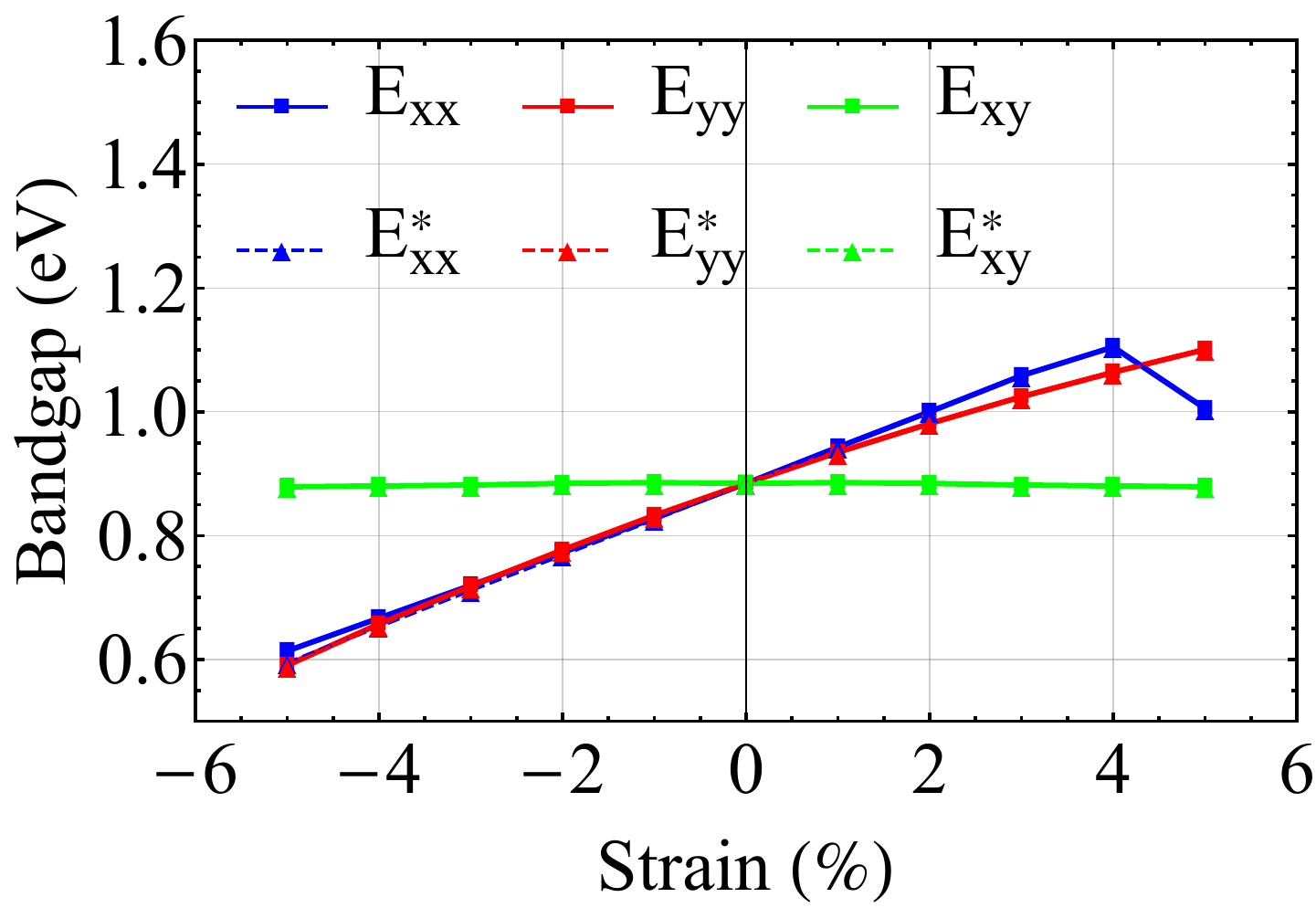}
  \label{fig:pmonobg}}
\end{subfloat}
\begin{subfloat}[Monolayer P $m_e/m_0$]{
  \centering
\includegraphics[width=0.3\textwidth]{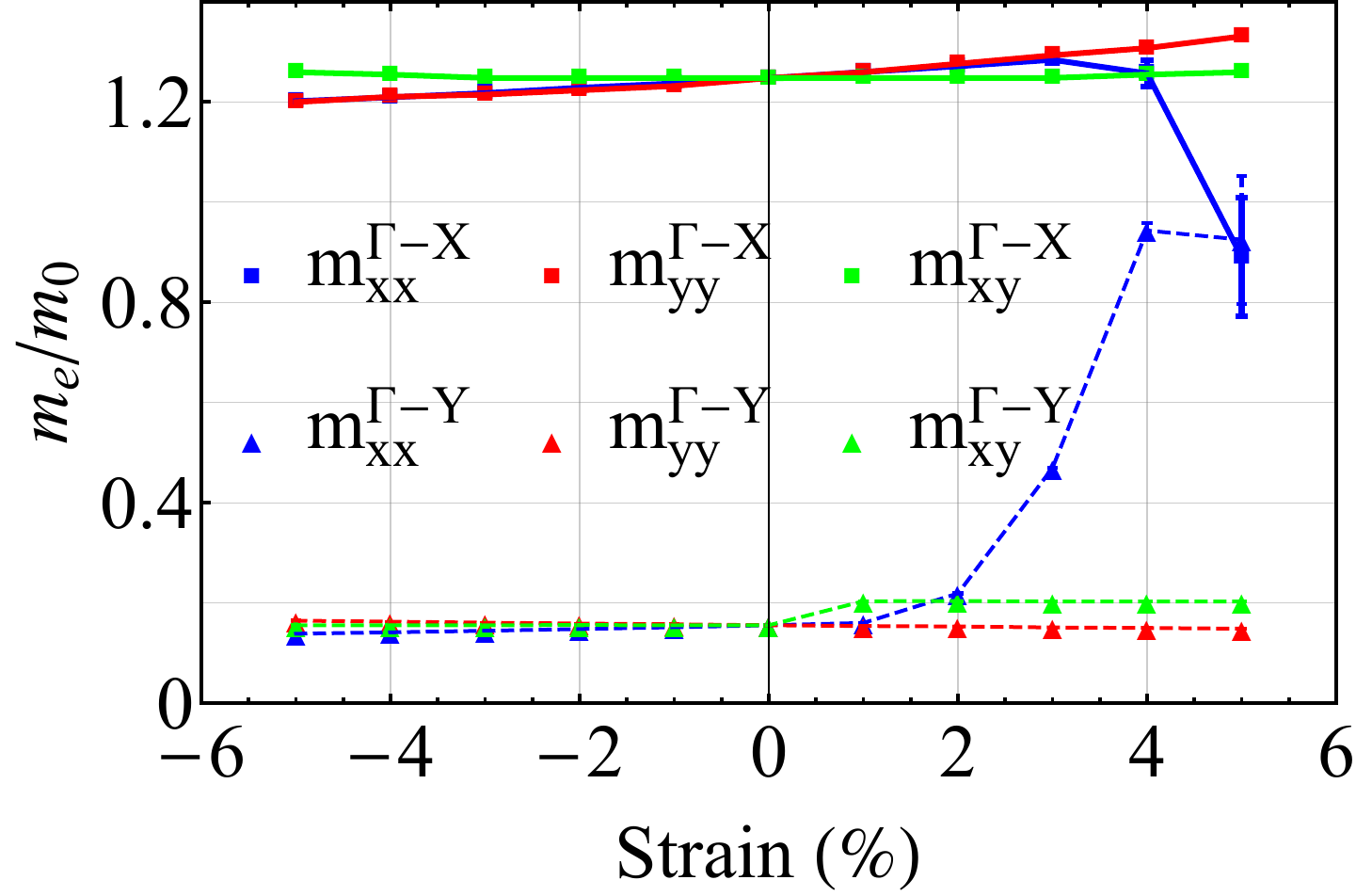}
  \label{fig:pmonomeff}}
\end{subfloat}
\begin{subfloat}[Monolayer P $m_h/m_0$]{
  \centering
\includegraphics[width=0.3\textwidth]{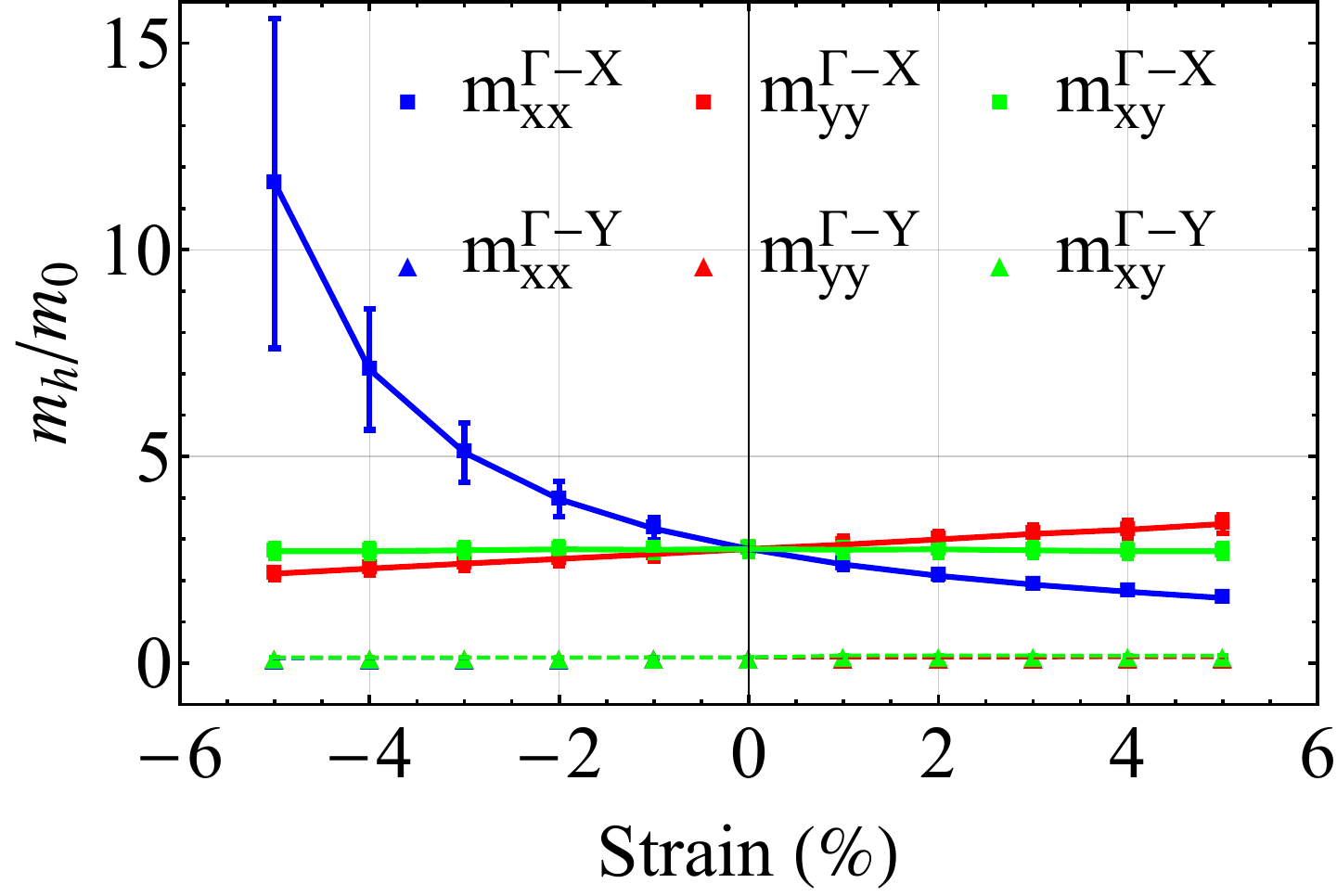}
  \label{fig:pmonomhff}}
\end{subfloat}
\begin{subfloat}[Bilayer P band gap (eV)]{
  \centering
\includegraphics[width=0.3\textwidth]{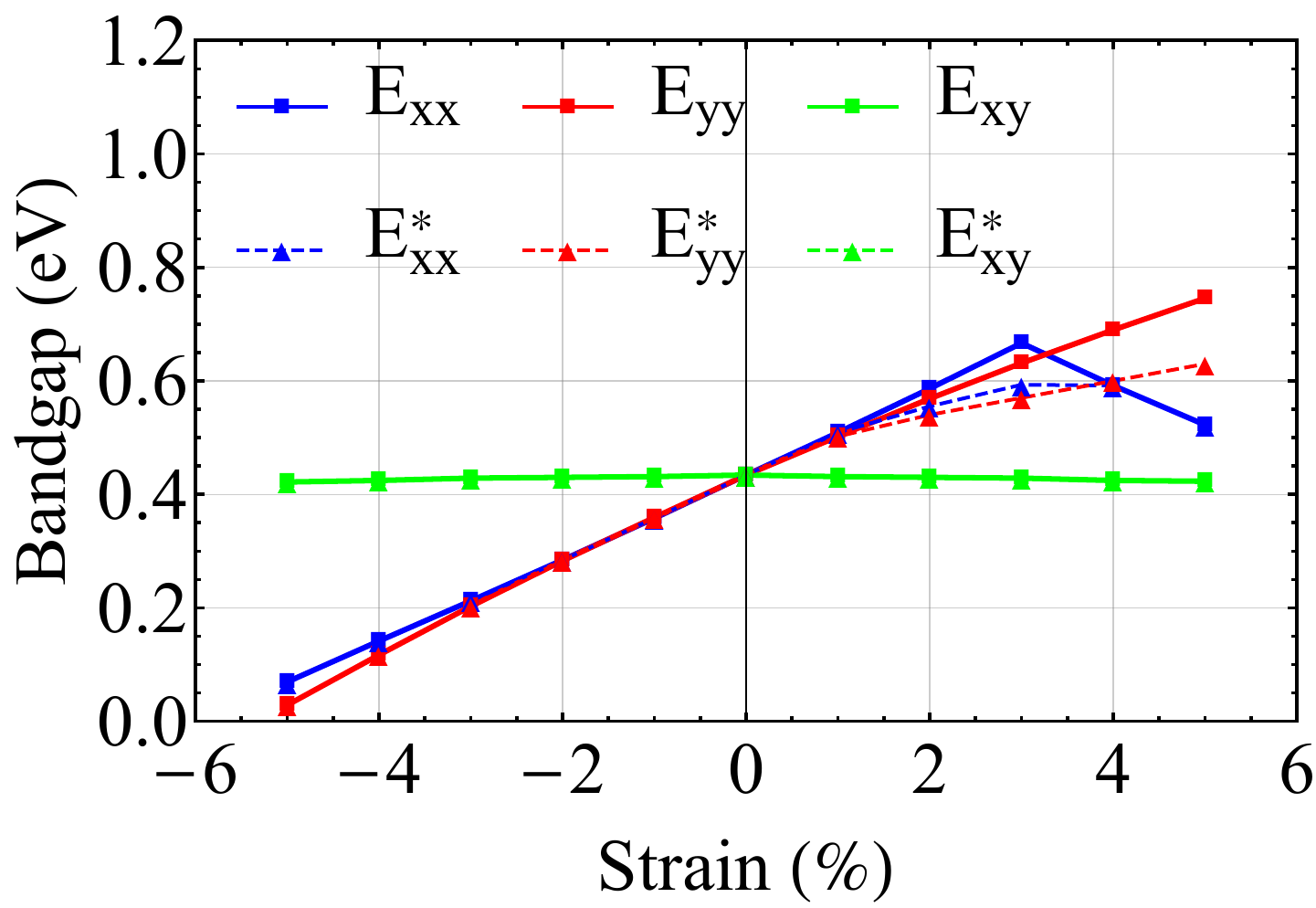}
  \label{fig:pbibg}}
\end{subfloat}
\begin{subfloat}[Bilayer P $m_e/m_0$]{
  \centering
\includegraphics[width=0.3\textwidth]{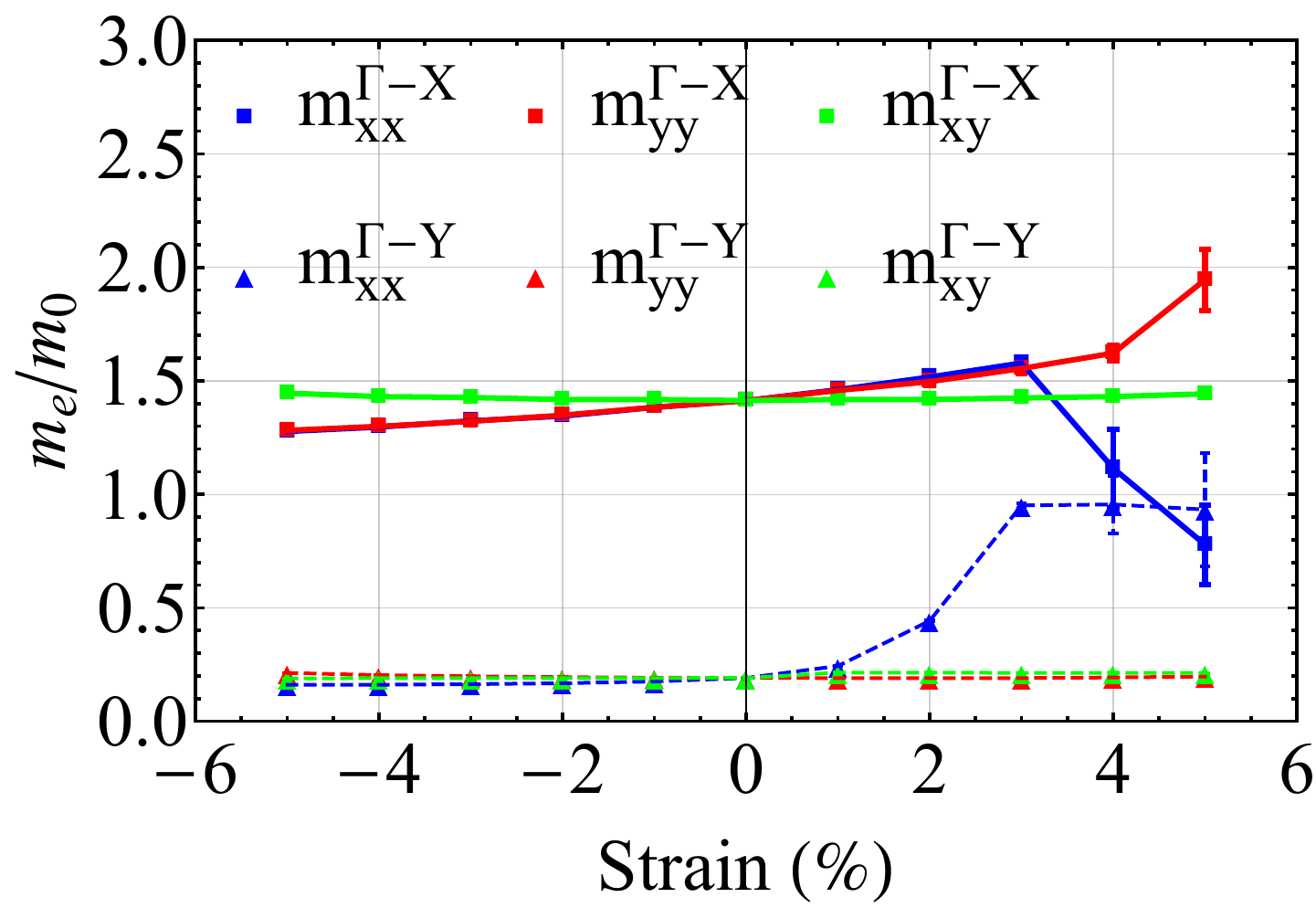}
  \label{fig:pbimeff}}
\end{subfloat}
\begin{subfloat}[Bilayer P $m_h/m_0$]{
  \centering
\includegraphics[width=0.3\textwidth]{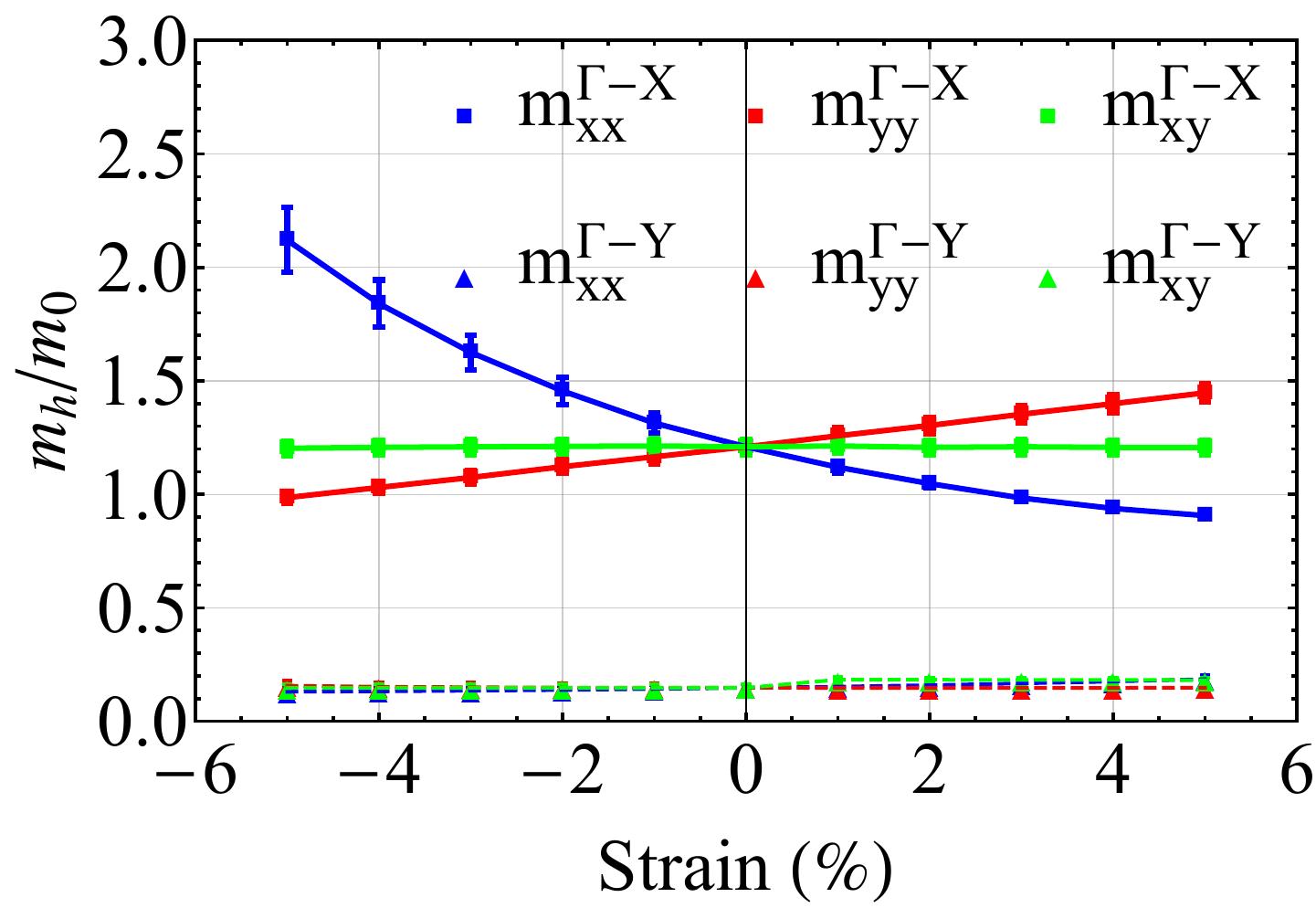}
  \label{fig:pbimhff}}
\end{subfloat}
\begin{subfloat}[Bulk P band gap (eV)]{
  \centering
\includegraphics[width=0.3\textwidth]{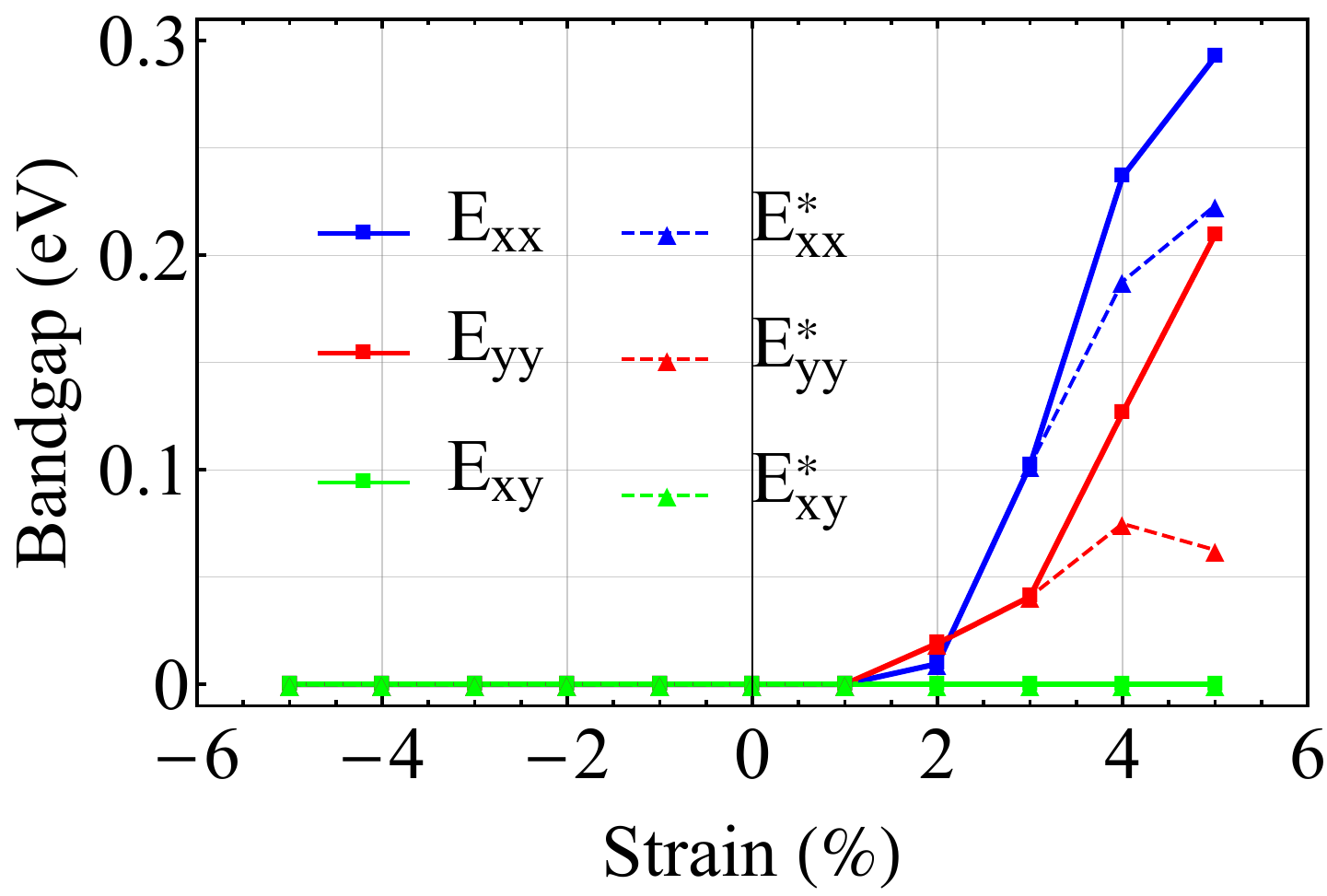}
  \label{fig:pbulkbg}}
\end{subfloat}
\begin{subfloat}[Bulk P $m_e/m_0$]{
  \centering
\includegraphics[width=0.3\textwidth]{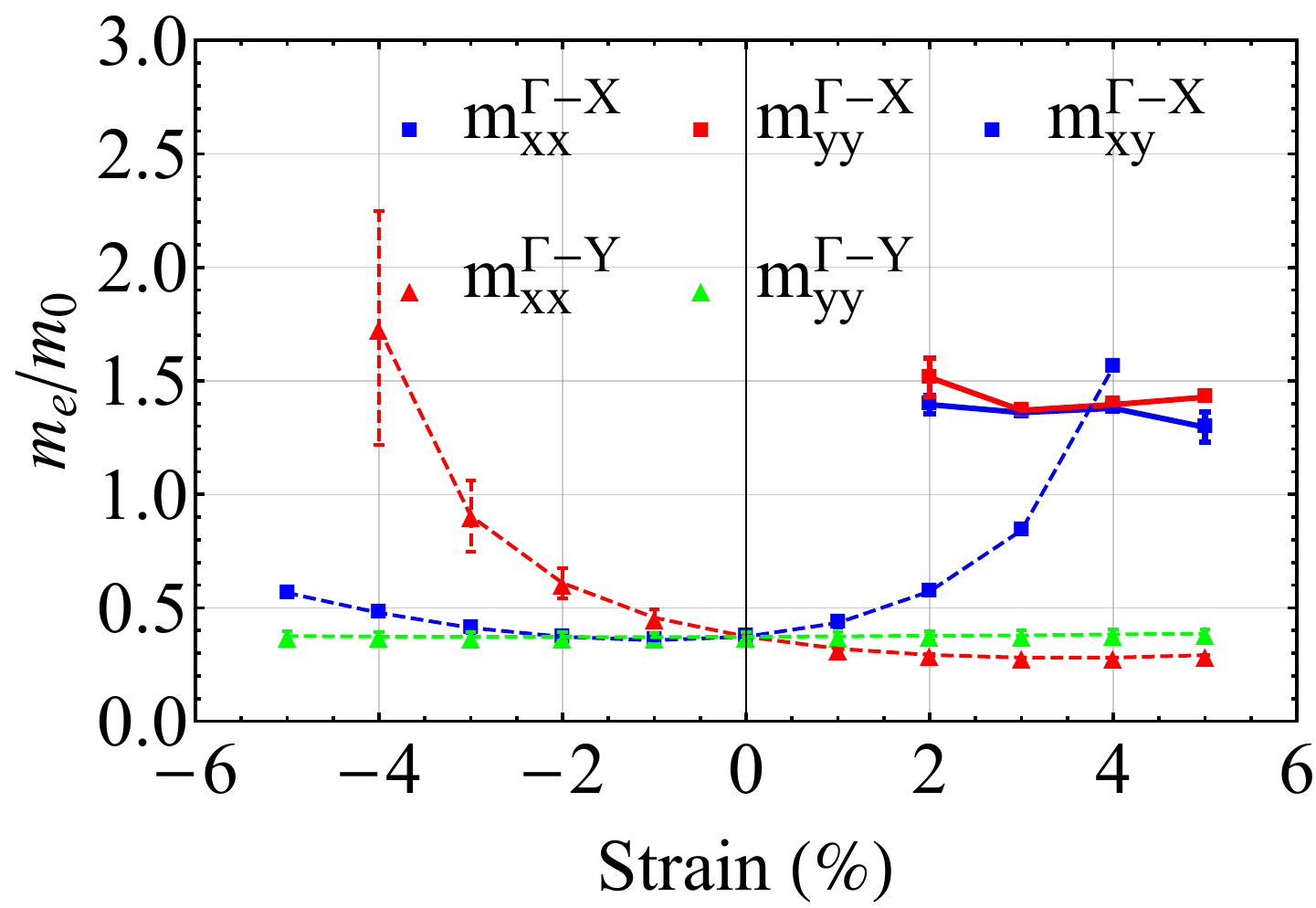}
  \label{fig:pbulkmeff}}
\end{subfloat}
\begin{subfloat}[Bulk P $m_h/m_0$]{
  \centering
\includegraphics[width=0.3\textwidth]{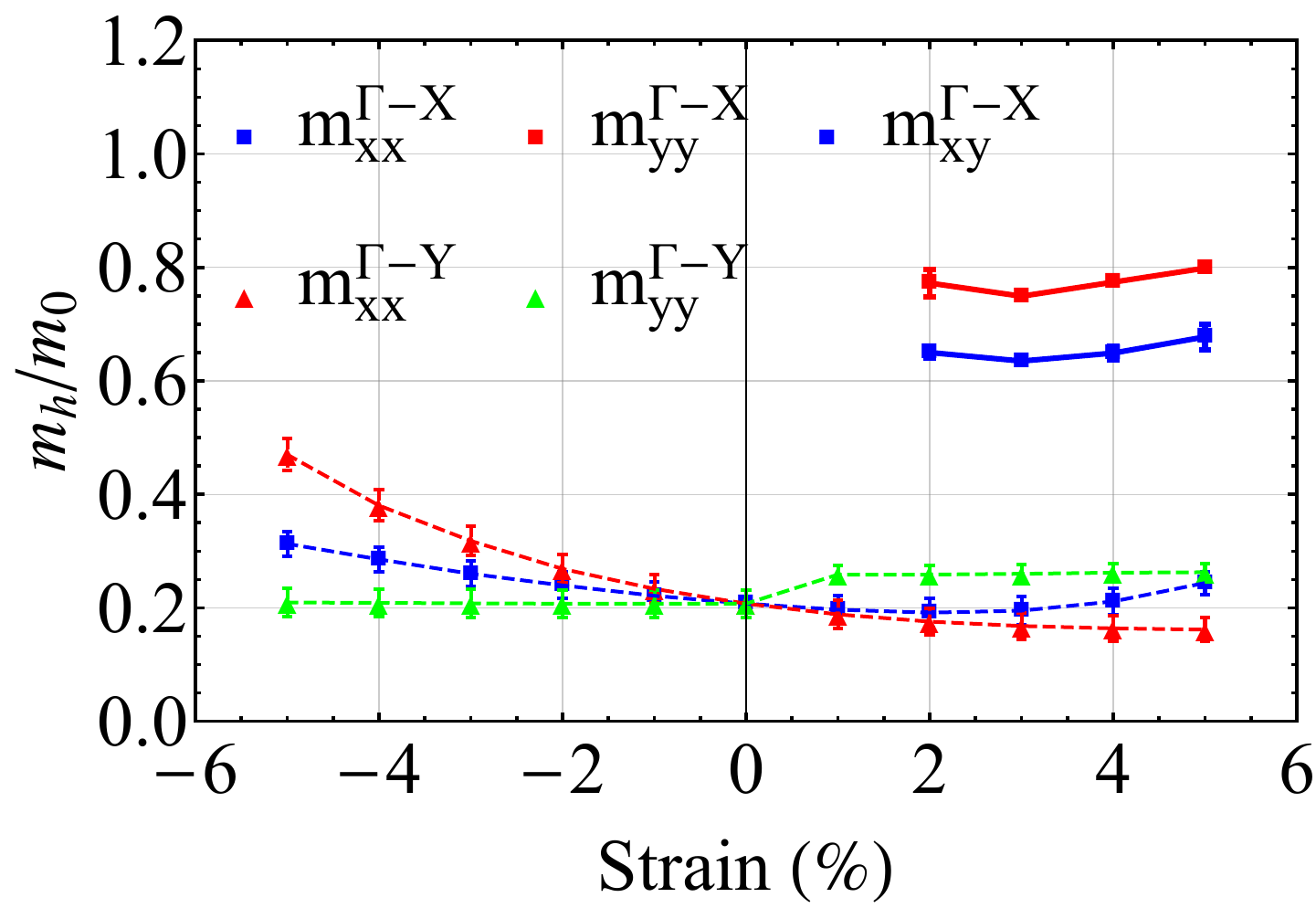}
  \label{fig:pbulkmhff}}
\end{subfloat}
\caption{(Color online) The relationships between 
the applied in-plane strains $\varepsilon_{xx}$ (blue),
$\varepsilon_{yy}$ (red) and $\varepsilon_{xy}$ (green) against 
[\protect\subref{fig:pmonobg}, \protect\subref{fig:pbibg}, \protect\subref{fig:pbulkbg}]
the direct $E$ (solid  squares) and 
indirect $E^\star$ (dashed triangles) band gaps (eV) ;
[\protect\subref{fig:pmonomeff}, \protect\subref{fig:pbimeff}, \protect\subref{fig:pbulkmeff}] 
the effective electron masses $m_e/m_0$ along $\Gamma-X$ (solid squares) and 
$\Gamma-Y$ (dashed triangles) ;
[\protect\subref{fig:pmonomhff}, \protect\subref{fig:pbimhff}, \protect\subref{fig:pbulkmhff}] 
the effective hole masses $m_h/m_0$ along $\Gamma-X$ (solid squares) and 
$\Gamma-Y$ (dashed triangles); for each phase of P.}
\label{fig:p_elec_properties}
\end{figure*}

As shown in Fig.~\ref{fig:pmonobg}, 
our calculations reproduce 
the direct band gap of 0.88~eV 
at the $\Gamma$-point 
in the relaxed P monolayer,  
which falls within range of the 
reported gap between 0.7~eV (DFT-PBEsol~\cite{PhysRevLett.112.176801})
and 1.0~eV (DFT-HSE06~\cite{doi:10.1021/nn501226z}).
On the other hand, quasi-particle calculations 
predict a larger 2~eV  band gap~\cite{PhysRevB.89.235319} 
with significant exciton binding~\cite{PhysRevB.90.205421} 
(between 0.4-0.83~eV).
However, it is well understood that  
approximate semi-local functionals such as PBE 
suffer from a systematic  
band gap problem~\cite{PhysRevB.77.115123} 
that may also adversely affect the metal-insulator 
critical strains.
Nevertheless, 
it is important to \edit{emphasize} that
band alignments and rates of change 
are quite often reliably reproduced~\cite{PhysRevB.90.155405}, 
as are the direct-indirect transitions~\cite{PhysRevB.90.085402} 
in two-dimensional materials. 
\edit{While absolute band gaps 
are therefore not expected to be exactly reproduced,  
we can expect reasonable agreement 
with trends in electronic and 
mechanical behavior~\cite{1402-4896-2004-T109-001,doi:10.1021/cr200107z}.}
The application of uniaxial in-plane strain 
is found to open the band gap for tensile strain 
and diminish it for compressive one,  
while shear strain has a negligible effect.
The electron and hole effective masses 
(Figs.~\ref{fig:pmonomeff}~\&~\ref{fig:pmonomhff}), 
compare well to the figures computed in Ref.~\cite{Wang2015},
where, at $\varepsilon_{xx}=+5\%$ tensile strain,  
the electron and hole 
effective masses   
coincide at 0.9~$m_0$ 
as higher energy bands 
fall below the conduction band.
For compressive strains, 
the hole effective mass along 
$\Gamma-X$ rises significantly 
as the valence band flattens.

Bilayer P is also found 
to have a direct band gap of 
0.43~eV (Fig.~\ref{fig:pbibg}) 
in the relaxed state, 
and broadly the same behavior as the monolayer, 
in which case the band gap closes 
at around $-5\%$ uniaxial compressive strain 
with a predicted Dirac \edit{state} 
at the $\Gamma$-point. 
\edit{The effect of SOC 
on the band structure 
(Fig.~\ref{fig:pbixxdiracbs}) 
induces no qualitative difference 
and the three-dimensional 
bands plotted about the Dirac points 
(Fig.~\ref{fig:pbixx3ddiracbs})
confirm the linear-dispersion,  
albeit in only one direction.
A linear fit to the 
surface of the bands 
(Fig.~\ref{fig:pbixxcutdiracbs}) 
 returns a maximum charge velocity 
of $v=3.80(1)\times 10^6$~$\textrm{ms}^{-1}$ 
along $\Gamma-Y$, 
while, in the orthogonal direction, 
the bands are flat with a  
a charge velocity that is relatively negligible.
This high anisotropy in charge velocities, 
dominated by ballistic conduction along $\Gamma-Y$, 
is indicative of effective one-dimensional conductivity 
and is further supported by 
the large disparity in effective 
masses at $\varepsilon_{xx}=-5\%$, 
evident in Figs.~\ref{fig:pbimeff}~\&~\ref{fig:pbimhff}.}
\edit{
The same analysis for $\varepsilon_{yy}=-5\%$, 
for which the Dirac states are due to band inversion 
and consequently occur off the $\Gamma-Y$ symmetry line
at a point $X^\prime$, 
can be found in Figs.~\ref{fig:pbixxdiracbs}~-~\ref{fig:pbixxcutdiracbs}.
Here the maximum charge velocity is
$v=3.22(1)\times 10^6$~$\textrm{ms}^{-1}$.}
\edit{
These results are further supported by 
the work of Doh \emph{et al.}~\cite{2053-1583-4-2-025071}, 
who demonstrated the effect of strain 
on hopping parameters 
can lead to a Dirac semi-metallic state in bilayer P.
}
\edit{
Similarly, Baik \emph{et al.}~\cite{doi:10.1021/acs.nanolett.5b04106} 
found that the SOC 
did not induce a band gap 
in potassium-doped multi-layer P,  
but did lift the spin-degeneracy of the Dirac points.}
A direct-indirect band gap transition 
is also observed at 
$+2\%$ uniaxial tensile strain. 
The effective masses 
(Figs.~\ref{fig:pbimeff}~\&~\ref{fig:pbimhff}), 
also exhibit broadly the same behavior as 
the monolayer, 
due to band-flattening at $\Gamma-X$ 
and the falling conduction bands along $\Gamma-Y$, 
which lead to 
the charge carrier effective  
masses along $\Gamma-X$ 
converging at $+4\%$ strain 
and an increasing hole effective mass 
for compressive strains.

In the bulk, however, 
we find that the band gap is completely 
closed (Fig.~\ref{fig:pbulkbg}), 
i.e. that the material is metallic.
After investigation, 
we concluded that this was an effect  
of the smearing functionality~\cite{PhysRevLett.82.3296} 
in the relaxation procedure 
and that it contradicts 
numerous experiments~\cite{Morita1986,doi:10.1063/1.1729699,NARITA1983422,MARUYAMA198199}
that have measured a direct gap in 
the range of 0.31-0.36~eV. 
When relaxed under fixed-occupancy 
conditions, instead, a band gap of 
$\sim$0.35~eV was produced.
While the PBE gap remains  
closed in the relaxed state, 
under uniaxial tensile strain it 
briefly becomes a single-point semi-metal 
at $+2\%$.
At such strains a possible Weyl \edit{state} is observed, 
before a direct gap opens 
that subsequently transitions to an 
indirect gap at $+3\%$.
\edit{
Shown in Fig.~\href{http://iopscience.iop.org/2053-1583/4/4/045018/media/Supplementary_Information.pdf}{\bl{S10}} 
in the Supplemental Material 
is the three-dimensional band structure with SOC 
in which a pair of potential Weyl points occur  
on an off-symmetry point $X^\prime$ 
along $\Gamma-X$.
Here, the SOC slightly 
reduces the band gap by $\sim0.05$~eV 
and does not qualitatively affect the overall results.}
\edit{
The maximum charge velocity is 
$v=2.40(1)\times 10^6$~$\textrm{ms}^{-1}$  
for both $\varepsilon_{xx}=-5\%$
and  $\varepsilon_{yy}=-5\%$ 
and occurs along a line parallel to $\Gamma-Y$.}
Under greater compression 
this band-inversion may also lead to further Weyl \edit{states},   
which have been experimentally observed 
at similar pressures~\cite{PhysRevLett.115.186403,PhysRevB.93.195434,PhysRevB.91.195319}.
Again, shear strain is seen  
to have a negligible effect on the gap.
The electron effective masses 
are quite responsive to strain 
(Fig.~\ref{fig:pbulkmeff}), 
where those along 
$\Gamma-Y$ rise for 
both tensile strain along $\varepsilon_{xx}$, 
due to falling conduction bands,  
and compressive strain along $\varepsilon_{yy}$
due to flattening bands along $\Gamma-Y$.
The effective masses along $\Gamma-X$ 
were necessarily not computed once 
the bands overlapped below $+2\%$ strain 
and the hole effective masses 
are found to vary with respect to the strain 
to a slightly lesser extent
(Fig.~\ref{fig:pbulkmhff}).

To summarize, 
we predict the onset of $\Gamma$-point 
Dirac \edit{states} in bilayer P 
at -5\% uniaxial compressive strain, 
with effective one-dimensional conductivity 
at $\varepsilon_{xx}=-5\%$, 
and a direct-indirect band gap transition at 
$+2\%$ tensile strain. 
We also predict the existence of a possible 
Weyl \edit{states} at $+2\%$ tensile strain in bulk P, 
followed by a direct-indirect 
band gap transition at +3\%.
Finally, effective masses are found to 
be particularly responsive to 
$\varepsilon_{xx}$ uniaxial strain. 

\subsubsection{Arsenic}
\begin{figure*}
\begin{subfloat}[Monolayer As band gap (eV)]{
  \centering
\includegraphics[width=0.3\textwidth]{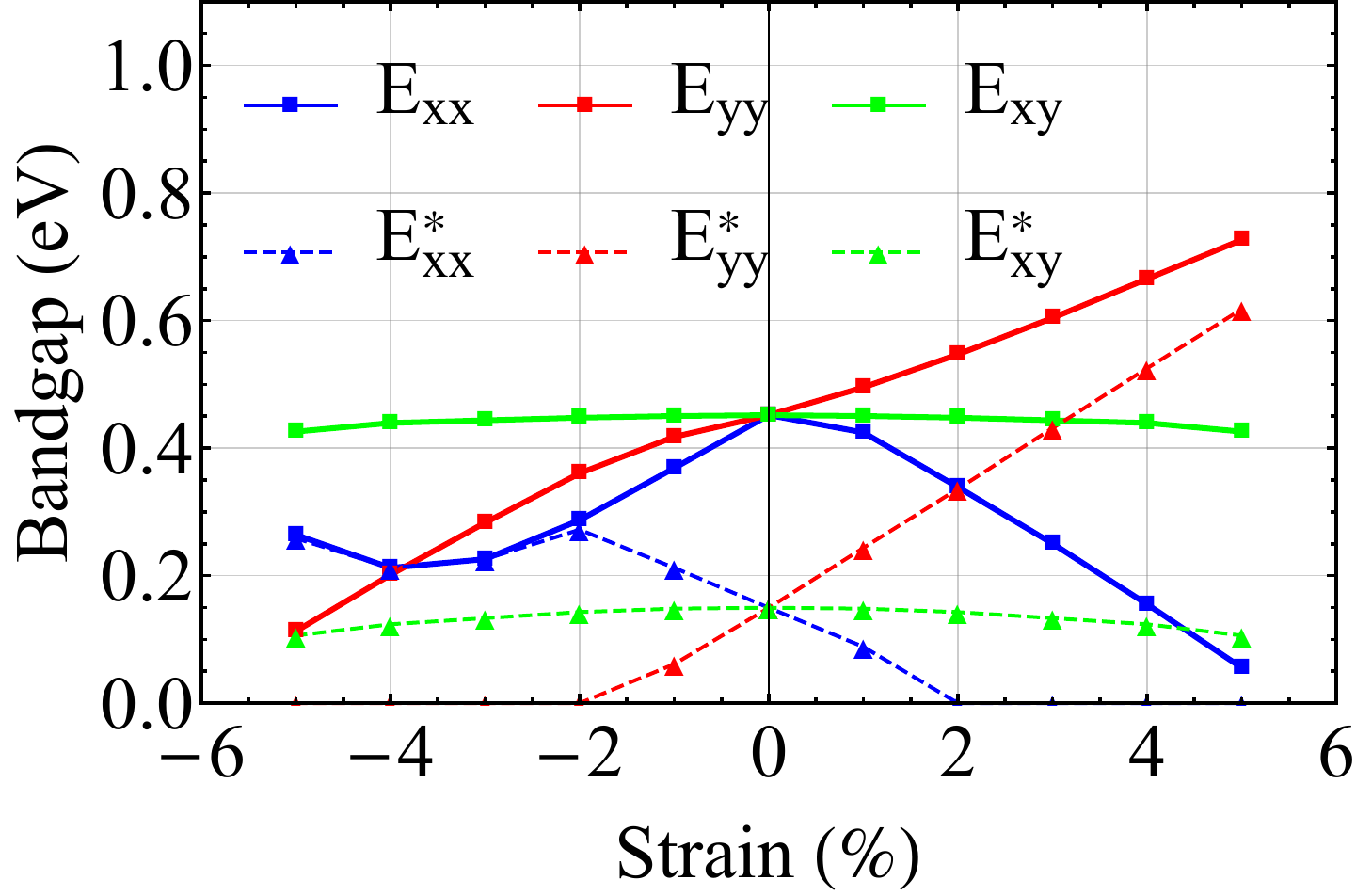}
  \label{fig:asmonobg}}
\end{subfloat}
\begin{subfloat}[Monolayer As $m_e/m_0$]{
  \centering
\includegraphics[width=0.3\textwidth]{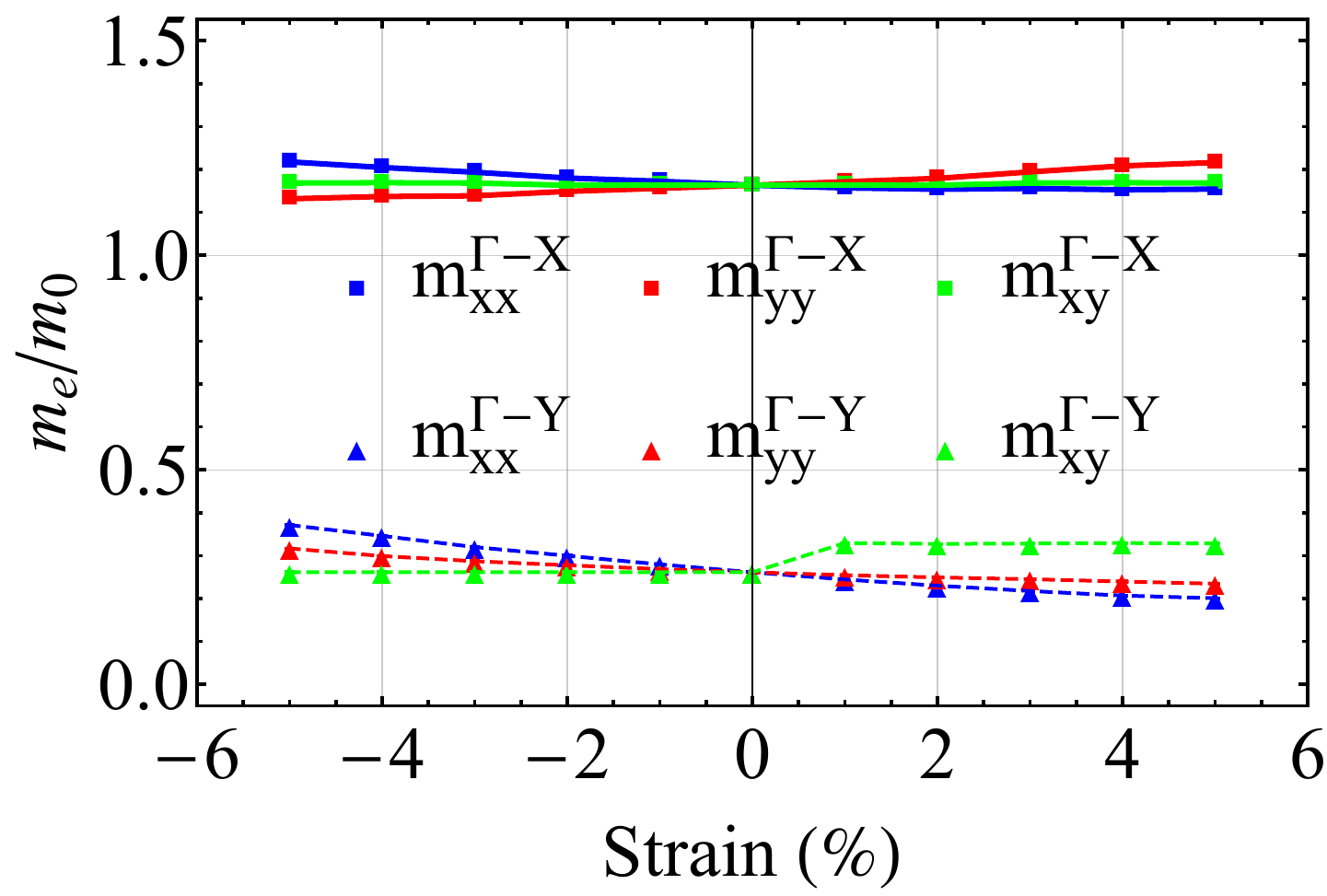}
  \label{fig:asmonomeff}}
\end{subfloat}
\begin{subfloat}[Monolayer As $m_h/m_0$]{
  \centering
\includegraphics[width=0.3\textwidth]{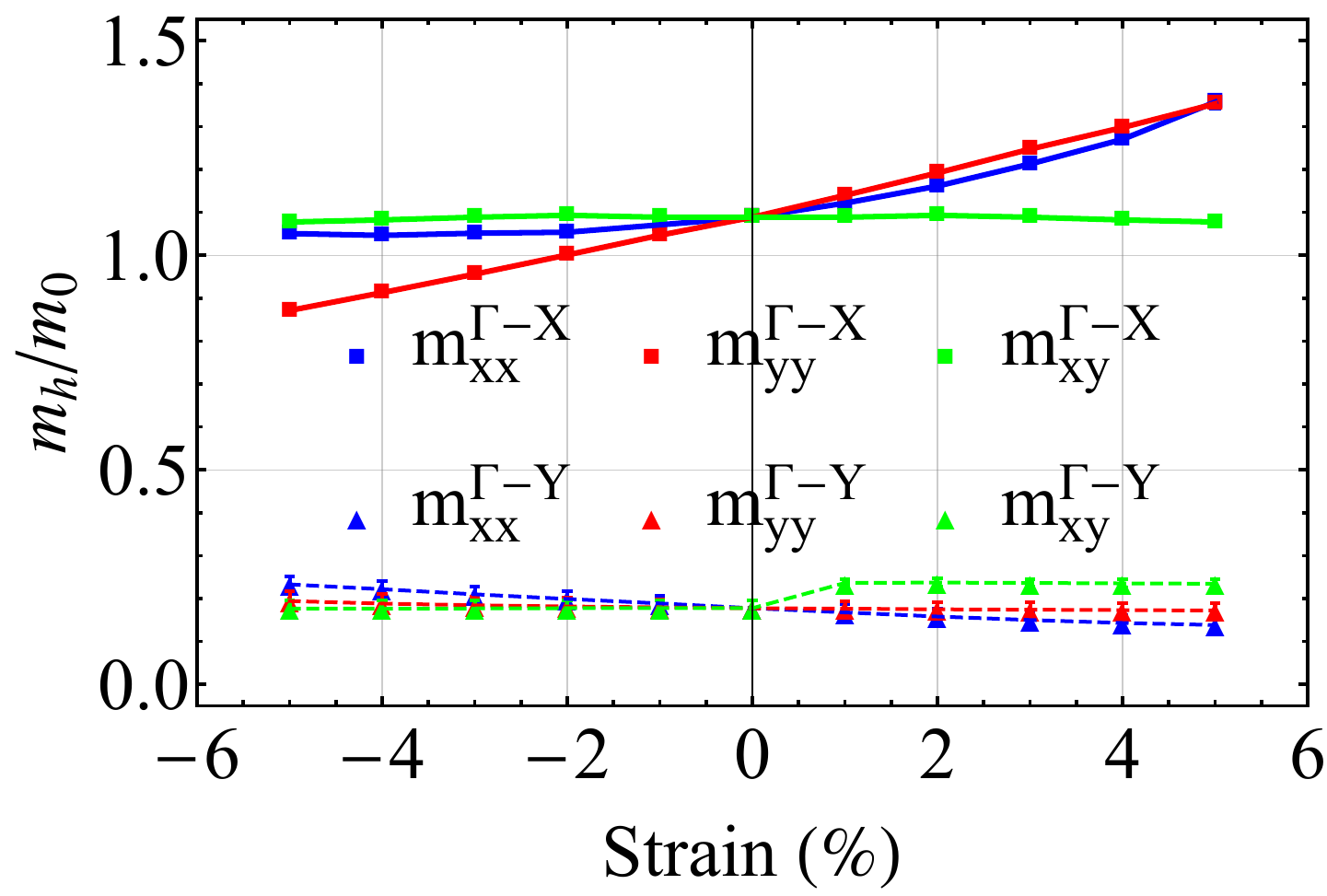}
  \label{fig:asmonomhff}}
\end{subfloat}
\begin{subfloat}[Bilayer As band gap (eV)]{
  \centering
\includegraphics[width=0.3\textwidth]{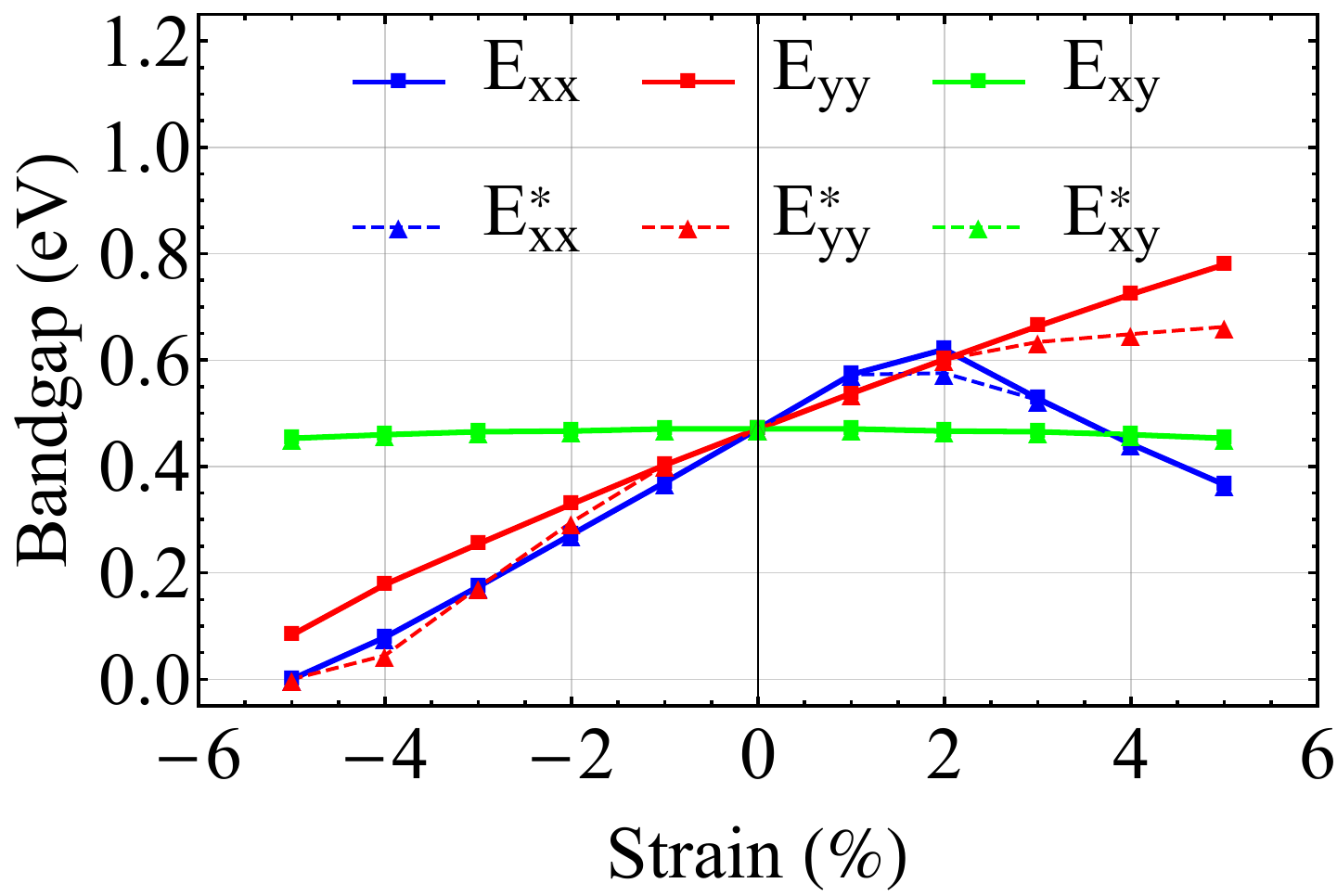}
    \label{fig:asbibg}}
\end{subfloat}
\begin{subfloat}[Bilayer As $m_e/m_0$]{
  \centering
\includegraphics[width=0.3\textwidth]{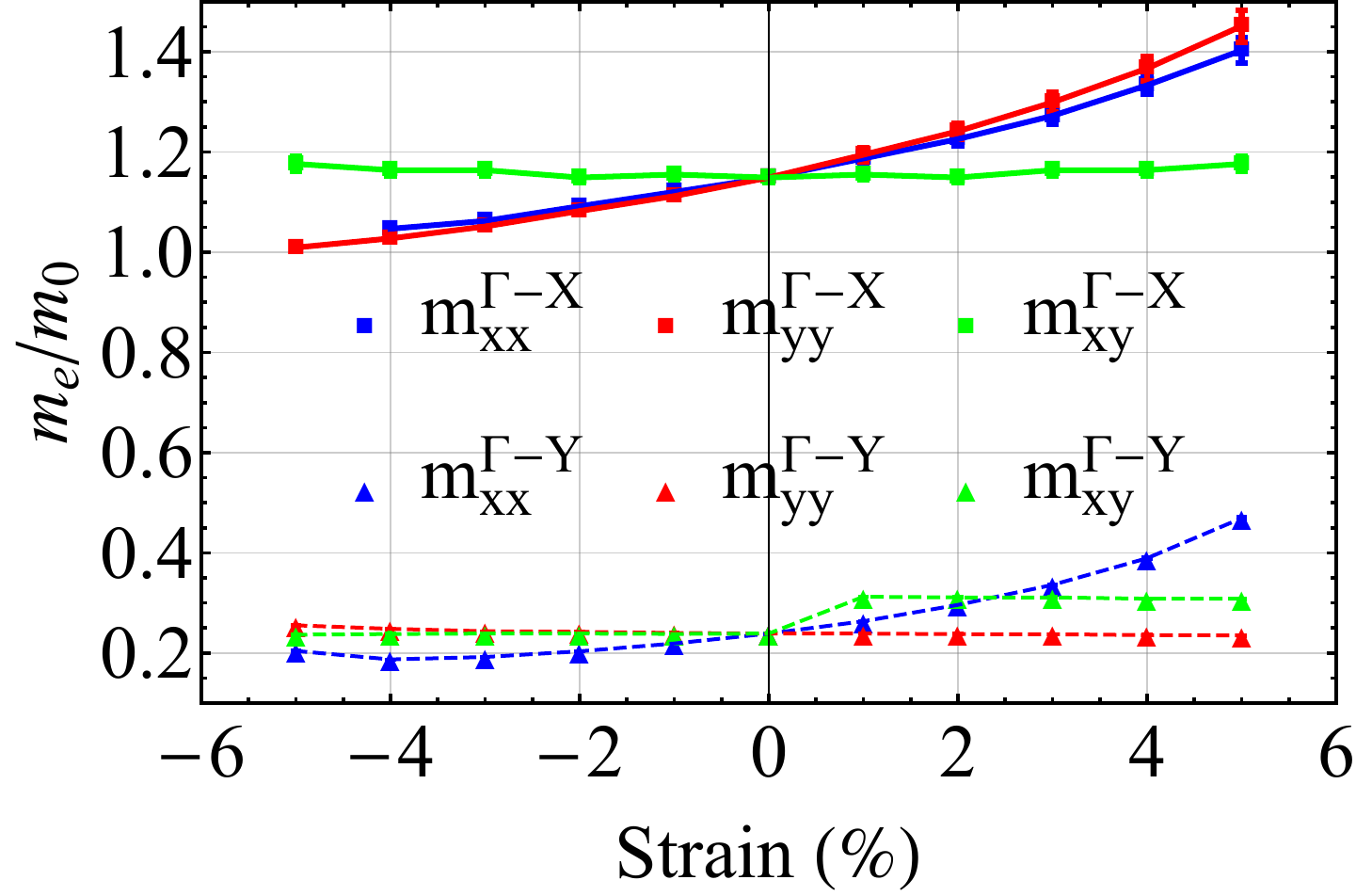}
  \label{fig:asbimeff}}
\end{subfloat}
\begin{subfloat}[Bilayer As $m_h/m_0$]{
  \centering
\includegraphics[width=0.3\textwidth]{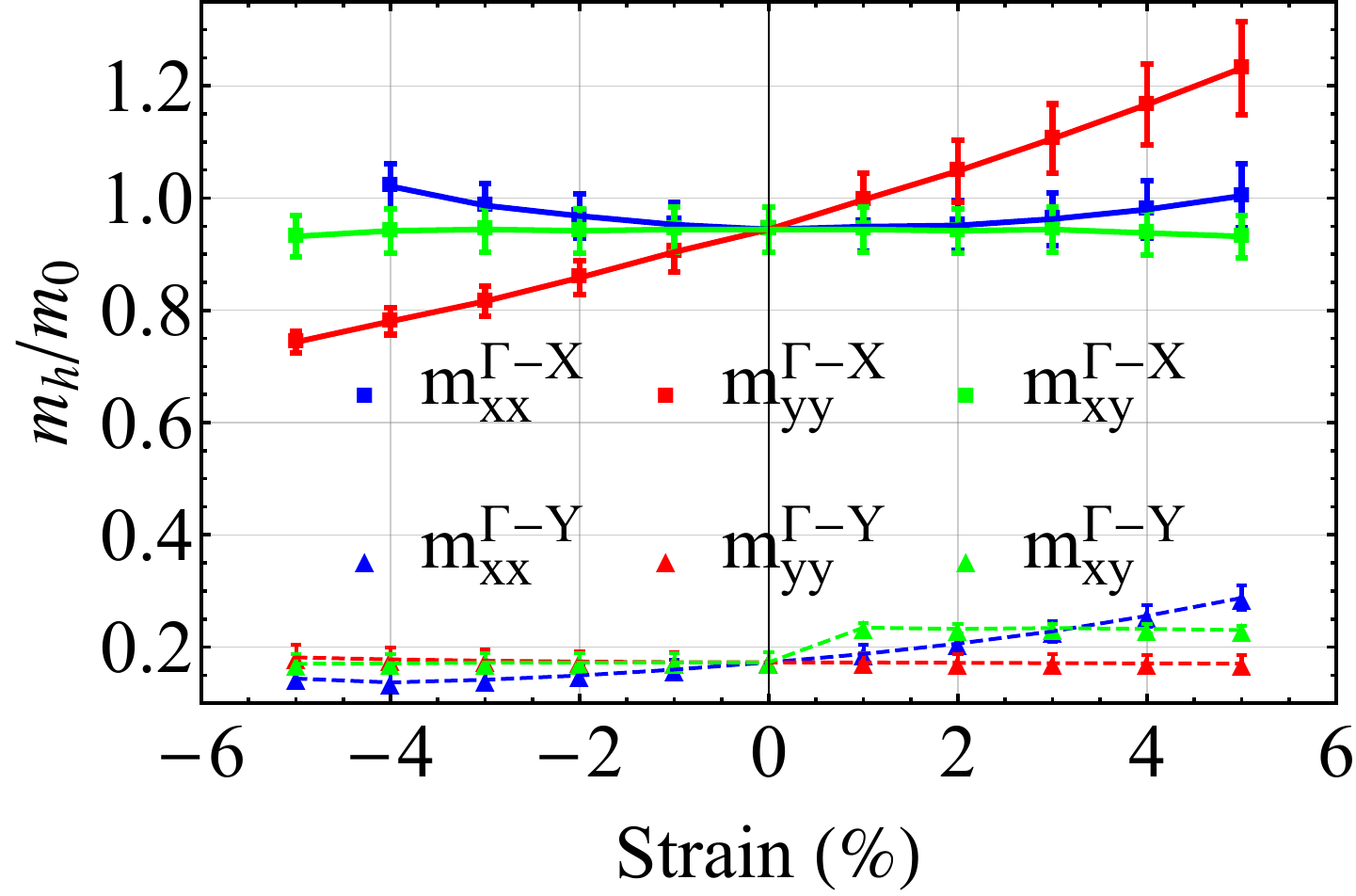}
  \label{fig:asbimhff}}
\end{subfloat}
\begin{subfloat}[Bulk As band gap (eV)]{
  \centering
\includegraphics[width=0.3\textwidth]{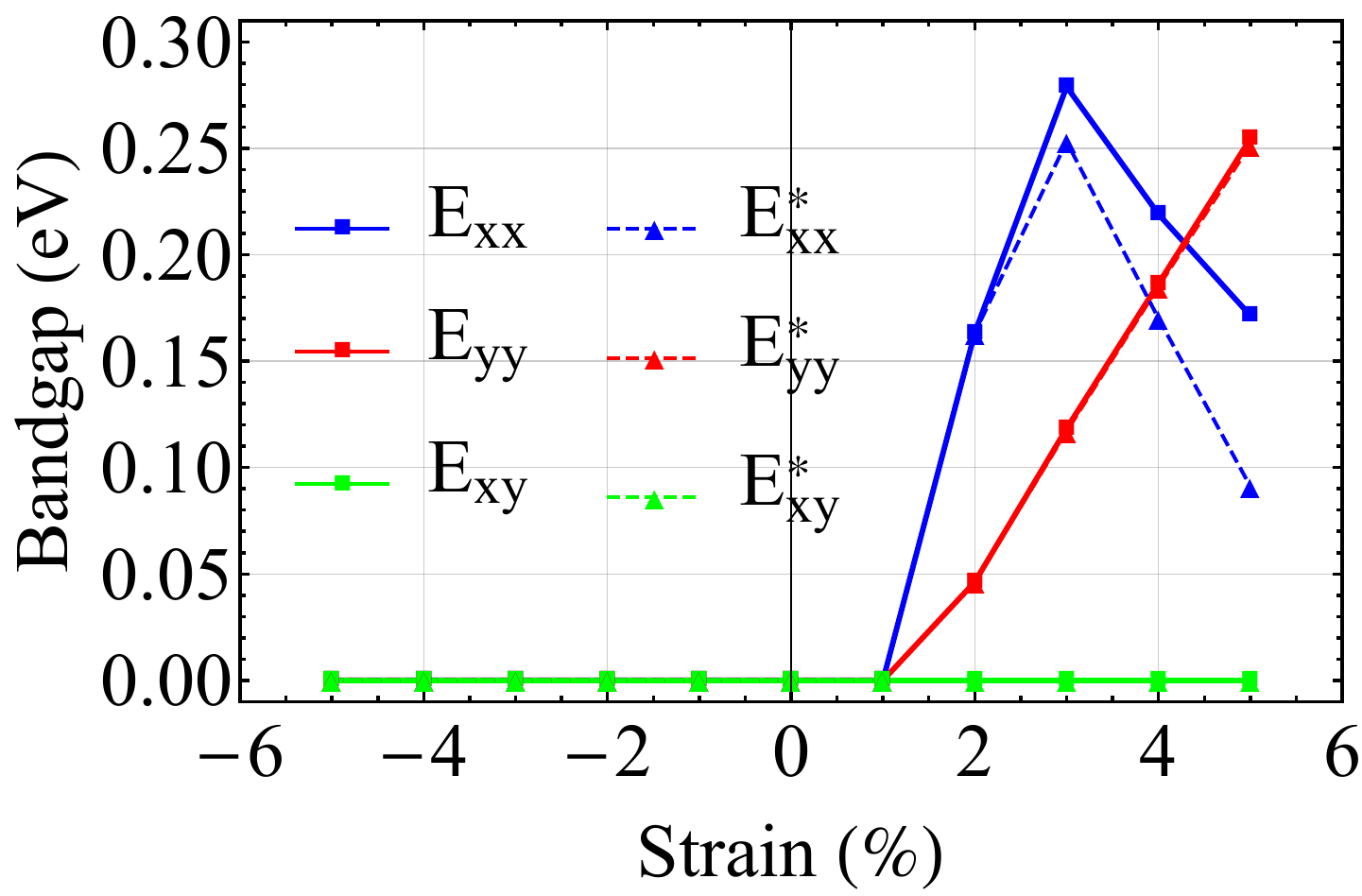}
  \label{fig:asbulkbg}}
\end{subfloat}
\begin{subfloat}[Bulk As $m_e/m_0$]{
  \centering
\includegraphics[width=0.3\textwidth]{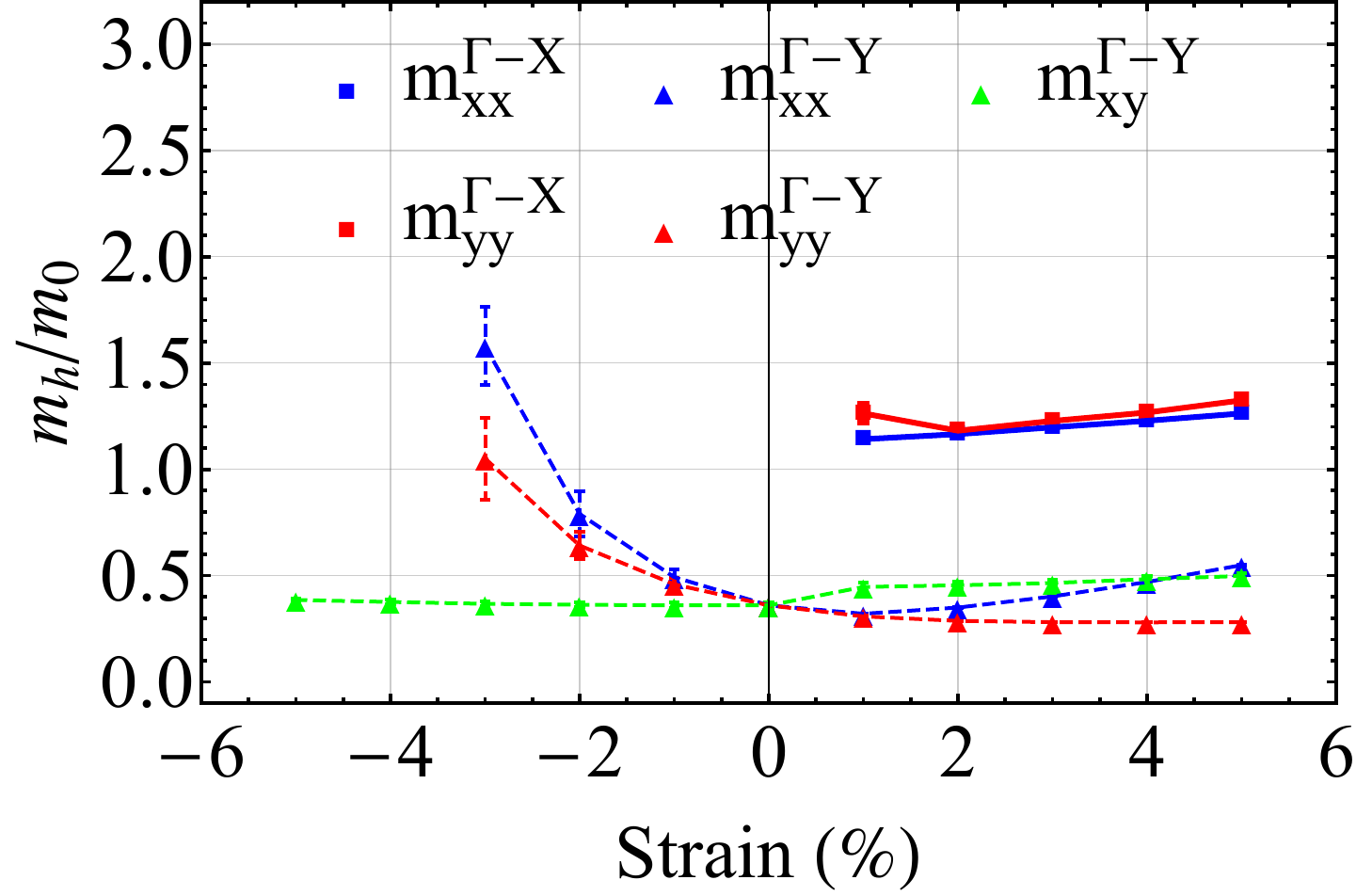}
  \label{fig:asbulkmeff}}
\end{subfloat}
\begin{subfloat}[Bulk As $m_h/m_0$]{
  \centering
\includegraphics[width=0.3\textwidth]{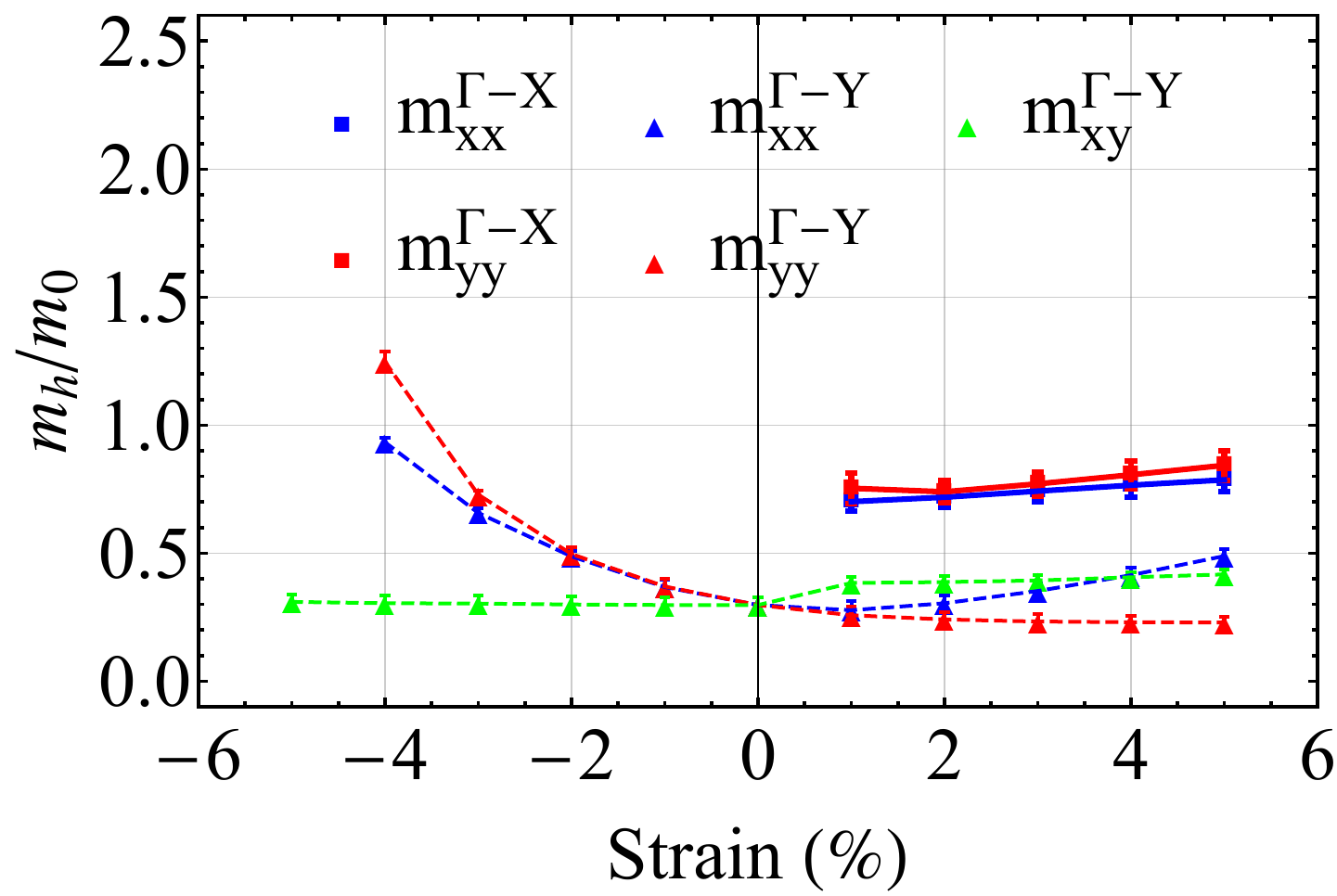}
  \label{fig:asbulkmhff}}
\end{subfloat}
\caption{(Color online) The relationships between 
the applied in-plane strains $\varepsilon_{xx}$ (blue),
$\varepsilon_{yy}$ (red) and $\varepsilon_{xy}$ (green) against 
[\protect\subref{fig:pmonobg}, \protect\subref{fig:pbibg}, \protect\subref{fig:pbulkbg}]  
the direct (solid  squares) and indirect (dashed triangles) band gaps (eV) ;
[\protect\subref{fig:pmonomeff}, \protect\subref{fig:pbimeff}, \protect\subref{fig:pbulkmeff}] 
the effective electron masses $m_e/m_0$ along $\Gamma-X$ (solid squares) and 
$\Gamma-Y$ (dashed triangles) ;
[\protect\subref{fig:pmonomhff}, \protect\subref{fig:pbimhff}, \protect\subref{fig:pbulkmhff}] 
the effective hole masses $m_h/m_0$ along $\Gamma-X$ (solid squares) and 
$\Gamma-Y$ (dashed triangles); for each phase of As.}
\label{fig:as_elec_properties}
\end{figure*}

In contrast to P, 
we identify an indirect 
band gap of 0.15~eV along the 
$\Gamma-Y$ direction
in the relaxed As monolayer (Fig.~\ref{fig:asmonobg}),
which is significantly lower than the  
predicted DFT-HSE06 gap~\cite{1882-0786-8-5-055201} 
of 0.83~eV.
However, 
the relaxed band structure and band gap profiles 
closely resemble those in 
Refs.~\cite{PhysRevB.91.085423,doi:10.1063/1.4943548}.
The band gap diminishes for tensile  
strain along $\varepsilon_{xx}$ 
and at $+2\%$ 
the material becomes semi-metallic 
with a Dirac \edit{state} at the $\Gamma$-point  
emerging at $\varepsilon_{xx}=+5\%$ 
\edit{accompanied by an electron pocket 
above the Fermi-level 
(Fig.~\href{http://iopscience.iop.org/2053-1583/4/4/045018/media/Supplementary_Information.pdf}{\bl{S11}} 
in the Supplemental Material), 
which is unaffected by the SOC.}
\edit{
The maximum charge velocity here 
is $v=3.01(1)\times10^6$~$\textrm{ms}^{-1}$ 
and lies along $\Gamma-Y$.
In the orthogonal direction 
the bands are flat, 
similarly to monolayer P, 
with a relatively small charge velocity.
This high anisotropy in charge velocities, 
dominated by the ballistic conduction along $\Gamma-Y$, 
is again indicative of effective one-dimensional conductivity 
and is further supported by 
the large disparity in effective 
masses at $\varepsilon_{xx}=5\%$, 
shown in Figs.~\ref{fig:asmonomeff}~\&~\ref{fig:asmonomhff}.}

For compressive strain along $\varepsilon_{xx}$  
the indirect band gap 
closes along $\Gamma-Y$ at $\varepsilon_{xx}=-2\%$. 
For tensile strain along $\varepsilon_{yy}$
the indirect band gap opens where 
an indirect-direct transition~\cite{PhysRevB.91.085423} 
occurs at $\varepsilon_{xx}=-3\%$.
Similar to monolayer phosphorus, 
there is no appreciable effect due to shear-strain.
Meanwhile, the charge-carrier effective masses 
(Figs.~\ref{fig:asmonomeff}~\&~\ref{fig:asmonomhff}) 
respond linearly to uniaxial strain and compare 
well to other works~\cite{doi:10.1080/14786437508229285}, 
where, in particular, valence band broadening along $\Gamma-X$
leads to an increasing hole effective mass.

For bilayer As, we identify a 
direct band gap of $0.45$~eV 
(Fig.~\ref{fig:asbibg}), 
in contrast to the indirect band gap 
observed in the monolayer.
Here, the band gap opens 
for uniaxial tensile strain 
and diminishes for compressive strain.
The direct band gap transitions
to an indirect gap at both 
$\varepsilon_{yy}=-3\%$ and 
$\varepsilon_{yy}=+2\%$, 
while at 
$\varepsilon_{xx}=+2\%$ 
it also transitions to an indirect gap 
before resuming to a 
direct gap again at $\varepsilon_{xx}=+3\%$.
\edit{
Moreover, we predict a Dirac state 
at the $\Gamma$-point 
at a compressive strain of 
$\varepsilon_{xx}=-4\%$ 
(Fig.~\href{http://iopscience.iop.org/2053-1583/4/4/045018/media/Supplementary_Information.pdf}{\bl{S11}} 
in the Supplemental Material)
for which the maximum charge velocity 
is $v=2.62(2)\times10^6$~$\textrm{ms}^{-1}$ 
along $\Gamma-Y$.
Along $\Gamma-X$
the bands are also flat, 
similar to the monolayer,
and have a relatively negligible charge velocity.
The high anisotropy in charge velocities,  
is again indicative of effective one-dimensional conductivity, 
dominated by the ballistic conduction along $\Gamma-Y$,
and is further supported by 
the large disparity in effective 
masses at $\varepsilon_{xx}=5\%$, 
shown in Figs.~\ref{fig:asbimeff}~\&~\ref{fig:asbimhff}.}
\edit{
Here again, the SOC has no appreciable 
effect on the bands.
}
The electron and hole effective masses 
(Figs.~\ref{fig:asbimeff}~\&~\ref{fig:asbimhff})
respond approximately linearly 
to the applied strain, 
where conduction band broadening 
leads to increased 
effective electron masses, 
and valence band flattening at $\Gamma$
leads to increasing hole effective masses 
for strain along $\varepsilon_{yy}$. 

Finally, no band gap is determined 
in the relaxed bulk phase (Fig.~\ref{fig:asbulkbg}), 
again contrary to experiments~\cite{doi:10.1080/00018737900101355}, 
where a small direct band gap of $\sim0.3$~eV 
is observed.
However, at $\varepsilon_{xx}=+1\%$ strain, 
a potential Weyl \edit{state} is briefly observed 
on an off-symmetry point $X^\prime$
\edit{
(Fig.~\href{http://iopscience.iop.org/2053-1583/4/4/045018/media/Supplementary_Information.pdf}{\bl{S12}} 
in the Supplemental Material)}
before a direct gap opens 
that subsequently transitions 
to an indirect one 
at $\varepsilon_{xx}=+3\%$, 
after which it reduces again.
Another potential Weyl \edit{state} 
around the same off-symmetry point $X^\prime$
is also predicted to occur between 
$+1\%\leq\varepsilon_{yy}\leq+2\%$ 
after which a direct band gap also appears.
\edit{
The recalculated band structure 
with the SOC for $\varepsilon_{yy}=+1\%$
(Fig.~\href{http://iopscience.iop.org/2053-1583/4/4/045018/media/Supplementary_Information.pdf}{\bl{S12}}) 
confirms the linear-dispersion.
The maximum charge velocity in both cases  
occurs along a line parallel to $\Gamma-Y$ 
and is $v=1.38(1)\times10^6$~$\textrm{ms}^{-1}$.}
Meanwhile, the electron and 
hole effective masses 
along $\Gamma-Y$ 
(Figs.~\ref{fig:asbulkmeff}~\&~\ref{fig:asbulkmhff}) 
increase rapidly for compressive strains 
as the band peaks rapidly flatten at 
the $\Gamma$-point.

In summary, 
we predict  
$\Gamma$-point Dirac \edit{states} 
in the monolayer and  bilayer of As, 
\edit{which support one-dimensional ballistic conduction}, 
as well as possible 
Weyl \edit{states} 
on off-symmetry points in the bulk 
at moderate levels of in-plane stress 
\edit{
that are unaffected by the SOC}.
We also observe several 
band gap transitions, 
in particular in the monolayer phase, 
which also include 
semi-conducting-metallic transitions.
Finally, the effective masses 
respond approximately linearly 
with respect to uniaxial strain, 
except in the bulk,
which exhibits quadratic behavior.

\subsubsection{Antimony}
\begin{figure*}
\begin{subfloat}[Monolayer Sb band gap (eV)]{
  \centering
\includegraphics[width=0.3\textwidth]{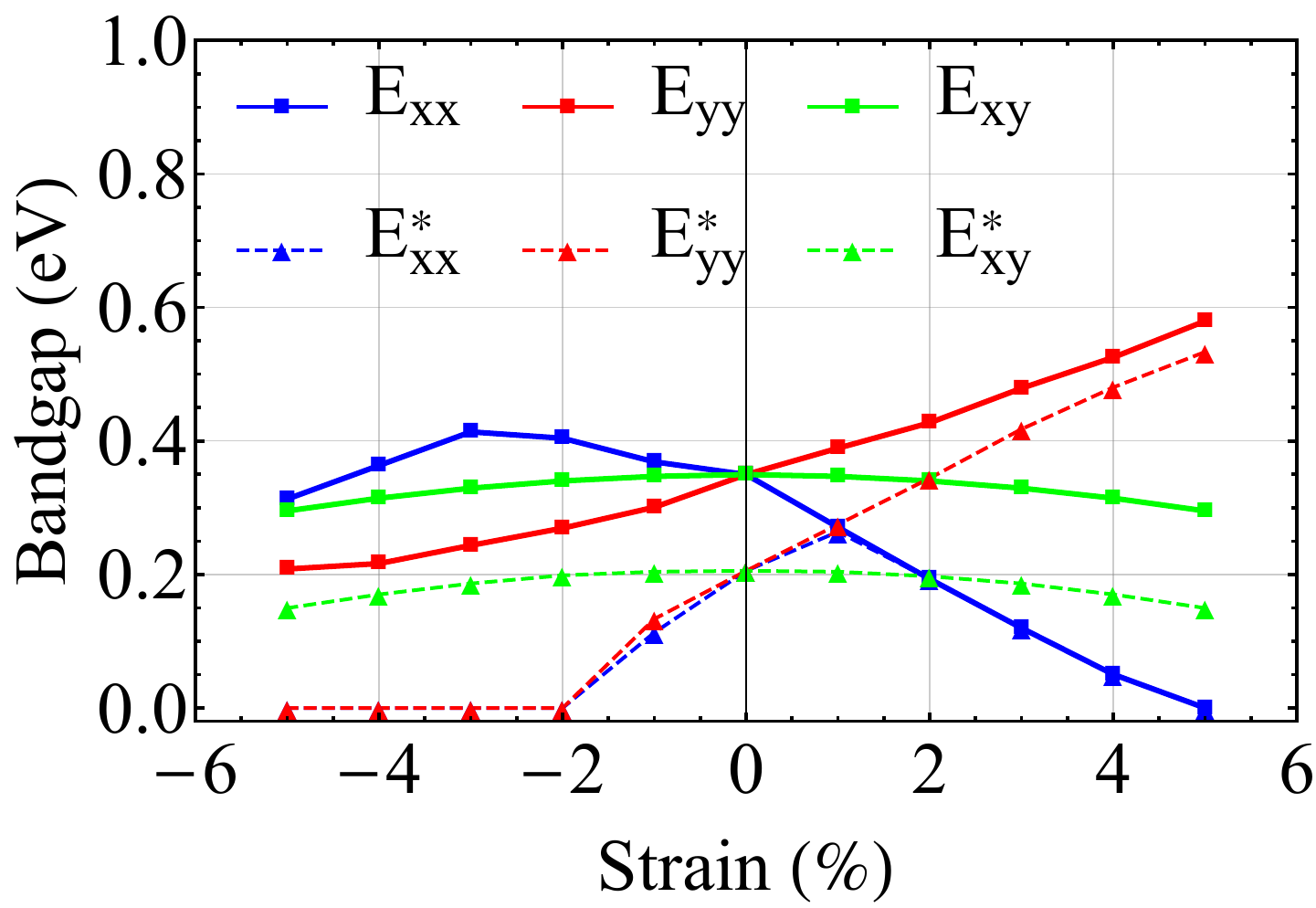}
  \label{fig:sbmonobg}}
\end{subfloat}
\begin{subfloat}[Monolayer Sb $m_e/m_0$]{
  \centering
\includegraphics[width=0.3\textwidth]{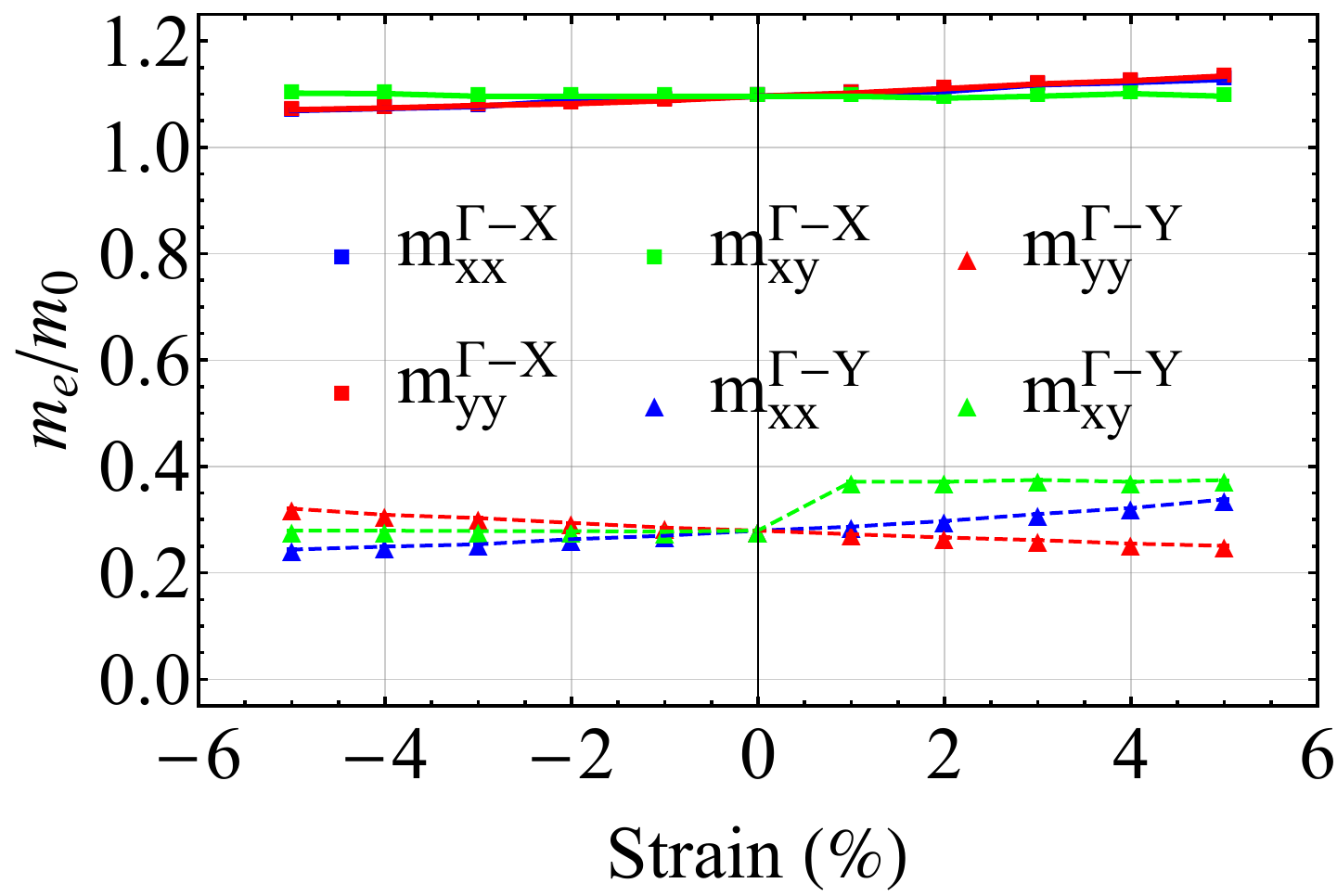}
  \label{fig:sbmonomeff}}
\end{subfloat}
\begin{subfloat}[Monolayer Sb $m_h/m_0$]{
  \centering
\includegraphics[width=0.3\textwidth]{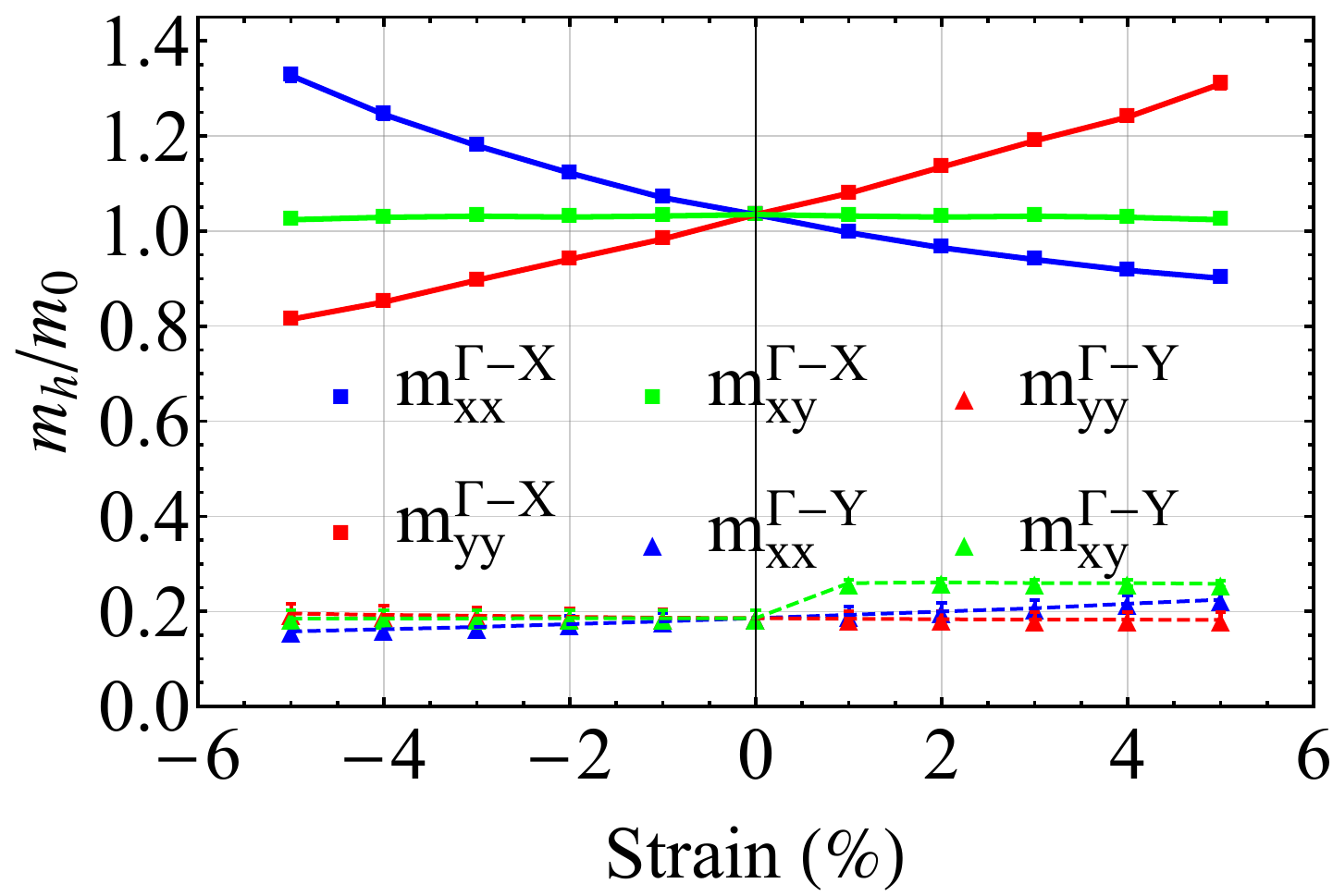}
  \label{fig:sbmonomhff}}
\end{subfloat}
\begin{subfloat}[Bilayer Sb band gap (eV)]{
  \centering
\includegraphics[width=0.3\textwidth]{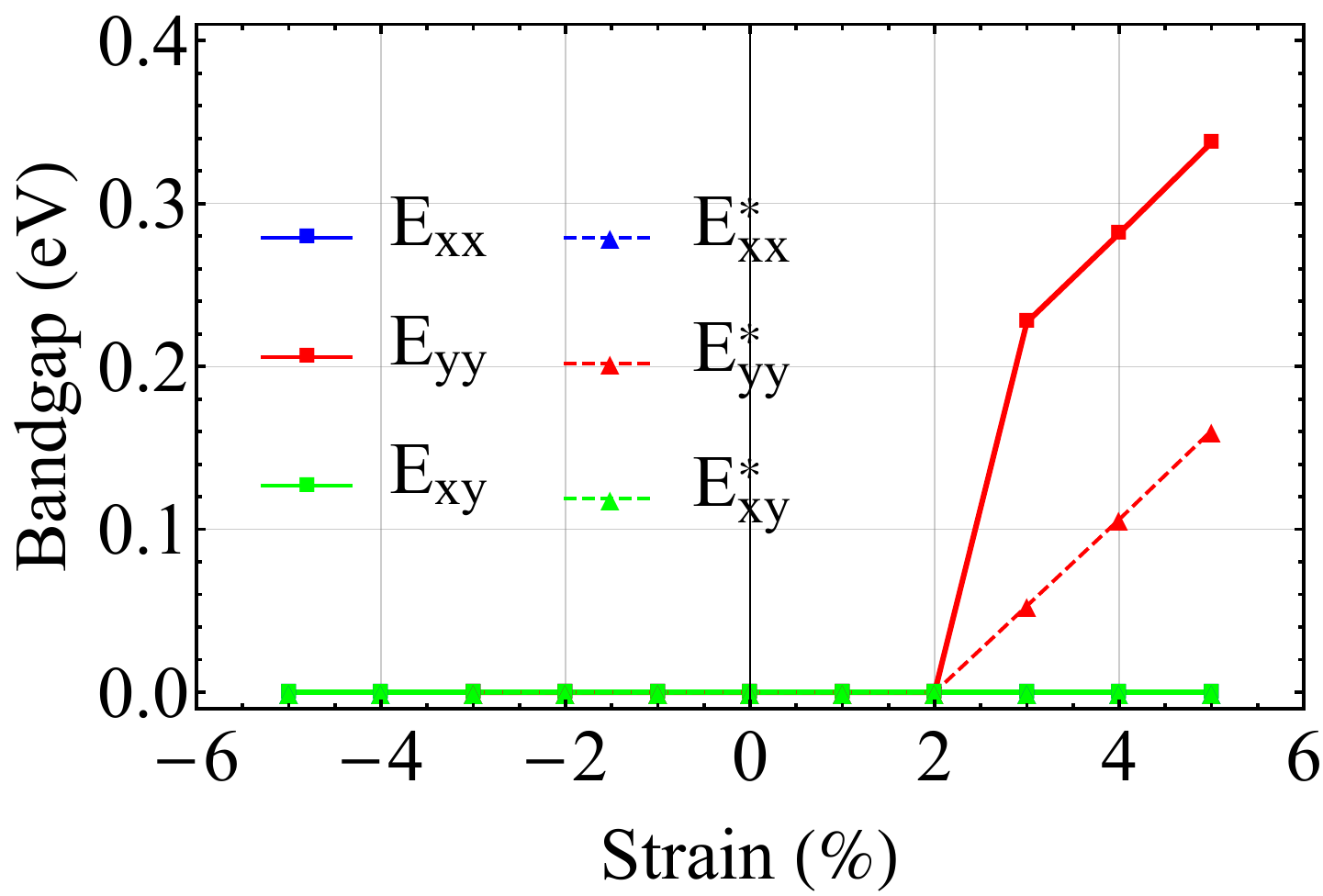}
  \label{fig:sbbibg}}
\end{subfloat}
\begin{subfloat}[Bilayer Sb $m_e/m_0$]{
  \centering
\includegraphics[width=0.3\textwidth]{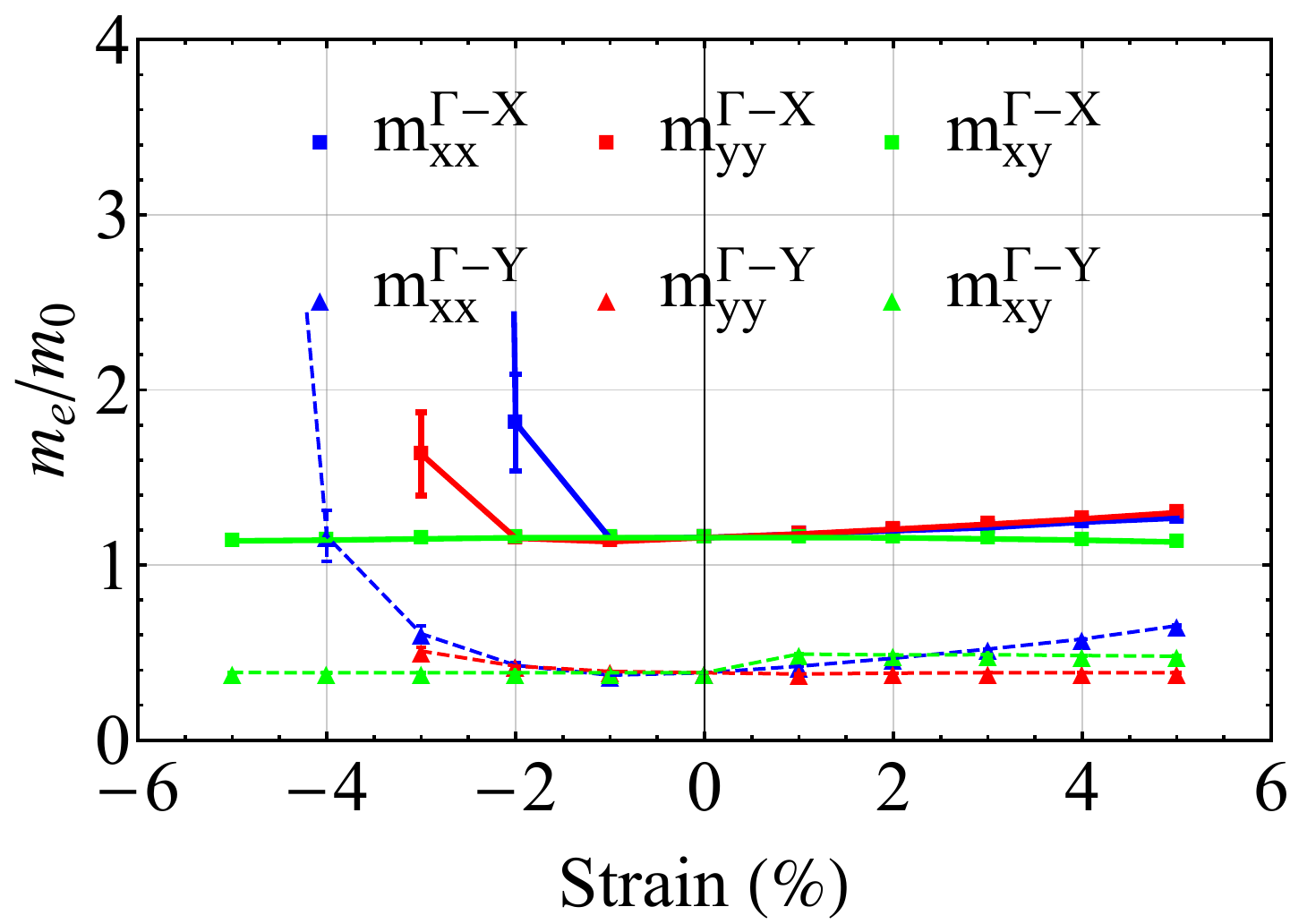}
  \label{fig:sbbimeff}}
\end{subfloat}
\begin{subfloat}[Bilayer Sb $m_h/m_0$]{
  \centering
\includegraphics[width=0.3\textwidth]{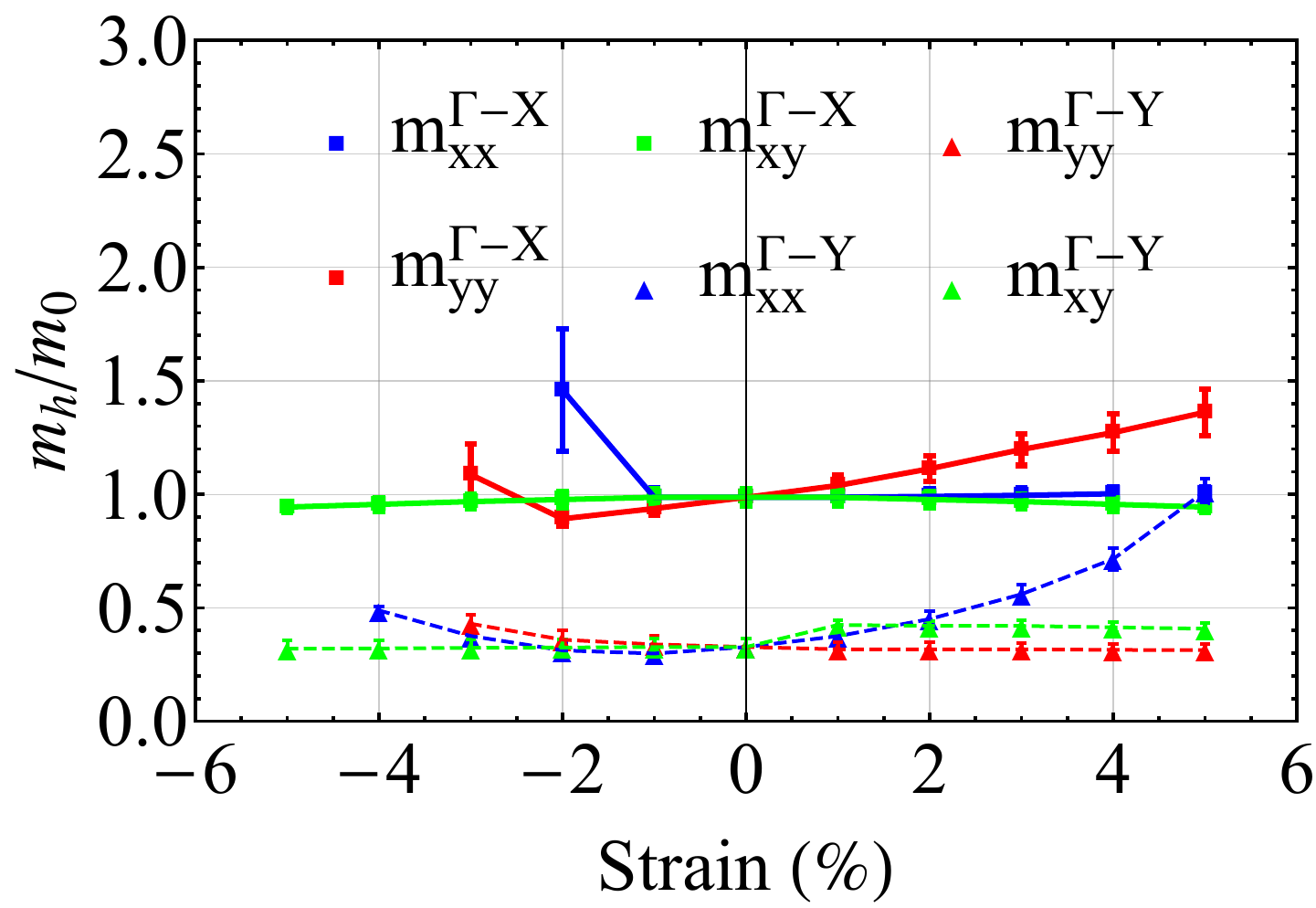}
  \label{fig:sbbimhff}}
\end{subfloat}
\caption{(Color online) The relationships between 
the applied in-plane strains $\varepsilon_{xx}$ (blue),
$\varepsilon_{yy}$ (red) and $\varepsilon_{xy}$ (green) against 
[\protect\subref{fig:pmonobg}, \protect\subref{fig:pbibg}, \protect\subref{fig:pbulkbg}]  
the direct (solid  squares) and indirect (dashed triangles) band gaps (eV) ;
[\protect\subref{fig:pmonomeff}, \protect\subref{fig:pbimeff}, \protect\subref{fig:pbulkmeff}] 
the effective electron masses $m_e/m_0$ along $\Gamma-X$ (solid squares) and 
$\Gamma-Y$ (dashed triangles) ;
[\protect\subref{fig:pmonomhff}, \protect\subref{fig:pbimhff}, \protect\subref{fig:pbulkmhff}] 
the effective hole masses $m_h/m_0$ along $\Gamma-X$ (solid squares) and 
$\Gamma-Y$ (dashed triangles); for each monolayer and bilayer Sb.}
\label{fig:sb_elec_properties}
\end{figure*}

\begin{figure*}
\begin{tabular}{ccc}
\begin{subfloat}[Bilayer P band structure $\varepsilon_{xx}=-5\%$]{
  \centering
\includegraphics[width=0.4\textwidth]{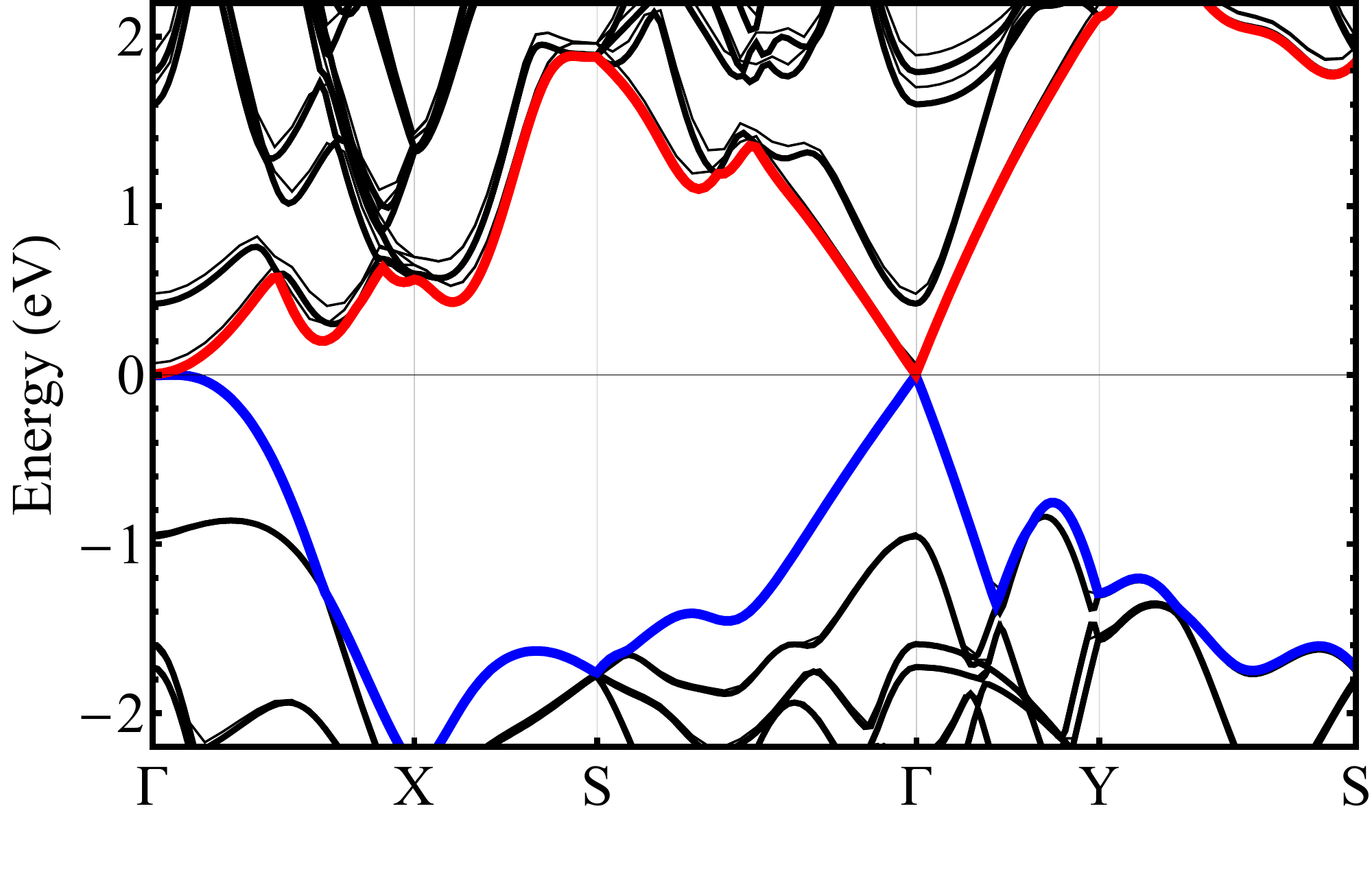}
  \label{fig:pbixxdiracbs}}
\end{subfloat}
&
\begin{subfloat}[]{
  \centering
\includegraphics[width=0.18\textwidth]{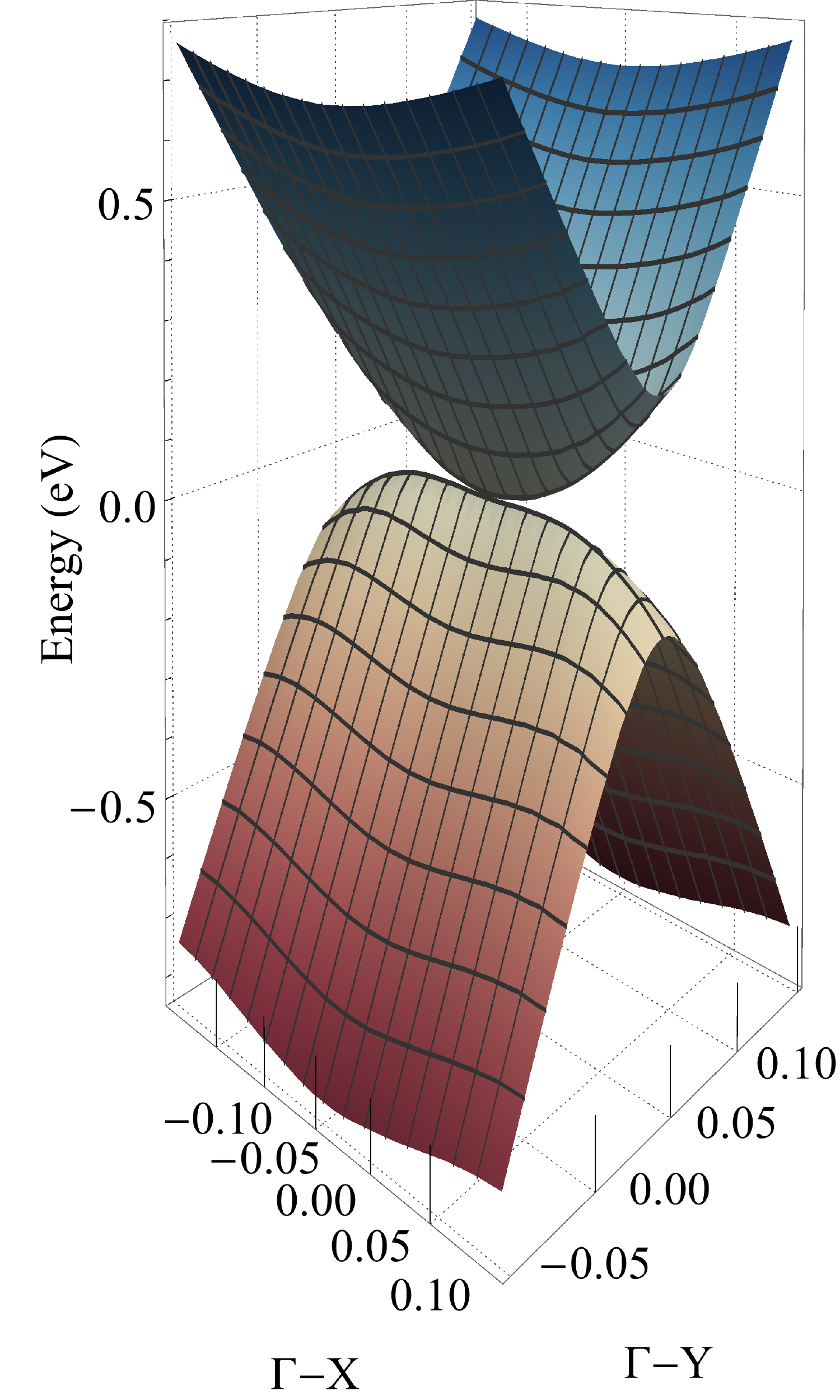}
  \label{fig:pbixx3ddiracbs}}
\end{subfloat}
&
\begin{subfloat}[$\Gamma$-point Dirac state]{
  \centering
\includegraphics[width=0.3\textwidth]{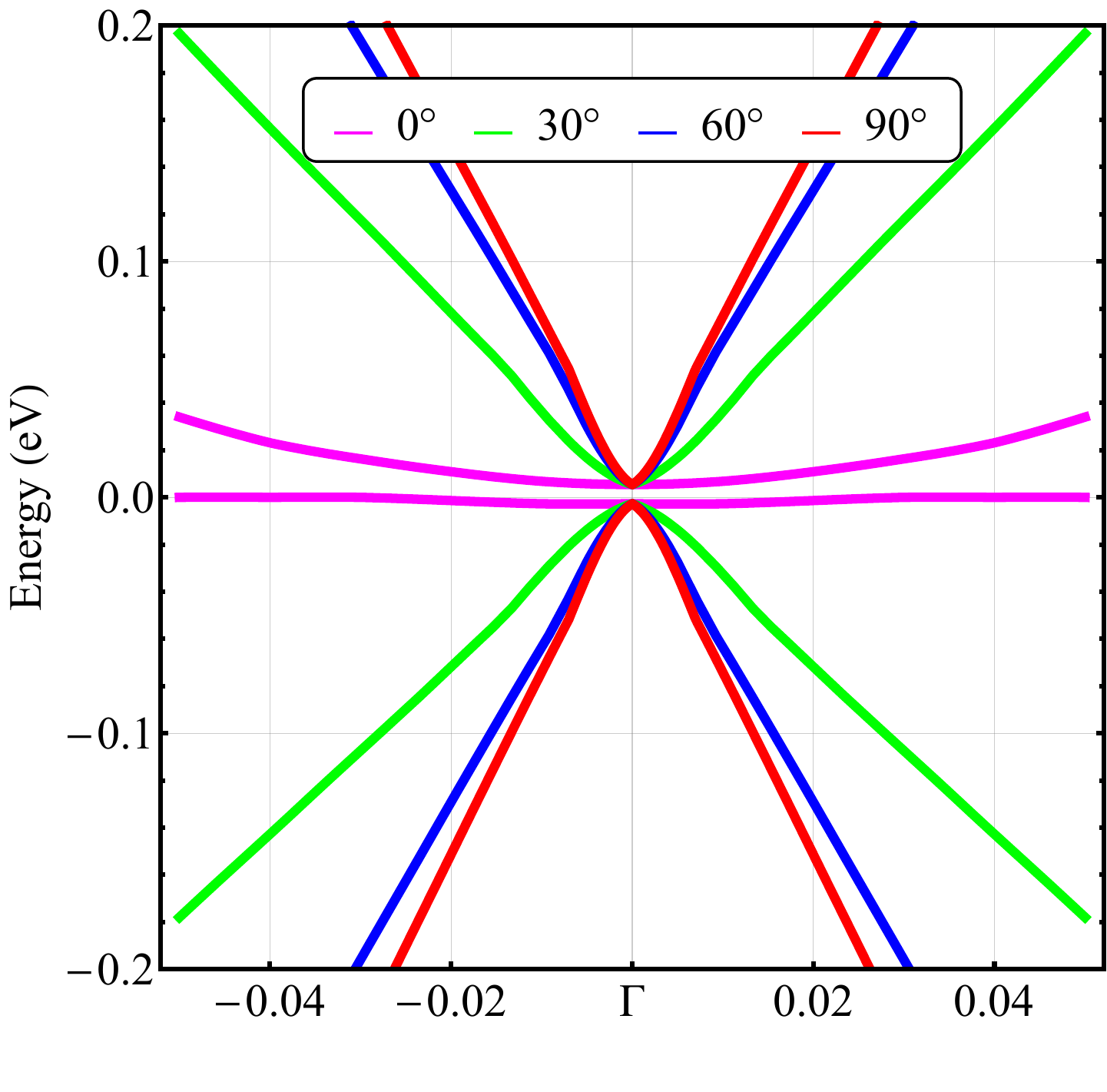}
  \label{fig:pbixxcutdiracbs}}
\end{subfloat}
\\
\begin{subfloat}[Bilayer P band structure $\varepsilon_{yy}=-5\%$]{
  \centering
\includegraphics[width=0.4\textwidth]{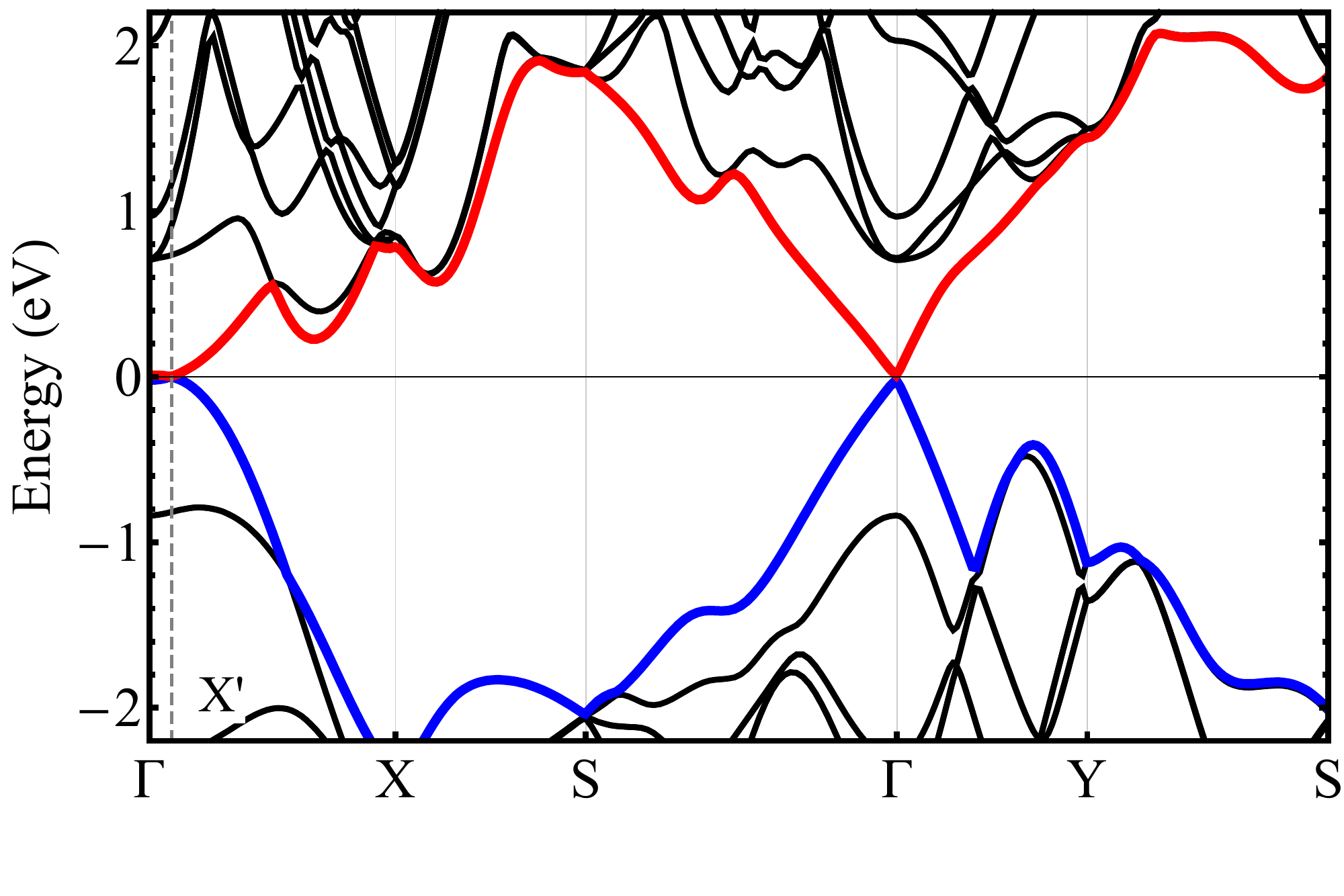}
    \label{fig:pbiyydiracbs}}
\end{subfloat}
&
\begin{subfloat}[]{
  \centering
\includegraphics[width=0.18\textwidth]{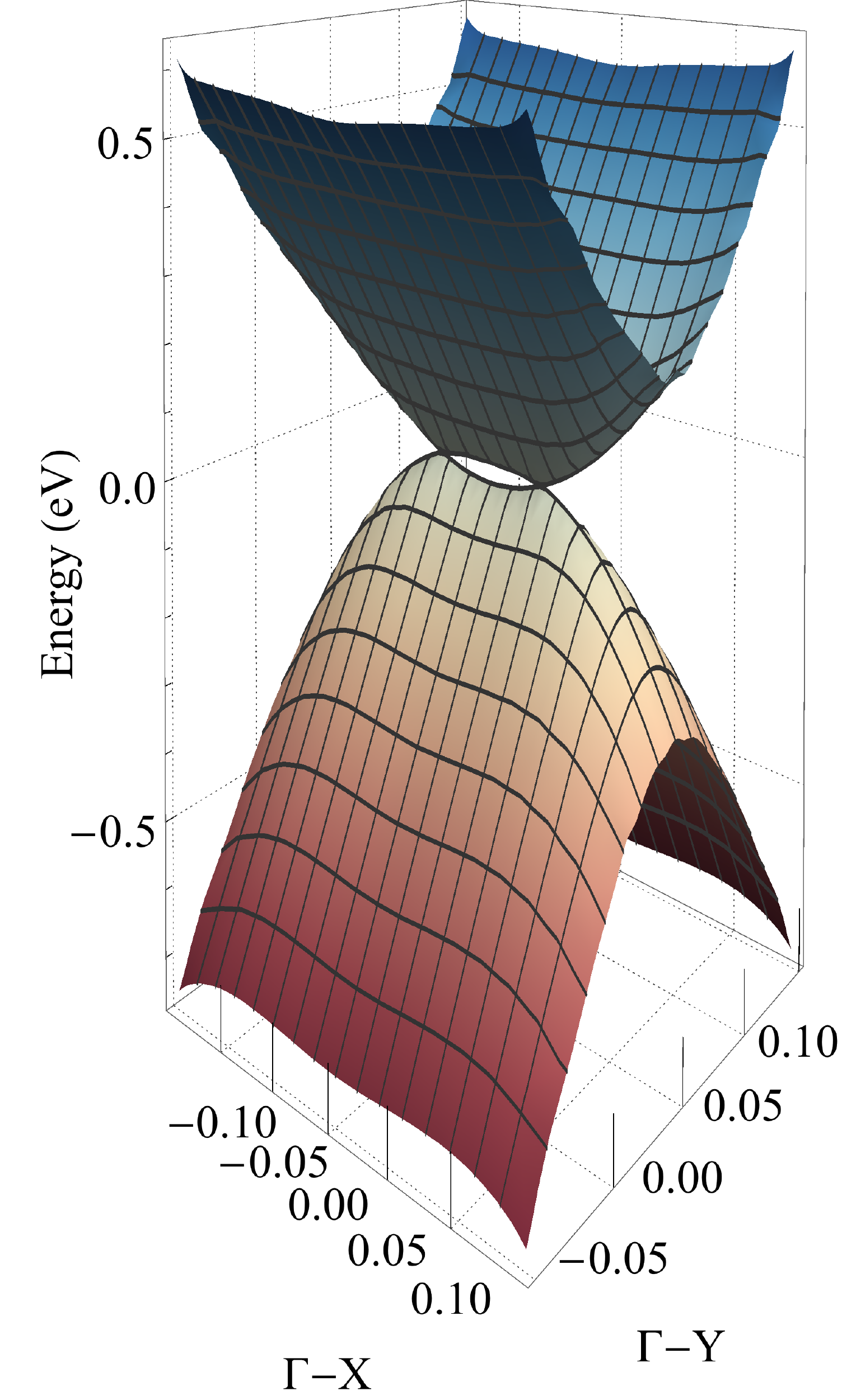}
  \label{fig:pbiyy3ddiracbs}}
\end{subfloat}
&
\begin{subfloat}[$X^\prime$-point Dirac state]{
  \centering
\includegraphics[width=0.3\textwidth]{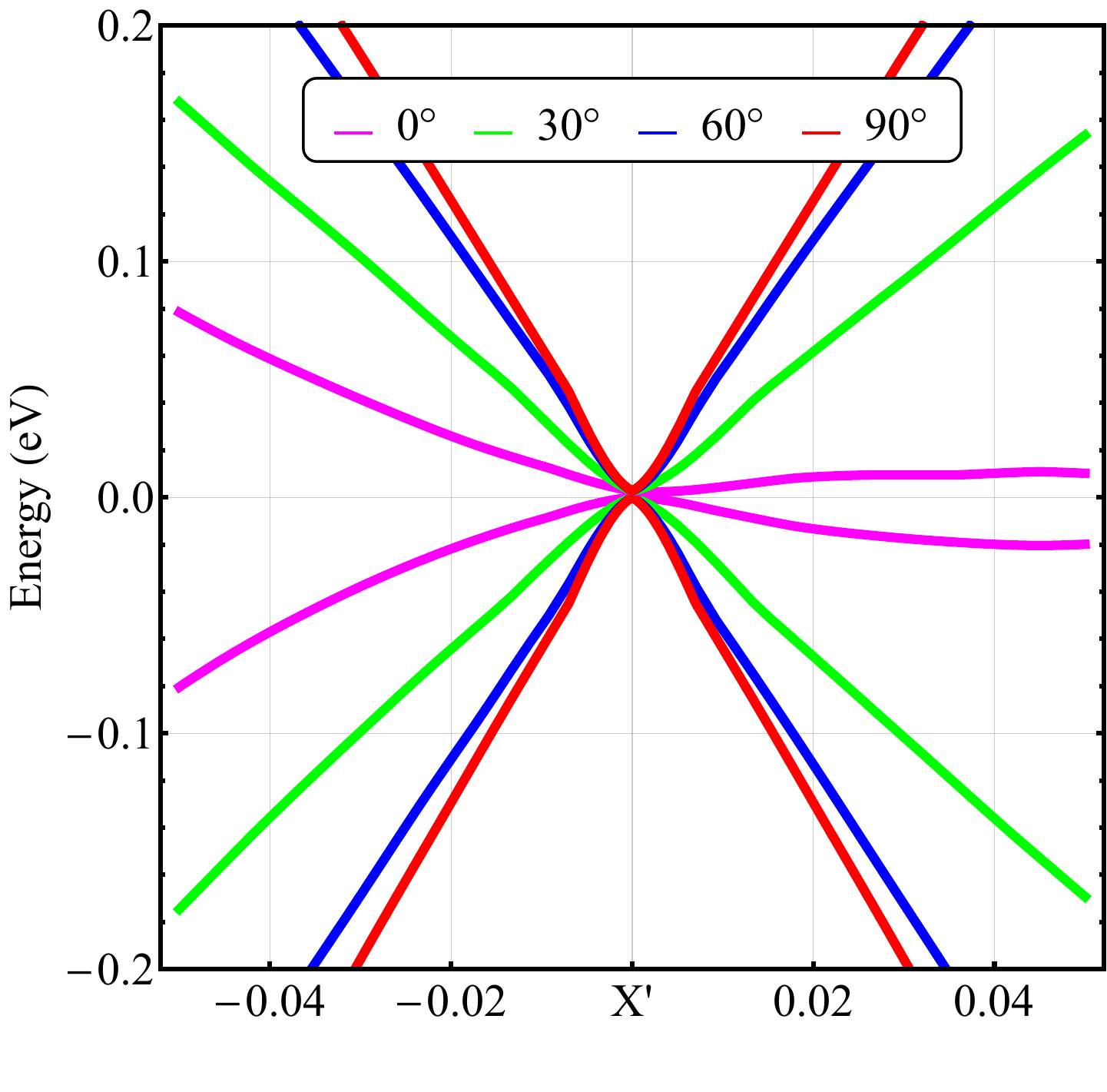}
  \label{fig:pbiyycutdiracbs}}
\end{subfloat}
\\
\begin{subfloat}[Monolayer Sb band structure $\varepsilon_{xx}=3\%$]{
  \centering
\includegraphics[width=0.4\textwidth]{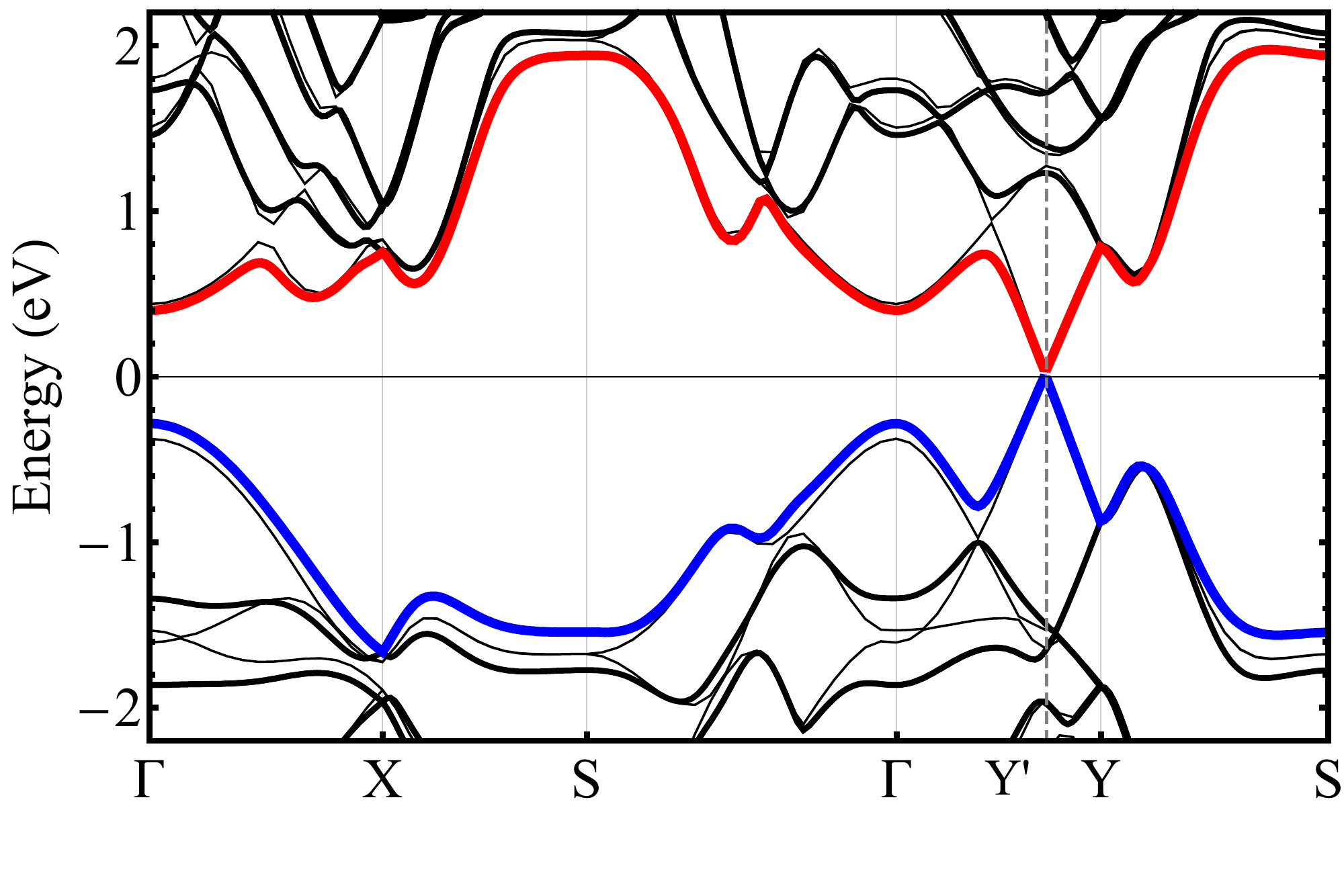}
  \label{fig:sbmonoxxdiracbs}}
\end{subfloat}
&
\begin{subfloat}[]{
  \centering
\includegraphics[width=0.18\textwidth]{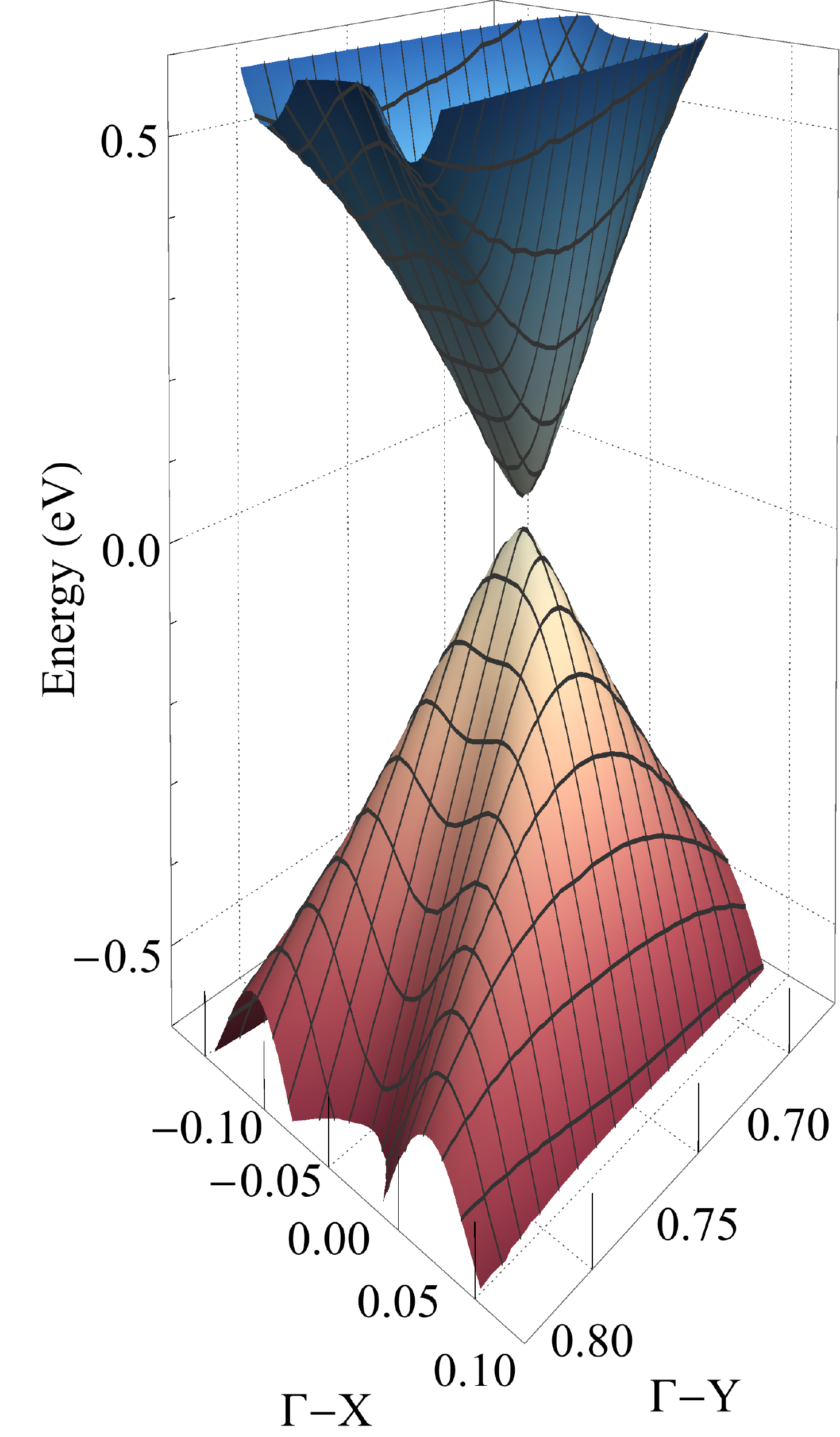}
  \label{fig:sbmonoxx3ddiracbs}}
\end{subfloat}
&
\begin{subfloat}[$Y^\prime$-point Dirac state]{
  \centering
\includegraphics[width=0.3\textwidth]{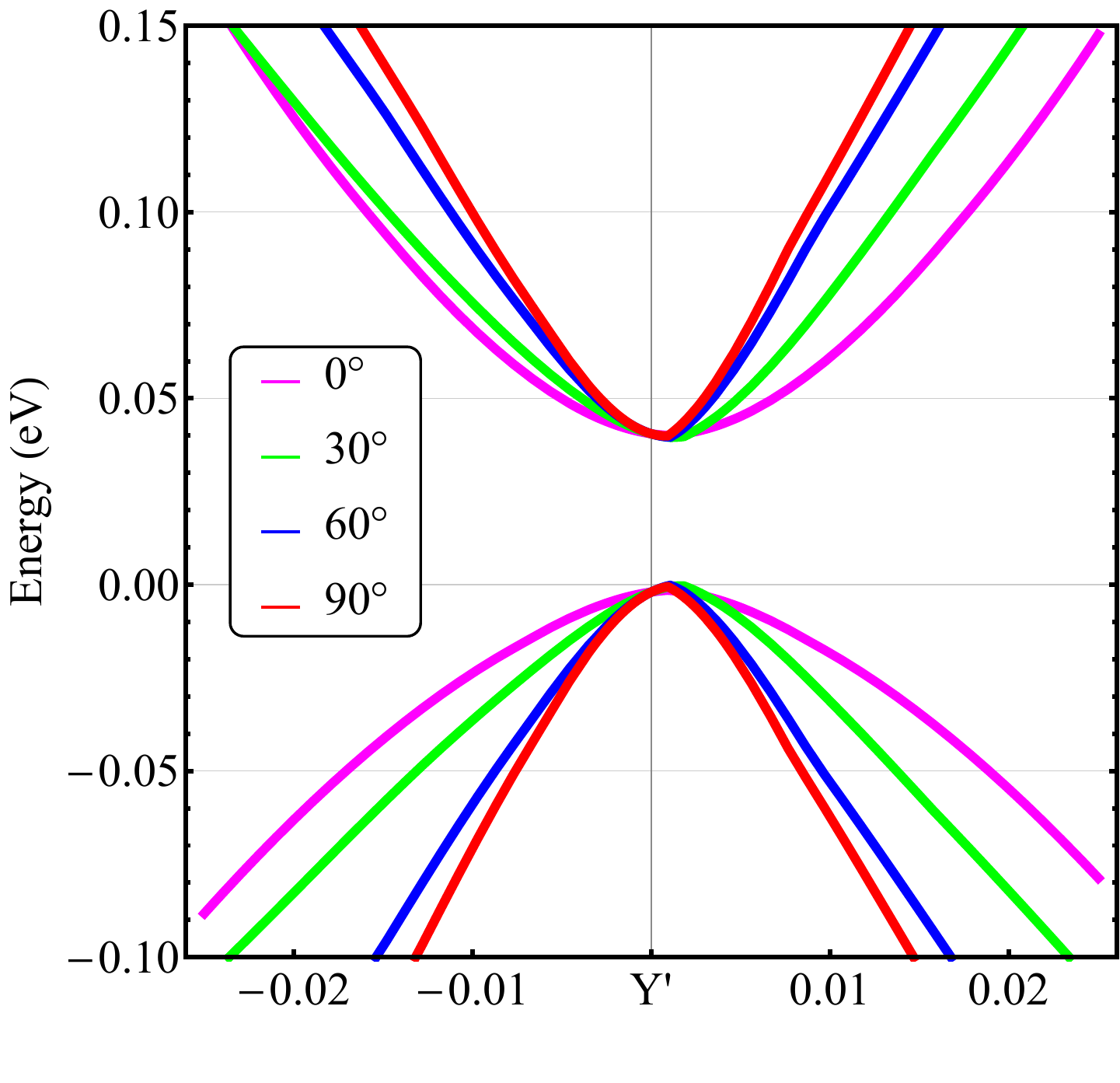}
  \label{fig:sbmonoxxcutdiracbs}}
\end{subfloat}
\end{tabular}
\caption{\edit{(Color online) 
 a) Band structure of bilayer P at $\varepsilon_{xx}=-5\%$ 
with SOC (thick lines), 
and without SOC (thin lines) 
b) three-dimensional bands about 
the predicted Dirac point at $\Gamma$
c) slices through the Dirac point 
at $0^\circ$, $30^\circ$, $60^\circ$ and $90^\circ$ 
relative to the $\Gamma-X$ line 
that indicate highly anisotropic conduction. 
[\ref{fig:pbiyydiracbs},\ref{fig:pbiyy3ddiracbs},\ref{fig:pbiyycutdiracbs}] 
Illustrates the same for bilayer P at $\varepsilon_{yy}=-5\%$, 
where the Dirac state occurs at the non-symmetry point $X^\prime$ 
along $\Gamma-X$.
[\ref{fig:sbmonoxxdiracbs},\ref{fig:sbmonoxx3ddiracbs},\ref{fig:sbmonoxxcutdiracbs}] 
Illustrates the same for monolayer Sb at $\varepsilon_{xx}=3\%$, 
where the possible Dirac state occurs at the non-symmetry point $Y^\prime$ 
along $\Gamma-Y$.
}}
\label{fig:dirac_states}
\end{figure*}

The relaxed Sb monolayer  
is found to possess 
an indirect band gap of $0.21$~eV 
along $\Gamma-Y$ (Fig.~\ref{fig:sbmonobg}), 
which is reasonably comparable to other 
PBE values 
0.28~\cite{doi:10.1021/acsami.5b02441}-0.37~\cite{PhysRevB.91.235446}~eV, 
although these have been obtained 
by including SOC.
For tensile strain along $\varepsilon_{yy}$
the band gap opens, 
suggesting an indirect-direct 
transition for strains above $6-7\%$,  
and it diminishes for compressive strains 
before finally closing at $\varepsilon_{yy}=-2\%$, 
where the material becomes a semi-metal.
Similarly, the indirect gap closes along $\Gamma-X$ 
at a compressive strain of 
$\varepsilon_{xx}=-2\%$ 
 at which monolayer Sb again becomes semi-metallic.
The indirect gap transitions to a direct gap
at $\varepsilon_{xx}=+1\%$ tensile strain 
and remains so until finally 
closing at $\varepsilon_{xx}=+4\%$, 
at which point we predict a \edit{potential}
Dirac \edit{state} along $\Gamma-Y$ 
at an off-symmetry point~\cite{PhysRevLett.102.166803}  
$Y^\prime$
(Fig.~\ref{fig:sbmonoxxdiracbs})
that has also been predicted 
in Ref.~\cite{Lu2016}.
Fig.~\ref{fig:sbmonoxxdiracbs} depicts the 
calculated band structure, 
in which it is shown that SOC 
preserves the Dirac \edit{state}  
but not does not open the band gap.
\edit{
The three-dimensional band structure 
about the Dirac point 
is shown in Fig.~\ref{fig:sbmonoxx3ddiracbs} 
in which the maximum charge velocity 
is  $v=4.31(1)\times10^6$~$\textrm{ms}^{-1}$ 
and occurs along a line parallel to the $\Gamma-Y$ direction 
(Fig.~\ref{fig:sbmonoxxcutdiracbs}).}
Moreover, 
the valence band at the $X$-point 
undergoes a Rashba 
splitting~\cite{1367-2630-17-5-050202} 
due to SOC, 
which is also predicted to occur in 
the monolayers of $\alpha$-P~\cite{PhysRevB.92.035135}, 
and $\beta$-Sb~\cite{Zhao2015,C6RA13101H}.
Finally, the electron effective masses  
experience a weak linear response to strain 
(Figs.~\ref{fig:sbmonomeff}~\&~\ref{fig:sbmonomhff}), 
while the hole effective masses along $\Gamma-X$ 
respond much more strongly 
to a rapid broadening or flattening 
of the valence band.

Furthermore, the relaxed bilayer phase 
is found to be semi-metallic 
where an indirect band gap 
opens at $\varepsilon_{yy}=+3\%$
tensile strain
and for uniaxial strains $<-1\%$ 
band-inversion at the $\Gamma$-point 
leads to to a 
fully-metallic state.
In addition, a possible Dirac \edit{state} emerges 
at a similar non-symmetry-point 
$Y^\prime$ along $\Gamma-Y$ 
for $\varepsilon_{xx}=+2\%$ 
tensile strain 
\edit{
(Fig.~\href{http://iopscience.iop.org/2053-1583/4/4/045018/media/Supplementary_Information.pdf}{\bl{S13}} 
in the Supplemental Material)}
and remains in place 
up to at least $+5\%$ strain.
\edit{
The maximum charge velocity 
$v=4.47(3)\times10^6$~$\textrm{ms}^{-1}$ 
is also along $\Gamma-Y$ 
and is approximately 
the same as that of the monolayer.
}
The effective masses  
(Figs.~\ref{fig:sbbimeff}~\&~\ref{fig:sbbimhff})
experience mild linear-response to strains 
prior to the transition to full metallicity, 
at which point a rapid flattening of the bands at the 
$\Gamma$-point suggesting strong electron localization
Beyond a compressive 
strain of  $\varepsilon_{yy}=-3\%$, 
however, 
at a stress of $\sim$0.3~GPa, 
the bilayer undergoes a structural 
transition and 
buckles in the puckered ($\vec{y}$) direction.
This buckled structure 
has a total energy 1.7~meV/atom 
lower than that of the relaxed state 
of the unperturbed $\alpha$-bilayer 
and 3.0~meV/atom lower when 
allowed to fully-relax,  
as shown above in 
Fig.~\ref{fig:buckled}.
This suggests the 
possible existence of a 
new structure that   
is attainable via strain.

Finally, bulk Sb is 
found to be completely metallic 
for all levels of strain explored in this work.
However, shear strains in this case 
do appear to have a significant
effect on the bands despite not opening a gap.
In summary, 
we predict possible non-symmetry-point 
Dirac \edit{states} in the strained monolayer \edit{and bilayer} of Sb, 
which are qualitatively unaffected by SOC, 
as well as Rashba splitting at the $X$-point 
\edit{in the monolayer}.
We also predict 
indirect-direct and indirect-semi-metallic 
transitions in the monolayer phase 
and a band gap opening in the bilayer phase.
Finally, we observe a 
buckled state induced in bilayer Sb 
at $-4\%$ compressive strain.
Bulk Sb was found to be metallic 
at all levels of strain explored.

In Table~\ref{table:bandgaps} below 
we present a summary of the calculated 
band gaps and effective 
charge carrier masses 
for the relaxed phases of each structure, 
and in Table~\ref{table:transitions} 
we provide a synopsis 
of the band gap and phase transitions 
of interest.
\edit{Finally, in Fig.~\ref{fig:dirac_states}, 
we present the band structures of bilayer P and monolayer Sb, 
that form a representative sample of 
the different Dirac points predicted at 
$\Gamma$, $X^\prime$ and $Y^\prime$,
as well the three-dimensional band structures 
about the region of the points.}
\begin{table}[h!]
\begin{ruledtabular}
\begin{tabular}{lccccc}
&$E_g$ (eV) &$Me_{\Gamma-X}$&$Me_{\Gamma-Y}$&$Mh_{\Gamma-X}$&$Mh_{\Gamma-Y}$\\
\hline
P\textsubscript{mono}	&0.9			&1.25(1)&0.16(1)&2.8(2)&0.14(1)\\
As\textsubscript{mono}	&$0.15^\star$	&1.16(1)&0.26(1)&1.09(1)&0.18(2)\\
Sb\textsubscript{mono}	&$0.2^\star$	&1.10(1)&0.28(1)&1.04(1)&0.19(2)\\
\hline
P\textsubscript{bi}		&0.4			&1.41(1)&0.19(1)&1.21(3)&0.15(2)\\
As\textsubscript{bi}		&0.45		&1.15(1)&0.24(1)&0.94(4)&0.17(2)\\
Sb\textsubscript{bi}		&$0^\dagger$	&1.16(1)&0.39(1)&0.99(4)&0.33(4)\\
\hline
P\textsubscript{bulk}		&$0^\ddagger$	&-&0.37(2)&-&0.21(2)\\
As\textsubscript{bulk}	&$0^\ddagger$	&-&0.36(1)&-&0.30(3)\\
Sb\textsubscript{bulk}	&$0^\ddagger$	&-&-&-&-
\end{tabular}
\end{ruledtabular}
\caption{(Color online) 
Kohn-Sham band gaps (eV), 
indicating the indirect semiconducting ($\star$), 
semi-metallic ($\dagger$) 
and metallic ($\ddagger$) states, 
as well as the charge-carrier 
effective masses ($m_0$) 
for each phase of P, As, Sb.
}
\label{table:bandgaps}
\end{table}

\begin{table}[h!]
\begin{ruledtabular}
\begin{tabular}{lllc}
&Transition&Direction&Strain (\%)\\
\hline
P\textsubscript{bi}		&D Gap $\to$ SM\textsuperscript{$\triangle$} 	&XX,YY&-5\\
					&D Gap $\to$ ID Gap 	&XX,YY&+2\\
P\textsubscript{bulk}		&SM
					$\to$ D Gap $\to$ ID Gap &XX,YY&+1$\to$+3\\
					&\edit{SM$\to$SM\textsuperscript{$\triangledown$}}							&\edit{XX,YY}&\edit{+2}\\
\hline
As\textsubscript{mono}	&ID $\to$ D Gap		&XX&-3\\
					&ID Gap $ \to$ SM		&XX&+2\\
					&ID Gap $ \to$ SM		&XX&+2\\
					&SM $\to$ SM\textsuperscript{$\triangle$}		&XX&+5\\
As\textsubscript{bi}		&D Gap $\to$ ID Gap $\to$ D Gap	&XX&+2$\to$+3\\
					&D Gap $\to$ ID Gap 	&YY&-3,+2\\
As\textsubscript{bulk}	&SM 
					$\to$ D Gap $\to $ ID Gap&XX&0$\to$+3\\
					&SM  
					$\to$ D Gap&YY&0$\to$+3\\
					&\edit{SM$\to$SM\textsuperscript{$\triangledown$}}
					&\edit{XX,YY}&\edit{+1}\\
\hline
Sb\textsubscript{mono}	&ID Gap $\to$ SM		&XX&-2\\
					&ID Gap $\to$ D Gap $\to$ SM\textsuperscript{$\triangle$}&XX&+2$\to$+4\\
					&ID Gap $\to$ SM		&YY&-2\\
Sb\textsubscript{bi}		&SM $\to$ SM\textsuperscript{$\triangle$}	&XX&+4\\
					&Structural Transition	&YY&-3\\
					&SM $\to$ ID Gap		&YY&+3
\end{tabular}
\end{ruledtabular}
\caption{
Summary of the  
band gap transitions  
including direct (D); 
indirect (ID); metallic (M);
and semi-metallic (SM), 
in particular those that 
indicate potential Dirac \edit{states} ($\triangle$), 
\edit{Weyl states ($\triangledown$)}, 
and the structural phase transition.
}
\label{table:transitions}
\end{table}

\subsection{Isotropic bulk properties}
In order to obtain these electronic states, 
and to ensure accurate strain-engineering, 
knowledge of the mechanical properties is paramount.
Therefore, in this section we review the 
mechanical response of the few-layer and bulk 
phases in order to compute the elastic properties, 
both isotropically averaged 
and as a function of orientation of applied in-plane stress.

The computed elements of the stiffness tensor $C$ 
in GPa 
of each structure 
are given in Table~\href{http://iopscience.iop.org/2053-1583/4/4/045018/media/Supplementary_Information.pdf}{\bl{S1}} 
in the Supplemental Material, 
where those pertaining to bulk P 
compare well to experiments~\cite{doi:10.1143/JPSJ.55.1196,doi:10.1143/JPSJ.60.1612} 
and similarly computed values~\cite{PhysRevB.86.035105,Wang2015}.
For the elements related  
to in-plane strains 
($c_{11}$, $c_{22}$, $c_{66}$, $c_{12}$)
we observe an expected increase in stiffness 
as the layer number increases 
and for increasing atomic number.
However, for other elements 
relating to out-of-plane and shear stresses
($c_{33}$, $c_{44}$, $c_{55}$, $c_{23}$, $c_{13}$) 
the stiffness actually increases 
in the bulk phase from P to As to Sb.

The Hill-averaged bulk properties 
are presented in Table~\ref{table:3dhill_data}, 
which compare well to other DFT values~\cite{PhysRevB.86.035105,Wang2015}, 
while our calculated  
bulk modulus for bulk P (37.2~GPa) 
is also within reasonable range of 
 the experimental values
(32.32~\cite{doi:10.1063/1.438523} - 
36.02~GPa~\cite{doi:10.1080/08957958908201013}).
We also observe that the bulk properties 
remain largely comparable for all the species, 
but generally decrease from P to As to Sb 
(except for the Poisson's ratio and bulk modulus, 
which are largest for As).
We also note that 
while bulk P has the largest in-plane responses, 
As and Sb have larger out-of-plane 
and shear responses, 
which enable the net isotropic properties 
for all the species to remain comparable overall.

\begin{table}
\begin{ruledtabular}
\begin{tabular}{lllll}
&$Y_H$ (GPa) &$G_H$ (GPa)&$\nu_H$&$\mathcal{B}_H$ (GPa)\\
\hline
P\textsubscript{bulk} &61.1 (70.3~\footnote{\label{f3:3dhill_data}Ref.~\cite{PhysRevB.86.035105}})	
&24.9 (29.4~\textsuperscript{\ref{f3:3dhill_data}})	&0.23 (0.30~\textsuperscript{\ref{f3:3dhill_data}})	&37.2 (38.5~\textsuperscript{\ref{f3:3dhill_data}}, 32.32~\footnote{\label{f1:3dhill_data}Ref.~\cite{doi:10.1063/1.438523}}, 36.02~\footnote{\label{f2:3dhill_data}Ref.~\cite{doi:10.1080/08957958908201013}})\\
As\textsubscript{bulk} &60.0	&23.8	&0.26	&41.4\\
Sb\textsubscript{bulk}&52.5	&21.1	&0.24	&33.8
\end{tabular}
\end{ruledtabular}
\caption{
The Hill-averaged 
Young's modulus $Y_H$,
shear modulus $G_H$ 
and bulk modulus $\mathcal{B}$ in GPa, 
and Poisson's ratio $\nu_H$ 
for bulk P, As and Sb 
compared to similarly calculated DFT~\cite{PhysRevB.86.035105} values 
and available experimental data~\cite{doi:10.1063/1.438523,doi:10.1080/08957958908201013}.
}
\label{table:3dhill_data}
\end{table}

\subsection{In-plane elastic properties}

\begin{figure*}
\begin{subfloat}[Monolayer P]{
\includegraphics[width=0.3\textwidth]{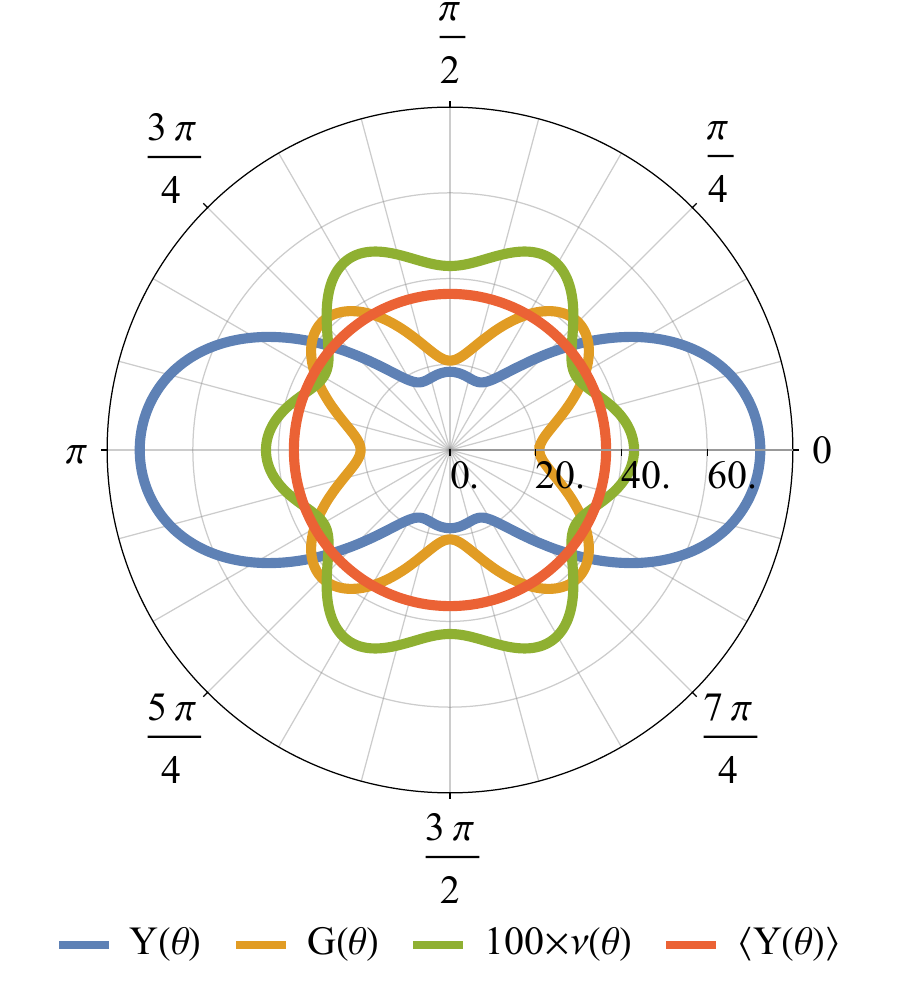}
  \label{fig:pmono1}}
\end{subfloat}
\begin{subfloat}[Monolayer As]{
\includegraphics[width=0.3\textwidth]{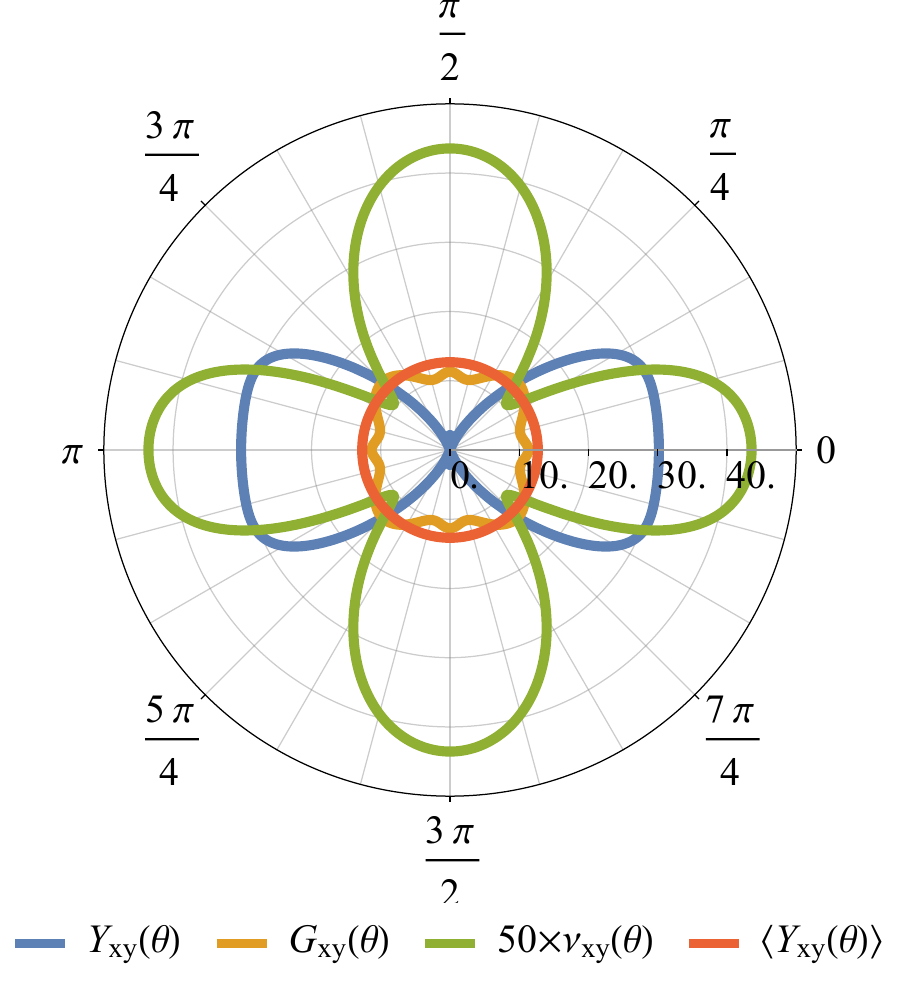}
  \label{fig:asmono1}}
\end{subfloat}
\begin{subfloat}[Monolayer Sb]{
\includegraphics[width=0.3\textwidth]{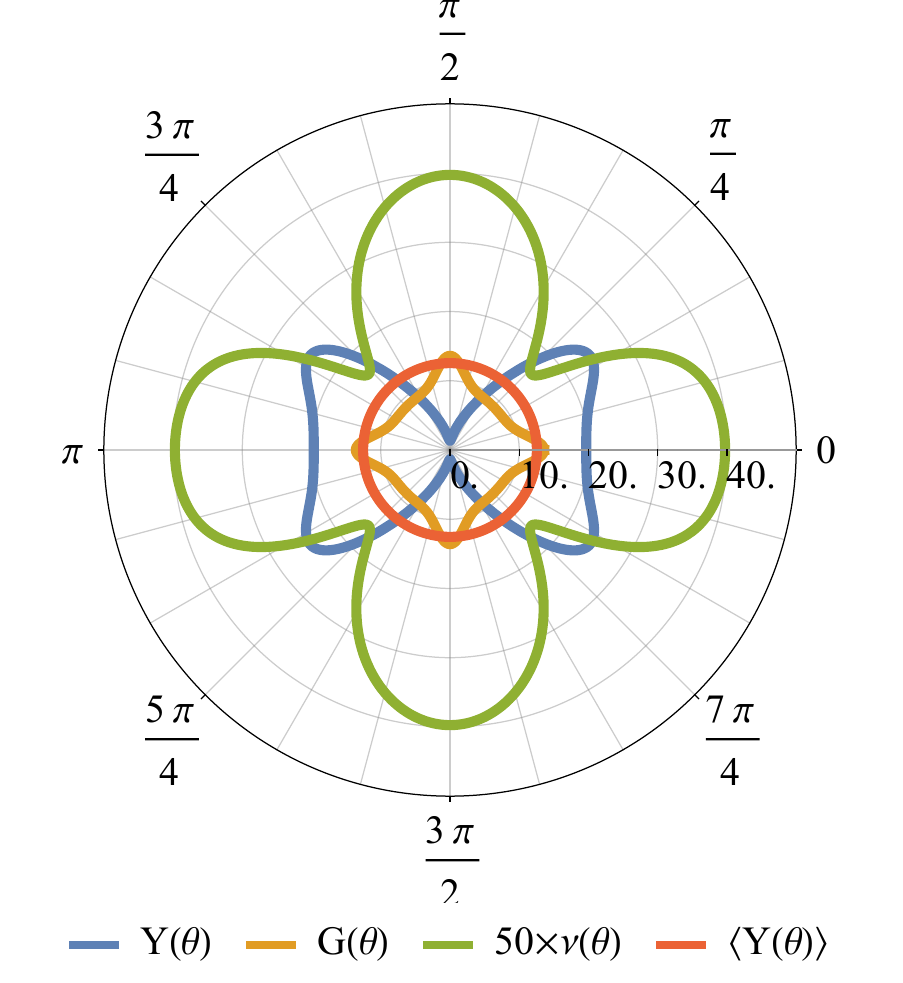}
  \label{fig:sbmono1}}
\end{subfloat}
\begin{subfloat}[Bilayer P]{
\includegraphics[width=0.3\textwidth]{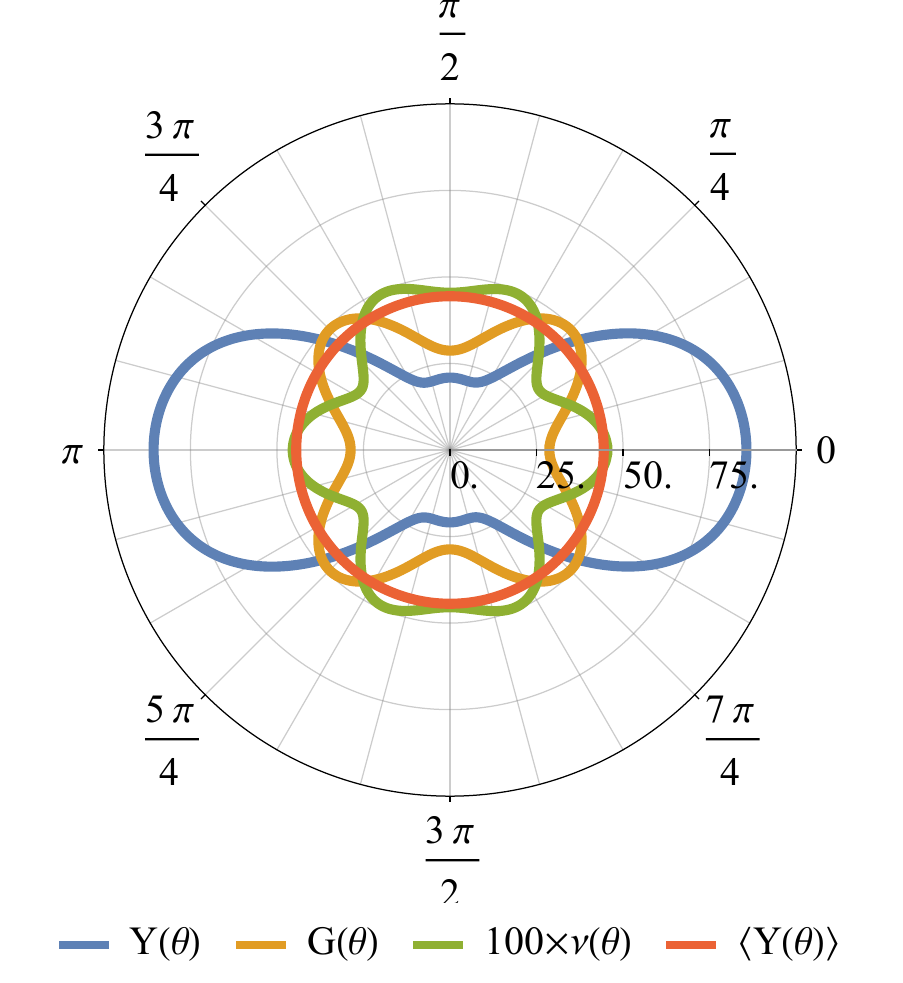}
  \label{fig:pbi1}}
\end{subfloat}
\begin{subfloat}[Bilayer As]{
\includegraphics[width=0.3\textwidth]{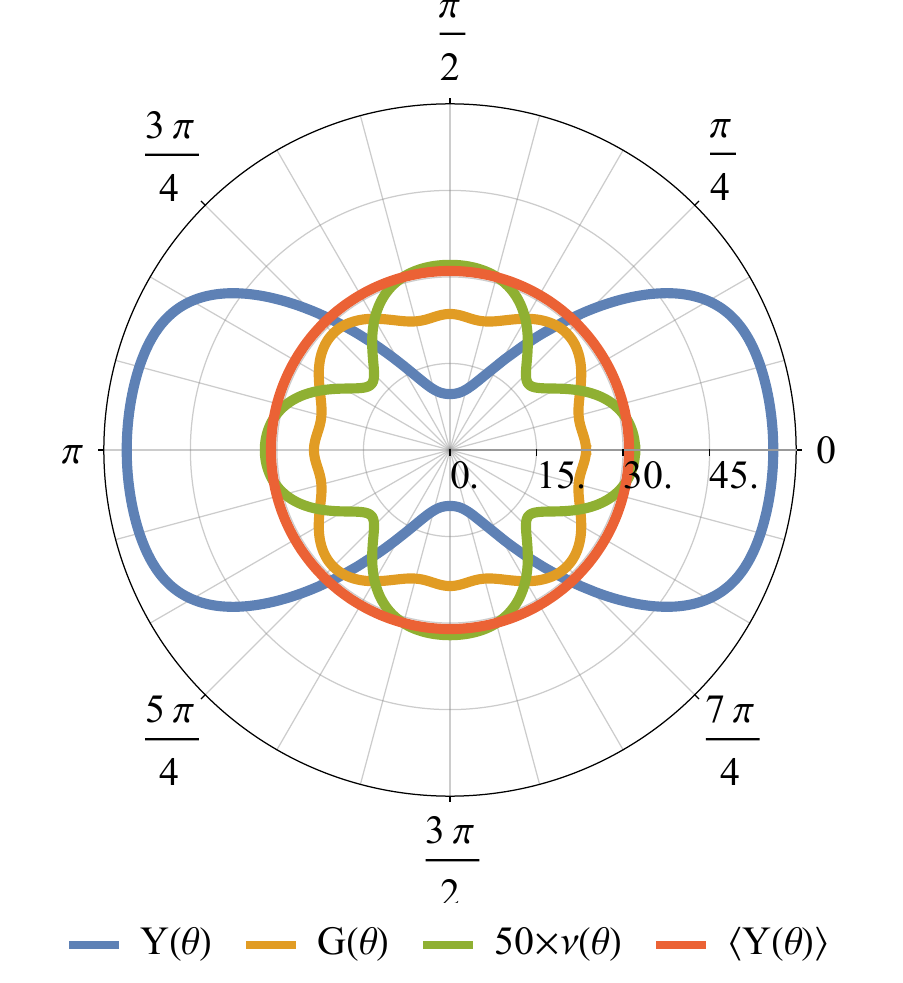}
  \label{fig:asbi1}}
\end{subfloat}
\begin{subfloat}[Bilayer Sb]{
\includegraphics[width=0.3\textwidth]{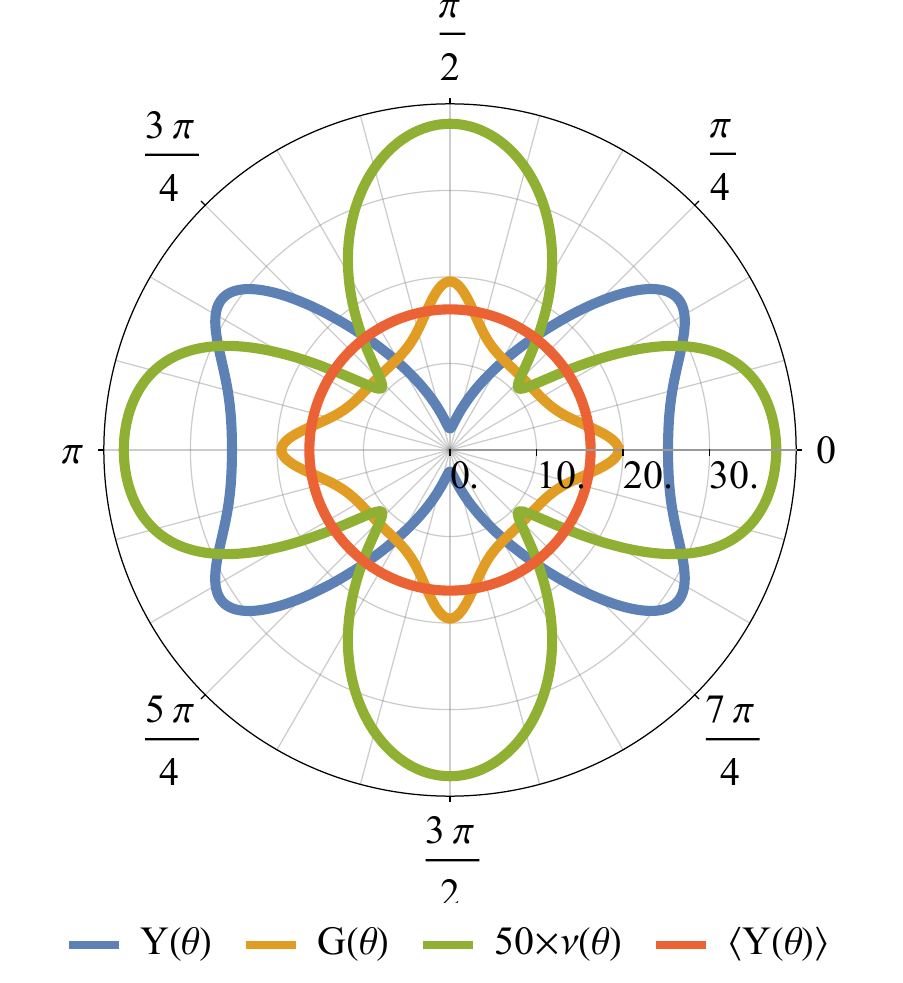}
  \label{fig:sbbi1}}
\end{subfloat}
\begin{subfloat}[Bulk P]{
\includegraphics[width=0.3\textwidth]{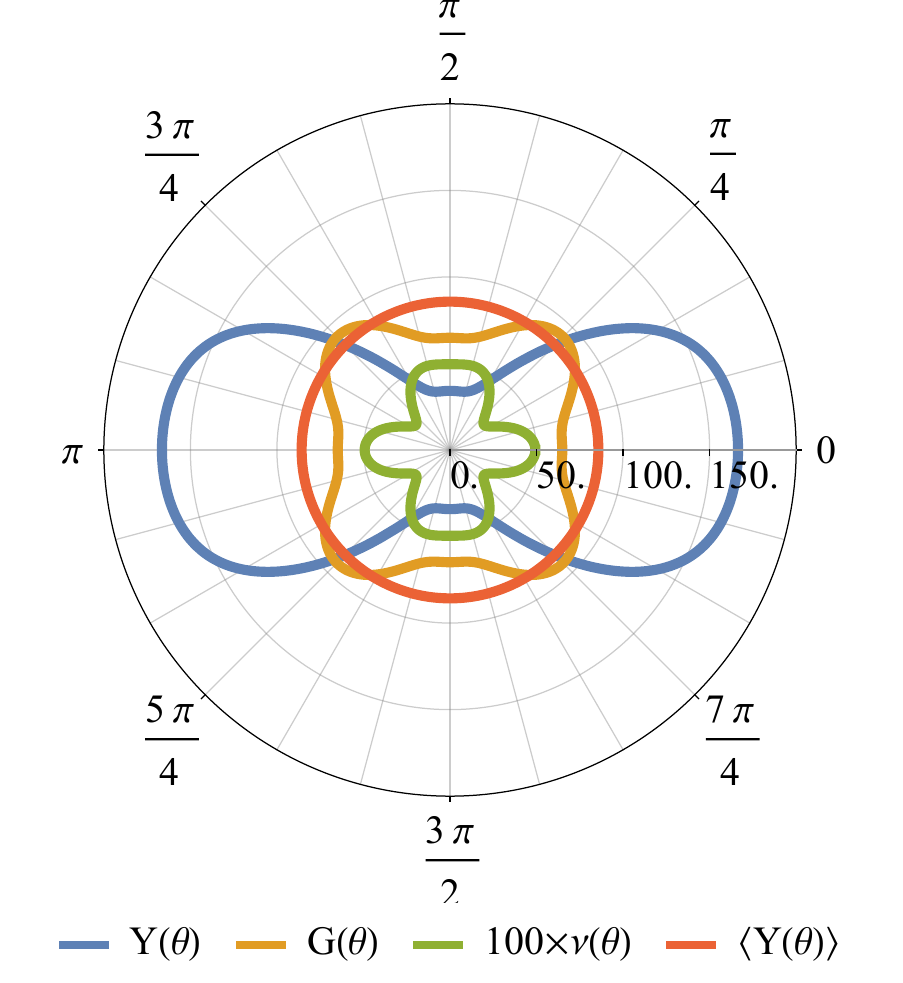}
  \label{fig:pbulk1}}
\end{subfloat}
\begin{subfloat}[Bulk As]{
\includegraphics[width=0.3\textwidth]{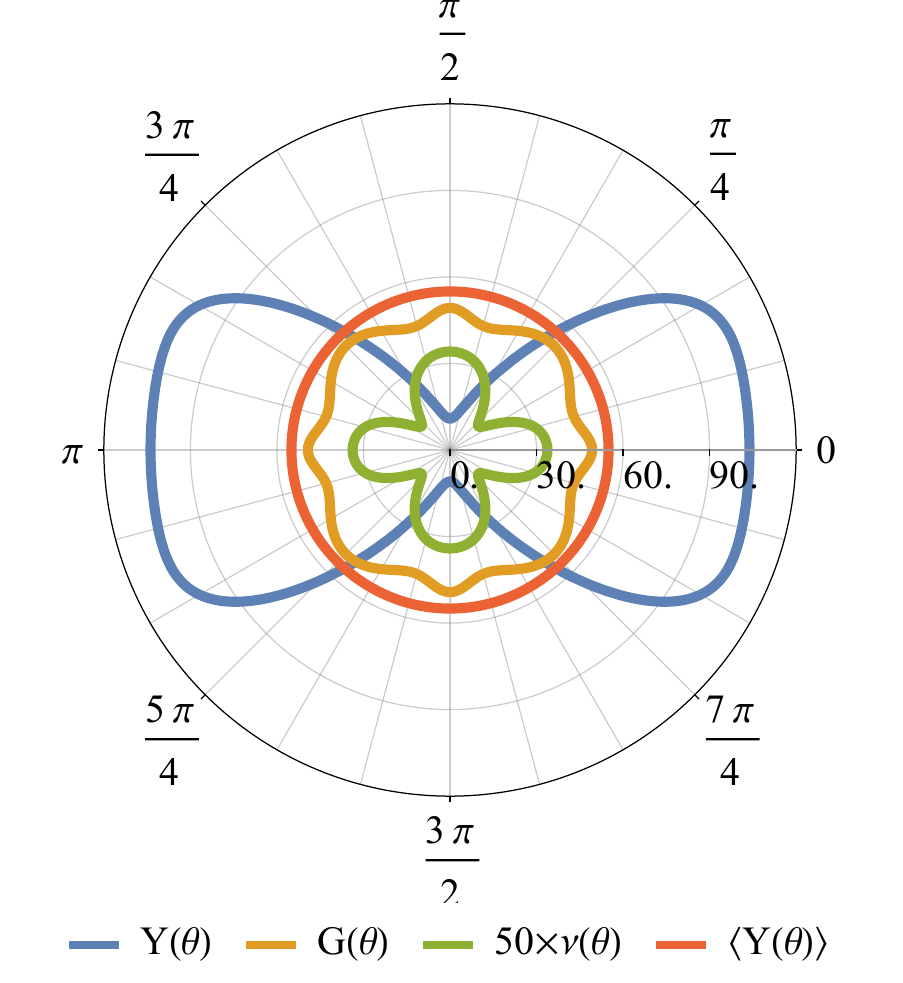}
  \label{fig:asbulk1}}
\end{subfloat}
\begin{subfloat}[Bulk Sb]{
\includegraphics[width=0.3\textwidth]{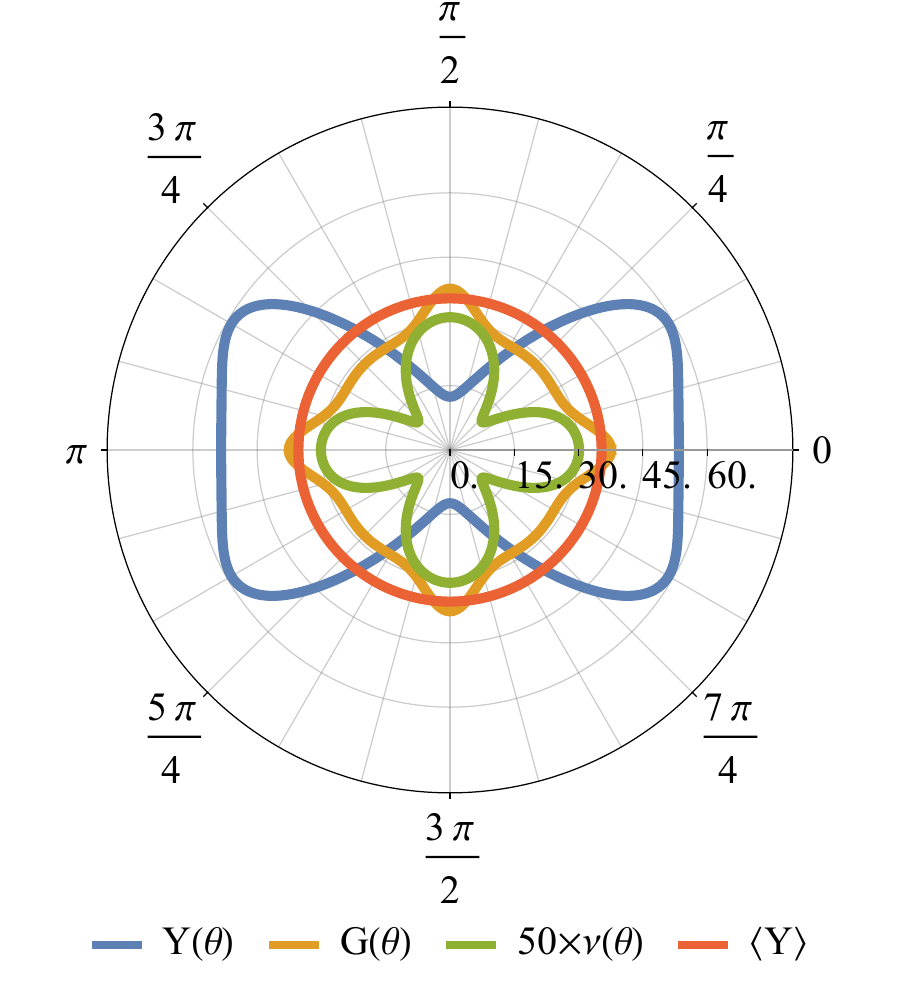}
  \label{fig:sbbulk1}}
\end{subfloat}
\caption{(Color online) 
In-plane functions for the 
Young's modulus $Y\left(\theta\right)$ (blue)
and its isotropic average $\left<Y\left(\theta\right)\right>$ 
(red) in units of GPa; 
shear modulus $G\left(\theta\right)$ in GPa (orange);
and Poisson's ratio $\nu\left(\theta\right)$ 
(green, scaled by 100 for P and 50 for As, Sb)
for each of the structures.
}
\label{fig:2d_in_plane_figs}
\end{figure*}

In section~\href{http://iopscience.iop.org/2053-1583/4/4/045018/media/Supplementary_Information.pdf}{\bl{S1}} of the 
Supplemental Material we re-derive the 
equations for the elastic properties as a 
function of the in-plane orientation angle $\theta$, 
defined in Fig.~\ref{fig:phos_structure}, 
as outlined in Ref.~\cite{jones1975mechanics}.
These functions are the plotted in 
Fig.~\ref{fig:2d_in_plane_figs} 
and include  
the Young's modulus $Y\left(\theta\right)$ 
and it's average $\left<Y\left(\theta\right)\right>$,
the shear modulus $G\left(\theta\right)$,
and Poisson's ratio $\nu\left(\theta\right)$.
The experimental Young's modulus of 130~GPa, 
determined in Ref.~\cite{PMID:26469634}
via the ROM of a nano-flake-polymer composite, 
lies precisely between the average in-plane 
bulk Young's modulus 85.7~GPa 
and the elastic stiffness in 
the zigzag ($\vec{x}$) direction 187.9~GPa, 
thus fitting the results of our model reasonably well. 

The anisotropy of the elastic properties  
is also apparent in the mechanical profiles, 
particularly with regard to the Young's modulus, 
which has a 2-fold symmetry 
about the $x$-axis, 
in contrast to the 
shear modulus and Poisson's ratios, 
which display 4-fold symmetry about both the axes 
(except for the Poisson's ratio of P, 
which remains 2-fold symmetric).
Indeed, for a given species, 
the general shape of each profile 
is approximately preserved 
with respect to the number of layers, 
while the range of each 
property tends to increase.
This discovery is advantageous 
in the strain-engineering of nano-flakes 
since one may now forecast in advance the 
response of a material to in-plane strain, 
once the underlying profile 
and number of layers are known.

Another interesting feature is that 
the extrema of the elastic functions 
do not necessarily coincide 
with the coordinate-axes.
For instance the Young's modulus  maximum 
for monolayer Sb occurs at $22^{\degree}$.
Table~\ref{table:2d_extrema} summarizes   
the global minima and maxima of each function 
and the angles at which they occur.
While most of the function extrema occur 
expectedly at $0^{\degree}$, $45^{\degree}$ or $90^{\degree}$, 
many are incident away from the coordinate-axes.
This result lends further insight into  
the mechanical anisotropy of the 
orthorhombic group-$V$ materials.

\begin{table*}
\begin{ruledtabular}
\begin{tabular}{lccccc|ccccc|ccccc}
&$Y_{\min}$&$\theta$&$Y_{\max}$&$\theta$&$\left<Y\right>$
&$G_{\min}$&$\theta$&$G_{\max}$&$\theta$&$\left<G\right>$
&$\nu_{\min}$&$\theta$&$\nu_{\max}$&$\theta$&$\left<\nu\right>$\\
\hline
P\textsubscript{mono}	&17.1 & $69^{\degree}$ &72.4 & $0^{\degree}$ &36.4
&20.9& $90^{\degree}$&42.2 &$45^{\degree}$&31.0&0.3& $31^{\degree}$&0.5 &$63^{\degree}$&0.5\\
As\textsubscript{mono}	&-2.1 &$90^{\degree}$&30.5& $18^{\degree}$&12.7
&10.4 &$77^{\degree}$&13.8 &$45^{\degree}$&11.2
&0.2 &$39^{\degree}$&0.9 &$90^{\degree}$&0.4\\
Sb\textsubscript{mono}&1.4 &$90^{\degree}$&24.4 &$22^{\degree}$&12.6
&8.9 &$45^{\degree}$&13.6 &$90^{\degree}$&9.4
&0.3 &$43^{\degree}$&0.8& $90^{\degree}$&0.6\\
\hline
Pb\textsubscript{bi}		&20.5 &$75^{\degree}$&85.6 &$0^{\degree}$&44.4
&28.1& $90^{\degree}$&49.2& $45^{\degree}$&37.8
&0.31& $34^{\degree}$&0.5& $68^{\degree}$&0.4\\
As\textsubscript{bi}	&9.7& $90^{\degree}$&56.0 &$0^{\degree}$&31.0
&23.0 &$78^{\degree}$&28.6& $45^{\degree}$&24.3
&0.4 &$40^{\degree}$&0.6& $90^{\degree}$&0.5\\
Sb\textsubscript{bi}		&2.6 &$90^{\degree}$&31.6&$33^{\degree}$&16.3
&11.8 &$45^{\degree}$&19.5 &$90^{\degree}$&12.8
&0.2&$42^{\degree}$&0.8& $90^{\degree}$&0.5\\
\hline
P\textsubscript{bulk}		&34.2& $87^{\degree}$&166.4& $0^{\degree}$&85.7
&64.9 &$87^{\degree}$&92.4 &$45^{\degree}$&74.5
&0.2& $36^{\degree}$ &0.5& $75^{\degree}$&0.3\\
As\textsubscript{bulk} 	&10.9 &$90^{\degree}$&105.1 &$20^{\degree}$ &55.0
&44.4&$73^{\degree}$&51.1& $45^{\degree}$&66.7
&0.3 &$40^{\degree}$&0.7& $90^{\degree}$ &0.4\\
Sb\textsubscript{bulk} 	&12.4 &$90^{\degree}$&58.9& $30^{\degree}$&35.4
&28.1& $59^{\degree}$&28.2& $45^{\degree}$&63.2
&0.2& $41^{\degree}$&0.6& $90^{\degree}$&0.4
\end{tabular}
\end{ruledtabular}
\caption{
Summary of the minima and maxima 
of the Hill-averaged in-plane 
Young's modulus $Y\left(\theta\right)$ (GPa),  
shear modulus $G\left(\theta\right)$ (GPa), 
and Poisson's ratio $\nu\left(\theta\right)$ 
as well as the angle $\theta$ 
with respect to the $\vec{x}$-direction (zigzag) 
at which they occur in degrees,
and their in-plane averages.
}
\label{table:2d_extrema}
\end{table*}

The emergence of a negative Young's modulus 
of $-2.1$~GPa in monolayer As (Fig.~\ref{fig:asmono1}), 
at first glance, may give cause for concern.
It arises due to a negative Voigt estimate for 
the Young's modulus at $90^{\degree}$ 
where  
$(c_{22}^2-c_{12}^2)/c_{22} = -8.9$~GPa 
(since $c_{12}>c_{22}$), 
which is larger in absolute magnitude 
than the Reuss estimate at $90^{\degree}$ 
given by $1/s_{22}=4.7$~GPa  
and results in a net negative Hill-average.
In this instance, we surmise that 
either the assumptions of the 
Voigt model break down, 
or the Hill-method
is not universally appropriate 
in arbitrary directions in-plane 
and a more robust averaging 
scheme must be employed.
Nevertheless, 
the qualitative in-plane 
functions and their isotropic averages 
remain physically meaningful 
and in general can provide valuable physical insight.

In contrast to the isotropic averages for 
the bulk properties in Table~\ref{table:3dhill_data}, 
which remained largely comparable for P, As and Sb, 
a much clearer trend across the species emerges 
once we have eliminated contributions from 
the out-of-plane and shear stresses.
In this instance, P clearly possesses superior 
in-plane mechanical strength in both moduli, 
which decrease with the number of layers as expected 
As and Sb are largely similar in the monolayer, 
though less so in the bilayer and bulk phases, 
where they are stronger in the $\vec{x}$-direction.
In contrast,  
the Poisson's ratio tends to remain relatively stable 
aside from generally decreasing with increasing 
number of layers.

In summary, there exists in 
the elastic properties 
a broad range of responses, 
profile shapes and behavior 
that is reflective of the underlying 
anisotropic crystal structure, 
which also strongly depend on 
the number of layers with 
P typically the stiffest 
and Sb the most flexible.
In general we find 
the shapes of the in-plane response profiles 
to be conveniently consistent, 
which implies that, for a given number of layers, 
the in-plane elastic response 
of a nano-flake can be 
reliably estimated {\it a priori}.

\section{Conclusion}
\label{sec:conclusion}
We have extensively explored 
the mechanical and electronic properties 
of P, As and Sb in their few-layer and bulk phases.
We have identified several band gap transitions 
in almost all of the structures.
\edit{The SOC tended to close the 
bands by $\sim0.05$~eV but did not   
alter any of our qualitative findings.}
We also predict the existence of Dirac \edit{states}  
in the strained phases of 
monolayer As and Sb,  
bilayer P, As and Sb 
as well as possible Weyl \edit{states} in 
bulk P and As 
for moderate levels of strain.
\edit{
The linear-dispersion was observed 
along $\Gamma-Y$ of each 
of the predicted Dirac or Weyl states, 
corresponding to the direction 
of softest mechanical response in the puckered direction.}
\edit{
The maximum charge velocity is calculated 
to be over 
$10^6$~$\textrm{ms}^{-1}$.
In particular, 
for bilayer P and few-layer As 
we predict highly anisotropic conductivity 
dominated by ballistic transport 
along the puckered direction  
that is indicative of effective, one-dimensional conduction.}
 We predict that an appropriate strain 
 could yield these effects in experiments.

We also observe 
the existence of a notable 
buckled state of compressed 
bilayer Sb at $-4\%$ strain.
Finally, the angular-resolved 
elastic properties 
as well as the  
stress-dependence of the 
Kohn-Shame band gaps 
and charge-carrier effective masses 
revealed highly anisotropic behavior, 
spanning a broad range of values, 
and angular-dependent behavior 
that has become characteristic of 
these group-V layered structures. 
Moreover, the critical stresses 
at which these transitions occur 
are expected to be experimentally 
accessible and highly switchable, 
paving the way for possible 
verification in the near future.
Thus, the group-V layered materials 
are poised to become central to 
the next generation of electronic devices with 
potential novel applications in field-effect transistors; 
batteries; gas-sensors and opto-electronic devices.

\acknowledgements 
The authors would like to sincerely thank 
Damien Hanlon, Claudia Backes, Conor Boland,  Jonathan Coleman, 
Beata Szydlowska, Gaozhong Wang, and Werner Blau 
for their helpful discussions relating to 
the work conducted for this Article.
This work was enabled by Science Foundation Ireland (SFI) 
funded centre AMBER (SFI/12/RC/2278).
All calculations were performed on the Kelvin cluster maintained by 
Trinity College Dublin Research IT and funded through grants from SFI.

%

\end{document}